\documentclass[twocolumn]{aastex63}

\usepackage[flushleft]{threeparttable}
\usepackage{amsmath}
\usepackage[T1]{fontenc}
\usepackage[utf8]{inputenc}

\graphicspath{{./}{figures/}}

\newcommand{\feh}{[Fe/H]}
\newcommand{\alphafe}{[$\alpha$/Fe]}
\newcommand{\kms}{km s$^{-1}$}
\newcommand{\teffphot}{$T_{\rm{eff,phot}}$}
\newcommand{\loggphot}{$\log$ $g$}
\newcommand{\fehphot}{[Fe/H]$_{\rm{phot}}$}
\newcommand{\teff}{$T_{\rm{eff}}$}
\newcommand{\logg}{$\log$ $g$}

\received{2020 May 2}
\revised{YYY}
\accepted{ZZZ}
\submitjournal{ApJ}

\shorttitle{M31's Inner Halo Abundances}
\shortauthors{Escala et al.}

\begin{document}
\raggedbottom

\title{Elemental Abundances in M31: Properties of the Inner Stellar Halo\footnote{The data presented herein were obtained at the W. M. Keck Observatory, which is operated as a scientific partnership among the California Institute of Technology, the University of California and the National Aeronautics and Space Administration. The Observatory was made possible by the generous financial support of the W. M. Keck Foundation.}}

\correspondingauthor{I. Escala}
\email{iescala@carnegiescience.edu}

\author[0000-0002-9933-9551]{Ivanna Escala}
\altaffiliation{Carnegie-Princeton Fellow}
\affiliation{Department of Astronomy, California Institute of Technology, 1200 E California Blvd, Pasadena, CA 91125, USA}
\affiliation{The Observatories of the Carnegie Institution for Science, 813 Santa Barbara St, Pasadena, CA 91101, USA}
\affiliation{Department of Astrophysical Sciences, Princeton University, 4 Ivy Lane, Princeton, NJ 08544, USA}

\author[0000-0001-6196-5162]{Evan N. Kirby}
\affiliation{Department of Astronomy, California Institute of Technology, 1200 E California Blvd, Pasadena, CA 91125, USA}

\author[0000-0003-0394-8377]{Karoline M. Gilbert}
\affiliation{Space Telescope Science Institute, 3700 San Martin Drive, Baltimore, MD 21218, USA}
\affiliation{Department of Physics \& Astronomy, Bloomberg Center for Physics and Astronomy, John Hopkins University, 3400 N. Charles St, Baltimore, MD 21218, USA}

\author[0000-0002-3233-3032]{Jennifer Wojno}
\affiliation{Department of Physics \& Astronomy, Bloomberg Center for Physics and Astronomy, John Hopkins University, 3400 N. Charles St, Baltimore, MD 21218, USA}

\author[0000-0002-6993-0826]{Emily C. Cunningham}
\affiliation{Center for Computational Astrophysics, Flatiron Institute, 162 5th Ave, New York, NY 10010, USA}

\author[0000-0001-8867-4234]{Puragra Guhathakurta}
\affiliation{UCO/Lick Observatory, Department of Astronomy \& Astrophysics, University of California Santa Cruz, 
 1156 High Street, 
 Santa Cruz, California 95064, USA}

\begin{abstract}
We present measurements of \feh\ and \alphafe\ for \replaced{129}{128} individual red giant branch stars (RGB) in the stellar halo of M31, including its Giant Stellar Stream (GSS), obtained using spectral synthesis of low- and medium-resolution Keck/DEIMOS spectroscopy ($R \sim 3000$ and 6000, respectively). We observed four fields in M31's stellar halo (at projected radii of 9, 18, 23, and 31 kpc), as well as two fields in the GSS (at 33 kpc). In combination with existing literature measurements, we have increased the sample size of \feh\ and \alphafe\ measurements \added{from 101} to a total of \replaced{230}{229} individual M31 RGB stars. From this sample, we investigate the chemical abundance properties of M31's inner halo, finding $\langle$\feh$\rangle$ = $-$1.08 $\pm$ 0.04 and $\langle$\alphafe$\rangle$ = 0.40 $\pm$ 0.03. Between 8--34 kpc, the inner halo has a steep \feh\ gradient ($-$0.025 $\pm$ 0.002 dex kpc$^{-1}$) and negligible \alphafe\ gradient, where substructure in the inner halo is systematically more metal-rich than the smooth component of the halo at a given projected distance. Although the chemical abundances of the inner stellar halo are largely inconsistent with that of present-day dwarf \added{spheroidal (dSph)} satellite galaxies of M31, we identified \replaced{37}{22} RGB stars kinematically associated with the smooth \added{component of the} stellar halo that have chemical abundance patterns similar to M31 \replaced{dwarf galaxies}{dSphs}.
\deleted{This stellar population is more metal-poor ($\langle$\feh$\rangle$ = $-$1.73 $\pm$ 0.05) and less $\alpha$-enhanced ($\langle$\alphafe$\rangle$ = 0.16 $\pm$ 0.07) than the inner stellar halo as a whole. Comparisons to abundances of the Milky Way (MW) halo indicate that M31 and the MW may both have populations of low-$\alpha$ stars with halo-like kinematics, whereas comparisons to the Magellanic Clouds suggest M31's low-$\alpha$ population had similarly low star formation efficiency}
We discuss formation scenarios for M31's halo, concluding that these \replaced{low-$\alpha$}{dSph-like} stars may have \replaced{an accretion origin}{been accreted from galaxies of similar stellar mass and star formation history, or of higher stellar mass and similar star formation efficiency}.
\end{abstract}



\section{Introduction} \label{sec:intro}

In the $\Lambda$CDM cosmological paradigm, $L_\star$ galaxies like the Milky Way (MW) and Andromeda (M31) form through hierarchical assembly (e.g., \replaced{\citealt{SearleZinn1978}}{\citealt{WhiteRees1978}}). Debris from mergers across cosmic time are deposited within the extended stellar halo, where they remain observationally identifiable \replaced{for many dynamical times}{because the phase-mixing timescales in the outskirts of galaxies are long compared to the age of the universe}
(e.g., \citealt{Helmi1999,BullockJohnston2005}). Simulations of stellar halo formation in MW-like galaxies have shown that the mass and accretion time distributions of progenitor dwarf galaxies can imprint strong chemical signatures in a galaxy's stellar population (e.g., \citealt{Robertson2005,Font2006b,Johnston2008,Zolotov2010,Tissera2012}), particularly in terms of \feh\ and \alphafe. Measurements of $\alpha$-element abundance (O, Ne, Mg, Si, S, Ar, Ca, and Ti) and iron (Fe) abundance encode information concerning the relative timescales of Type Ia and core-collapse supernovae (e.g., \citealt{GilmoreWyse1998}), such that galactic systems with different evolutionary histories will have distinct chemical abundance patterns. In this way, stellar halos serve as fossil records of a galaxy's accretion history. This theory has been extensively put into practice in the MW, where the differing patterns of \alphafe\ and \feh\ between its stellar halo and satellite dwarf galaxies have revealed their fundamentally incompatible enrichment histories \citep{Shetrone2001,Venn2004}.

Studies of the kinematics and chemical composition of individual stars in the MW have provided a detailed window into the formation of its stellar halo (e.g., \citealt{Carollo2007, Carollo2010, NissenSchuster2010,Ishigaki2012, Haywood2018, Helmi2018, Belokurov2018, Belokurov2020}). However, the MW is a single example of an $L_\star$ galaxy. Observations of MW-like stellar halos in the Local Volume have revealed a wide diversity in their properties, such as stellar halo fraction, mean photometric metallicity, and satellite galaxy demographics, which likely results from halo-to-halo variations in merger history \citep{Merritt2016,Monachesi2016,Harmsen2017,Geha2017,Smercina2019arXiv}. In these studies, both the MW and M31 have emerged as outliers at the quiescent and active ends, respectively, of the spectrum of accretion histories for nearby $L_\star$ galaxies.

Owing to its proximity (785 kpc; \citealt{McConnachie2005}), M31 is currently the only $L_\star$ galaxy that we can study at a level of detail approaching what is possible in the MW\@. M31's nearly edge-on orientation ($i=77^\circ$; \citealt{deVaucouleurs1958}) provides an exquisite view of its extended, highly structured stellar halo (e.g., \citealt{Ferguson2002,Guhathakurta2005,Kalirai2006b,Gilbert2007,Gilbert2009b,Ibata2007,Ibata2014,McConnachie2018}). Most notably, M31's stellar halo contains a prominent tidal feature known as the Giant Stellar Stream (GSS; \citealt{Ibata2001}), where the debris from this event litters the inner halo \citep{Brown2006,Gilbert2007}. Since the discovery of M31's halo \citep{Guhathakurta2005,Irwin2005,Gilbert2006}, its global metallicity, and kinematical properties have been thoroughly characterized from photometric and shallow ($\sim$1 hr) spectroscopic surveys \citep{Kalirai2006a,Ibata2007,Koch2008,McConnachie2009,Gilbert2012,Gilbert2014,Gilbert2018,Ibata2014}. However, it is only recently that \citet{Vargas2014a,Vargas2014b} made the first chemical abundance measurements beyond metallicity estimates in M31's halo and dwarf galaxies.

We have undertaken a deep ($\gtrsim$6 hour) spectroscopic survey using Keck/DEIMOS to probe the formation history of M31 from the largest sample of \feh\ and \alphafe\ measurements in M31 to date. This has resulted in the first \alphafe\ measurements in the GSS \citep{Gilbert2019}, the inner halo \citep{Escala2019,Escala2020}, and the outer disk \citep{Escala2020}, in addition to an expanded sample of \alphafe\ and \feh\ measurements in M31 satellite galaxies (\citealt{Kirby2020} for individual stars; \citealt{Wojno2020arXiv} for coadded groups of spectra) and the outer halo \citep{Gilbert2020}. Some of our key results include (1) evidence for a high efficiency of star formation in the GSS progenitor \citep{Gilbert2019,Escala2020} and the outer disk \citep{Escala2020}, (2) the distinct chemical abundance patterns of the inner halo compared to M31 satellite galaxies \citep{Escala2020,Kirby2020}, (3) support for chemical differences between the inner and outer halo \citep{Escala2020,Gilbert2020}, and
(4) chemical similarity between MW and M31 satellite galaxies \citep{Kirby2020,Wojno2020arXiv}. In this contribution, we analyze the global chemical abundance properties of the kinematically smooth component of M31's inner stellar halo.

This paper is organized as follows. In \S~\ref{sec:data}, we present our recently observed M31 fields. We provide a brief overview of our chemical abundance analysis in \S~\ref{sec:spec_synth} and develop a statistical model to determine M31 membership in \S~\ref{sec:membership}. We present our \alphafe\ and \feh\ measurements for \replaced{129}{128} M31 RGB stars and analyze the combined sample of inner halo abundance measurements (including those in the literature) in \S~\ref{sec:abund}. Finally, we compare \replaced{our measurements}{M31} to the MW and the Magellanic Clouds, and place our results in the context of stellar halo formation models in \S~\ref{sec:discuss}.

\section{Data} \label{sec:data}

\subsection{Spectroscopy and Data Reduction} \label{sec:spectra}

Table~\ref{tab:m31_obs} presents previously unpublished deep ($\gtrsim$ 5 hr) observations of six spectroscopic fields in M31. Fields f109\_1, f123\_1, f130\_1, a0\_1, a3\_1, and a3\_2 were observed in total for 7.0, 6.25, 6.74, 6.79, 6.44, and 6.60 hours, respectively. For five of these fields, we utilized the Keck/DEIMOS \citep{Faber2003} 600 line mm$^{-1}$ (600ZD) grating with the GG455 order blocking filter, a central wavelength of 7200 \AA, and 0.8'' slitwidths. We observed a single field (f123\_1) with the 1200 line mm$^{-1}$ (1200G) grating with the OG550 order blocking filter, a central wavelength of 7800 \AA, and 0.8'' slitwidths. We observed each spectroscopic field using two separate slitmasks that are identical in design, excepting slit position angles. This difference minimizes flux losses owing to differential atmospheric refraction at blue wavelengths via tracking changes in parallatic angle. The spectral resolution of the 600ZD (1200G) grating is approximately 2.8 (1.3) \AA \ FWHM, or R$\sim$3000 (6500) at the Ca II triplet region ($\lambda\sim8500$ \AA). Similarly deep observations of DEIMOS fields in M31, which we further analyze in this work, were previously published by \citet{Escala2020,Escala2019} (600ZD) and \citet{Gilbert2019} (1200G). 

One-dimensional spectra were extracted from the raw, two-dimensional DEIMOS data using the spec2d pipeline \citep{Cooper2012,Newman2013}, including modifications for bright, unresolved stellar sources \citep{SimonGeha2007}. \citet{Kirby2020} provides a comprehensive description of the data reduction process\deleted{, particularly for the case of spectra observed with the 1200G grating}.\deleted{For 600ZD spectra,} We included additional alterations \added{to the data reduction pipeline} to correct for \added{the effects of differential} atmospheric refraction, which preferentially affects bluer optical wavelengths \citep{Escala2019}. \added{This correction is equally applied to both 1200G and 600ZD spectra, although it is most significant for 600ZD spectra with $\lambda$ $\lesssim$ 5000 \AA.}

\begin{figure}
    \centering
    \includegraphics[width=\columnwidth]{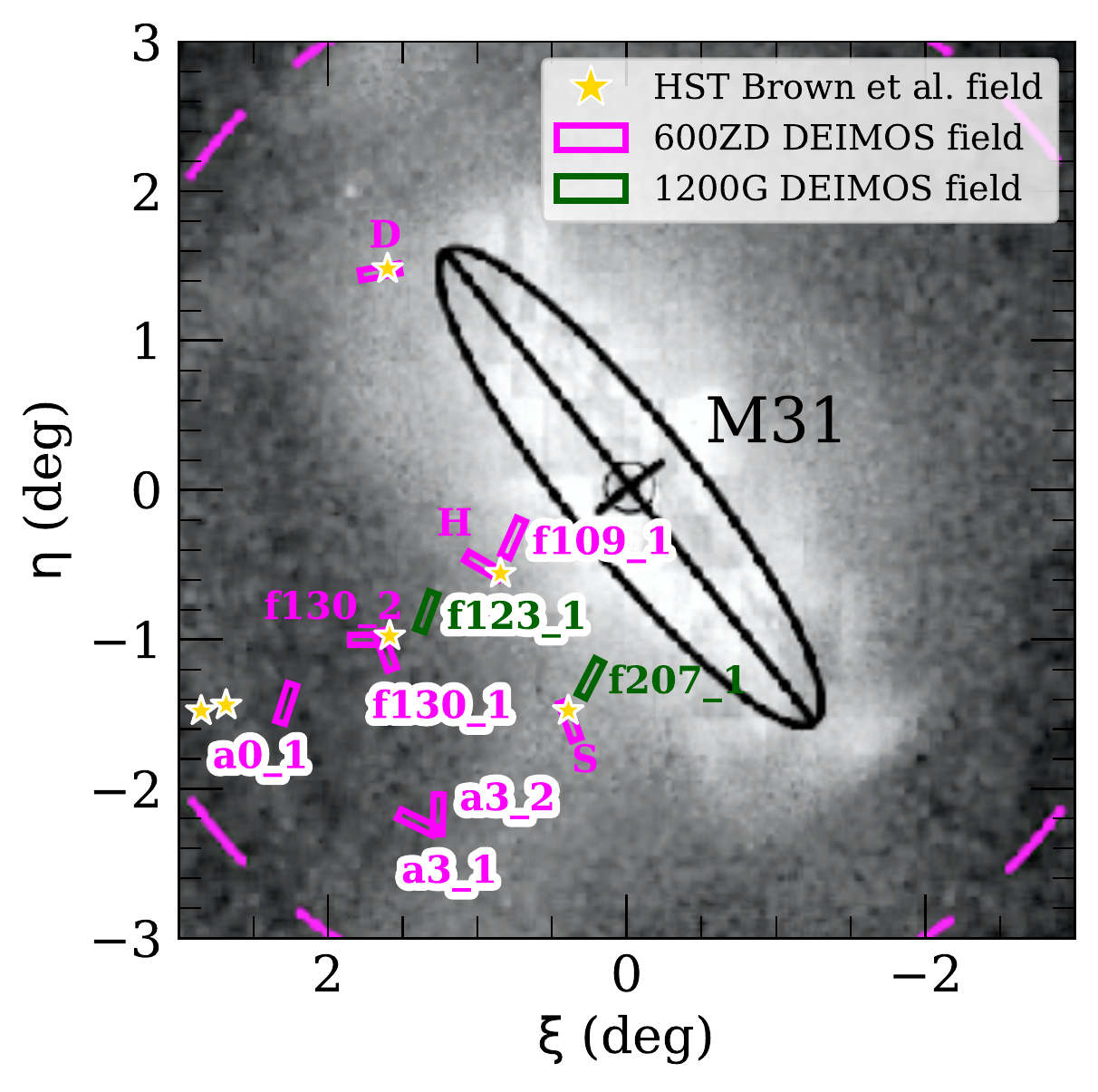}
    \caption{The location of deep ($\gtrsim$ 5 hr total exposure time) DEIMOS fields in M31-centric coordinates overlaid on the PAndAS red giant branch star count map \citep{McConnachie2018}. Rectangles represent the approximate size (16' x 4')  and orientation of the DEIMOS fields, whereas the dashed magenta line corresponds to 50 projected kpc. We include fields observed with both the low-resolution (600ZD; magenta) and medium-resolution (1200G; green) gratings on DEIMOS. Nearby {\it HST}/ACS fields (202''x 202''; \citealt{Brown2009}) are shown as gold stars. Observations first presented in this work are outlined in white (Table~\ref{tab:m31_obs}; \S~\ref{sec:data}). Our additional deep fields include 9 kpc, 18 kpc, 23 kpc, and 31 kpc in the stellar halo and 33 kpc in the Giant Stellar Stream.}
    \label{fig:m31_loc}
\end{figure}


\begin{table*}
\begin{threeparttable}
\caption{Deep DEIMOS Observations in M31}
\label{tab:m31_obs}
\begin{tabular*}{\textwidth}{l @{\extracolsep{\fill}} ccccccccc}
\hline
\hline
$\alpha_{\rm J2000}$ & $\delta_{\rm J2000}$ & P.A. & Grating & Slitmask\tnote{a} & Date & $\theta_s$ ($''$) & $X$ & $t_{\textrm{exp}}$ (hr) & $N$\\
\hline

\multicolumn{10}{c}{f109\_1 (9 kpc Halo Field)} \\\hline
00$^{\rm{h}}$45$^{\rm{m}}$47.02$^{\rm{s}}$ & +40$^{\circ}$56$^{'}$58.7$^{''}$ & 23.9 & 600ZD & f109\_1a & 2019 Oct 24 & 0.67 & 1.35 & 2.77 & 143\\
& & & & f109\_1a & 2019 Oct 25 & 0.61 & 1.32 & 2.00 & ...\\
& & & & f109\_1b & 2019 Oct 24 & 0.68 & 1.10 & 0.93 & ...\\ 
& & & & f109\_1b & 2019 Oct 25 & 0.73 & 1.10 & 1.30 & ...\\ \hline

\multicolumn{10}{c}{f123\_1 (18 kpc Halo Field)} \\\hline
00$^{\rm{h}}$48$^{\rm{m}}$05.83$^{\rm{s}}$ & +40$^{\circ}$27$^{'}$24.0$^{''}$ & $-$20 & 1200G & f123\_1a & 2017 Oct 23 & 0.88 & 1.52 & 2.83 & 136\\
& & & & f123\_1b & 2019 Sep 25 & 0.80 & 1.34 & 3.42 & ...\\ \hline

\multicolumn{10}{c}{f130\_1 (23 kpc Halo Field)} \\\hline
00$^{\rm{h}}$49$^{\rm{m}}$11.90$^{\rm{s}}$ & +40$^{\circ}$11$^{'}$50.3$^{''}$ & $-$20 & 600ZD & f130\_1b & 2019 Sep 25 & 0.43 & 1.07 & 1.15 & 93\\
& & & & f130\_1c & 2018 Sep 10 & 0.72 & 1.29 & 0.97 & ...\\
& & & & f130\_1c & 2018 Sep 11 & 0.80 & 1.25 & 2.07 & ...\\
& & & & f130\_1c & 2019 Sep 25 & 0.61 & 1.25 & 2.55 & ...\\ \hline

\multicolumn{10}{c}{a0\_1 (31 kpc Halo Field)} \\\hline
00$^{\rm{h}}$51$^{\rm{m}}$51.31$^{\rm{s}}$ & +39$^{\circ}$50$^{'}$26.9$^{''}$ & $-$17.9 & 600ZD & a0\_1a & 2019 Oct 24 & 0.68 & 1.07 & 0.73 & 67\\
& & & & a0\_1a & 2019 Oct 25 & 0.57 & 1.06 & 0.43 & ...\\
& & & & a0\_1b & 2019 Oct 24 & 0.66 & 1.24 & 2.84 & ...\\
& & & & a0\_1b & 2019 Oct 25 & 0.70 & 1.24 & 2.79 & ...\\ \hline

\multicolumn{10}{c}{a3\_1 (33 kpc GSS Field)} \\\hline
00$^{\rm{h}}$48$^{\rm{m}}$22.09$^{\rm{s}}$ & +39$^{\circ}$02$^{'}$33.1$^{''}$ & 64.2 & 600ZD & a3\_1a & 2018 Sep 10 & 0.80 & 1.49 & 2.00 & 84\\
& & & & a3\_1a & 2018 Sep 11 & 0.54 & 1.50 & 2.27 & ...\\
& & & & a3\_1a & 2019 Sep 26 & 0.55 & 1.44 & 1.67 & ...\\ 
& & & & a3\_1b & 2019 Sep 26 & 0.58 & 1.17 & 0.50 & ...\\ \hline

\multicolumn{10}{c}{a3\_2 (33 kpc GSS Field)} \\\hline
00$^{\rm{h}}$47$^{\rm{m}}$47.22$^{\rm{s}}$ & +39$^{\circ}$05$^{'}$50.7$^{''}$ & 178.2 & 600ZD & a3\_2a & 2018 Oct 02 & 0.60 & 1.16 & 2.10 & 80\\
& & & & a3\_2a & 2019 Sep 26 & 0.64 & 1.09 & 1.00 & ...\\
& & & & a3\_2b & 2019 Sep 26 & 0.61 & 1.18 & 3.50 & ...\\  \hline

\hline
\end{tabular*}
\begin{tablenotes}
\item Note. \textemdash\ The columns of the table refer to right ascension, declination, position angle in degrees east of north, grating, slitmask name, date of observation (UT), average seeing, average airmass, exposure time per slitmask, and number of stars targeted per slitmask. Additional deep DEIMOS observations of M31 fields utilized in this work were published by \citet{Escala2019,Escala2020} and \citet{Gilbert2019}.
\item[a] Slitmasks labeled as ``a'', ``b'', etc., are identical, except that the slits are tilted according to the parallactic angle at the approximate time of observation for the slitmask.
\end{tablenotes}
\end{threeparttable}
\end{table*}

\begin{figure*}
    \centering
    \includegraphics[width=0.8\textwidth]{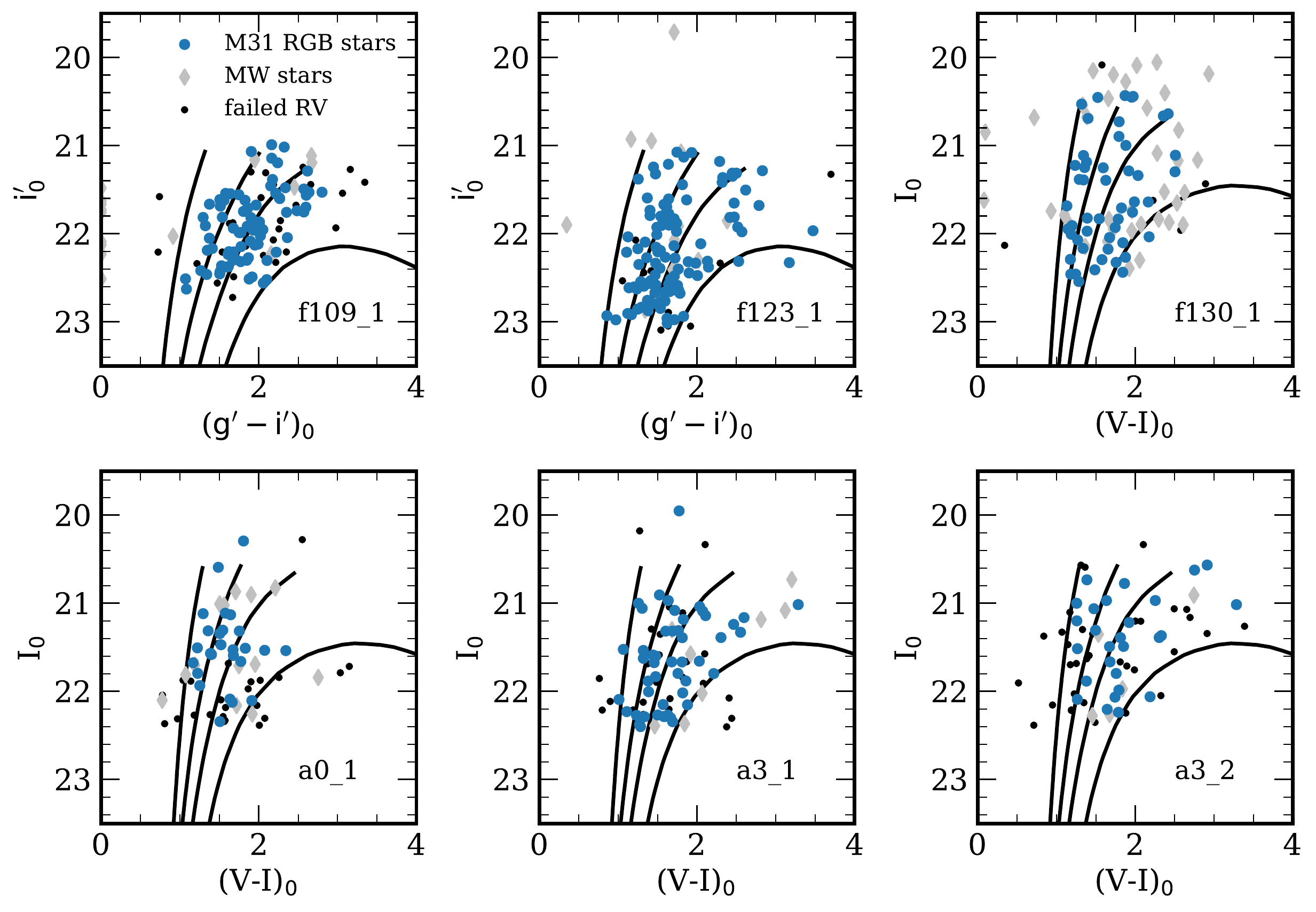}
    \caption{Extinction-corrected color-magnitude diagrams in the relevant photometric filters (Johnson-Cousins $V,I$ and CFHT/MegaCam $g,i$; \S~\ref{sec:phot}) for individual stars in each spectroscopic field (Table~\ref{tab:m31_obs}). For reference, we overplot PARSEC isochrones (\textit{black lines}; \citealt{Marigo2017}), assuming 9 Gyr ages and \alphafe\ = 0, with \feh\ = $-2.2$, $-1.0$, $-0.5$, and $0$ (from left to right). Stars without successful radial velocity measurements (\S~\ref{sec:spec_synth}) 
    are represented as black points. Stars that are likely to be MW foreground dwarfs are indicated as grey diamonds, whereas likely M31 RGB stars (\S~\ref{sec:prob_model}) are represented as \added{blue} circles. \deleted{color-coded according to their probability of belonging to kinematical substructure in the stellar halo (\S~\ref{sec:vel}).}
    \deleted{Open circles indicate M31 RGB stars with strong TiO absorption, which we excluded from our final sample (\S~\ref{sec:sample}).}}
    \label{fig:cmd}
\end{figure*}

\subsection{Field Properties}
\label{sec:fields}

Fields f109\_1, f123\_1, f130\_1, a0\_1, and a3 are located at 9, 18, 23, 31, and 33 kpc away from the galactic center of M31 in projected distance. We assumed that M31's galactocenter is located at a right ascension of 0.71 hours and a declination of 41.3 degrees. Table~\ref{tab:m31_obs} provides the mask center and mask position angle of each field on the sky. Figure~\ref{fig:m31_loc} illustrates the locations of our newly observed DEIMOS fields relative to the center of M31, including previously observed fields \citep{Escala2020,Escala2019,Gilbert2019} further analyzed in this work, overlaid on a RGB star count map from the Pan-Andromeda Archaeological Survey (PAndAS; \citealt{McConnachie2018}).  We also show the locations of pencil-beam {\it HST}/ACS fields \citep{Brown2009}, some of which overlap with DEIMOS fields \citep{Escala2020}, from which \citet{Brown2006,Brown2007,Brown2008} derived stellar age distributions. \deleted{The ACS images (202''$\times$202'') are effectively points on the sky compared to each DEIMOS field (approximately 16'$\times$4').}

The deep fields presented in this work were previously observed using shallow ($\sim$1 hr) DEIMOS spectroscopy with the 1200G grating to obtain kinematical information for each field \citep{Gilbert2007,Gilbert2009a}. Based on the shallow 1200G observations, \citet{Gilbert2007,Gilbert2009a} argued that
fields f109\_1, f123\_1, f130\_1, and a0\_1 probe the properties of the stellar halo of M31, whereas a3 probes the GSS\@. However, multiple kinematical components can be present in a given field. For example, f123\_1 contains substructure known as the Southeast shelf, which is likely associated with the GSS progenitor \citep{Gilbert2007,Fardal2007,Escala2020}. Unlike fields along the GSS located closer to M31's center, a3 does not show evidence for the secondary kinematically cold component (KCC) of unknown origin \citep{Kalirai2006a,Gilbert2009a,Gilbert2019}. We expect that f130\_1 is associated with the ``smooth'', relatively metal-poor halo of M31, based on the known properties of the overlapping DEIMOS field f130\_2 (Figure~\ref{fig:m31_loc}; \citealt{Escala2019}). Similarly, the velocity distributions of a0\_1 and f109\_1 are fully consistent with that of M31's stellar halo \citep{Gilbert2007}. Despite being the innermost M31 field in our sample ($r_{\rm proj} = 9$ kpc, or $r_{\rm disk} = 38$ kpc assuming $i$ = 77$^\circ$), f109\_1 does not show any kinematical evidence for significant contamination ($\gtrsim$10\%) by M31's extended disk \citep{Gilbert2007}. This also applies to the Southeast shelf, which was predicted to extend to the location of f109\_1 with a large velocity dispersion ($\sim$350 \kms; \citealt{Fardal2007,Gilbert2007}), such that any hypothetical SE shelf stars in this field could not be kinematically separated from the halo population. \deleted{We refer to the spectroscopic fields by their projected M31-centric distance and predominantly traced stellar structure (as in Table~\ref{tab:m31_obs}; e.g., f109\_1 is the 9 kpc halo field) where appropriate to emphasize their physical properties.}

\subsection{Photometry} \label{sec:phot}

The photometry for the majority of fields published in this work were obtained from MegaCam images in the $g^\prime, i^\prime$ filters using the 3.6 m Canada-France-Hawaii Telescope (CFHT). The MegaCam images were obtained by \citet{Kalirai2006a} and reduced with the CFHT MegaPipe pipeline 
\citep{Gwyn2008}. For 9 targets in field f109\_1 absent from our primary catalogs, we sourced g$^\prime$ and $i^\prime$ band photometry from the Pan-Andromeda Archaeological Survey (PAndAS) point source catalog \citep{McConnachie2018}.
For field f130\_1, the $g^\prime$, $i^\prime$ magnitudes were transformed to Johnson-Cousins $V,I$ using observations of Landolt photometric standard stars \citep{Kalirai2006a}. In the case of fields a0\_1 and a3, the original photometry was obtained in the Washington $M,T_2$ filters by \citet{Ostheimer2003} using the Mosaic camera on the 4 m Kitt Peak National Observatory (KPNO) telescope and subsequently transformed to the Johnson-Cousins $V,I$ bands using the relations of \citet{Majewski2000}. 

Figure~\ref{fig:cmd} presents the extinction-corrected color-magnitude diagrams (CMDs) for each field in the relevant photometric filters used to derive quantities based on the photometry such as the photometric effective temperature (\teffphot), surface gravity (\loggphot), and metallicity (\fehphot). We show all stars in a given field for which we extracted 1D spectra (M31 RGB stars, MW foreground dwarf stars, and stars for which we were unable to evaluate M31 membership owing to failed radial velocity measurements; \S~\ref{sec:membership}). \deleted{where each star is color-coded according to its probability of belonging to kinematically identified substructure in the stellar halo (\S~\ref{sec:membership}). That is, cyan points are likely to be associated with the dynamically hot, ``smooth'' stellar halo, whereas magenta points likely belong to substructure such as the GSS (a3 fields) or Southeast Shelf (f123\_1).} 

For fields f109\_1, f130\_1, a0\_1, and a3, for which stellar spectra were obtained using the 600ZD grating (\S~\ref{sec:spectra}), we calculated \teffphot, \loggphot, and \fehphot\ following the procedure described by \citet{Escala2020}. In summary, the color and magnitude of a star are compared to a grid of theoretical stellar isochrones to derive the above quantities.
We utilized the PARSEC isochrones \citep{Marigo2017}, which include molecular TiO in their stellar evolutionary modeling, and assumed a distance modulus to M31 of $m - M$ = 24.63 $\pm$ 0.20 \citep{Clementini2011}. For the 600ZD fields, we assumed 9 Gyr isochrones based on the \deleted{intermediate} mean stellar ages of the stellar halo and GSS, as inferred from {\it HST} CMDs \added{(9.7, 11.0, 10.5 Gyr for the 11, 21, and 35 kpc ACS fields, and 8.8 Gyr for the ACS stream field; \citealt{Brown2006,Brown2007,Brown2008}).}

The photometric quantities (\teffphot, \loggphot) for the single 1200G-based field, f123\_1, were derived following the procedure outlined by \citet{Kirby2008}, assuming an identical distance modulus and 14 Gyr isochrones from a combination of model sets \citep{Girardi2002,Demarque2004,VandenBerg2006}. As described in detail by \citet{Escala2019} and summarized in \S~\ref{sec:abund}, \teffphot\ and \loggphot\ are used as constraints in measuring \teff, \feh, and \alphafe\ from spectra of individual stars, where these measurements are insensitive to the employed isochrone models and assumed stellar age. 

\section{Chemical Abundance Analysis} \label{sec:spec_synth}

We use spectral synthesis of low- and medium-resolution stellar spectroscopy  to measure stellar parameters (\teff) and abundances (\feh\ and \alphafe) from our deep \added{($\gtrsim$5 hr)} observations of M31 RGB stars. For a detailed description of the low- and medium-resolution spectral synthesis methods, see \citet{Escala2019,Escala2020} and \citet{Kirby2008}, respectively. The low- and medium-resolution spectral synthesis procedures are nearly identical in principle, excepting differences in the continuum normalization given the differing wavelength coverage between the low- and medium-resolution spectra ($\sim$4500$-$9100 vs. 6300$-$9100 \AA). For 1200G spectra, the continuum is determined using ``continuum regions'' defined by \citet{Kirby2008}, whereas such regions would be unilaterally defined for 600ZD spectra owing to the high density of absorption features toward the blue wavelengths. Chemical abundances \deleted{(\feh\ and \alphafe)} for individual stars measured using each technique are generally consistent within the uncertainties \citep{Escala2020}. 

Prior to the chemical abundance analysis, the radial velocity of each star is measured via cross-correlation with empirical templates \citep{SimonGeha2007,Kirby2015,Escala2020} observed in the relevant science configuration (\S~\ref{sec:spectra}). Systematic radial velocity errors of 5.6 \kms\ \citep{Collins2011} and 1.49 \kms\ \citep{Kirby2015} from repeat measurements of identical stars are added in quadrature to the random component of the error for observations taken with the 600ZD and 1200G gratings, respectively. The spectral resolution is empirically determined as a function of wavelength using the width of sky lines \citep{Kirby2008}, and in the case of the bluer 600ZD spectra, arc lines from calibration lamps (\citealt{Escala2020}; McKinnon et al., in preparation).  The observed spectrum is then corrected for telluric absorption using a template of a hot star observed in the relevant science configuration \citep{Kirby2008,Escala2019}, shifted into the rest frame based on the measured radial velocity, and an initial continuum normalization is performed.

We measured the spectroscopic effective temperature, \teff, informed by photometric constraints, and fixed the surface gravity, \logg, to the photometric value (\S~\ref{sec:phot}). We simultaneously measured \feh \ and \alphafe \ from regions of the spectrum sensitive to Fe and $\alpha$-elements (Mg, Si, Ca -- with the addition of Ti for medium-resolution spectra), respectively, by comparing to a grid of synthetic spectra degraded to the resolution of the applicable DEIMOS grating (600ZD or 1200G) using Levenberg-Marquardt $\chi^2$ minimization. The grids of synthetic spectra utilized were generated for 4100$-$6300~\AA\ and 6300$-$9100~\AA, respectively, by \citet{Escala2019} and \citet{Kirby2008}. The continuum determination is refined throughout this process, where \teff, \feh, and \alphafe\ are measured iteratively until \teff\ changed by less than 1 K and \feh\ and \alphafe\ each changed by less than 0.001. Finally, systematic errors on the abundances are added in quadrature to the random component of the error from the fitting procedure. For 600ZD-based abundance measurements, $\delta$(\feh)$_{\rm sys}$ = 0.130 and $\delta$(\alphafe)$_{\rm sys}$ = 0.107, whereas for 1200G-based measurements, $\delta$(\feh)$_{\rm sys}$ = 0.101 and $\delta$(\alphafe)$_{\rm sys}$ = 0.084 \citep{Gilbert2019}.

\section{Membership Determination} 
\label{sec:membership}

Separating M31 RGB stars from the intervening foreground of MW dwarf stars has served as one of the primary challenges for spectroscopic studies of individuals stars in M31. The colors and heliocentric radial velocity distributions of MW and M31 stars exhibit significant overlap (e.g., \citealt{Gilbert2006}), thus the difficulty in disentangling the two populations when the distances to such faint stars are unknown. Early spectroscopic studies of M31 employed simple radial velocity cuts to exclude MW stars (e.g., \citealt{ReitzelGuhathakurta2002,Ibata2005,Chapman2006}), resulting in kinematically biased populations of M31 RGB stars and relatively uncontaminated, albeit incomplete samples of M31 stars. \citet{Gilbert2006} performed the first rigorous, probabilistic membership determination in M31 using various diagnostics, including (1) heliocentric radial velocity, (2) the strength of the surface-gravity sensitive Na I $\lambda\lambda$8190 doublet, (3) CMD position, and (4) the discrepancy between photometric and calcium triplet based metallicity estimates as a distance indicator.

Given the variety of photometric filters utilized, \citeauthor{Gilbert2006}'s method cannot be uniformly applied to all spectroscopic fields analyzed in this work. For inner halo fields ($r_{\rm proj}$ $\lesssim$ 30 kpc), \citet{Escala2020} illustrated that a binary membership determination using the aforementioned diagnostics is sufficient to recover the majority of stars classified as likely M31 RGB stars by the more sophisticated method of \citet{Gilbert2006} with minimal MW contamination. However, this binary determination excludes all stars with $v_{\rm helio}$ $>$ $-$150 \kms, where some of these stars may be M31 members at the positive tail of the stellar halo velocity distribution. It also does not allow us to assign a degree of certainty to our membership determination for each star. Thus, we used Bayesian inference to assign a membership probability to each observed star with a successful radial velocity determination (\S~\ref{sec:spec_synth}).

\subsection{Membership Probability Model}
\label{sec:prob_model}

We evaluated the probability of M31 membership for all stars with successful velocity measurements based on (1) a parameterization of color ($X_{\rm CMD}$), (2) the strength of the Na I absorption line doublet at $\lambda\lambda$8190 (EW$_{\rm Na}$), (3) heliocentric radial velocity (\S~\ref{sec:spec_synth}; $v_{\rm helio}$), and (4) an estimate of the spectroscopic metallicity based on the strength of the calcium triplet (\feh$_{\rm CaT}$).

Thus, according to Bayes' theorem, the posterior probability that a star labeled by an index $j$ is an M31 member---given independent measurements $\vec{x_j}$ = (EW$_{{\rm Na},j}$, X$_{{\rm CMD},j}$, $v_{{\rm helio},j}$, \feh$_{{\rm CaT}, j}$) with uncertainties $\delta\vec{x_j}$---is proportional to,
\begin{equation}
P({\rm M31} | \vec{x_j} ) \propto \\
P({\rm M31})_j \times P( \vec{x_j} | {\rm M31}),
\label{eq:bayes}
\end{equation}
where $P$(M31) is the prior probability that a star observed on a given DEIMOS slitmask is a member of M31, and $P(\vec{x_j}|$M31) is the likelihood of measuring a given set of membership diagnostics, $\vec{x_j}$, assuming the star is a M31 member. Analogously, we can also construct $P({\rm MW}|\vec{x_j})$, the posterior probability that a star belongs to the MW foreground population given a set of diagnostic measurements, where the sum of $P({\rm MW}|\vec{x}_j)$ and $P({\rm M31}|\vec{x}_j)$ is unity.

\subsubsection{Measuring Membership Diagnostics}
\label{sec:diagnostics}

\begin{figure*}
    \centering
    \includegraphics[width=0.85\textwidth]{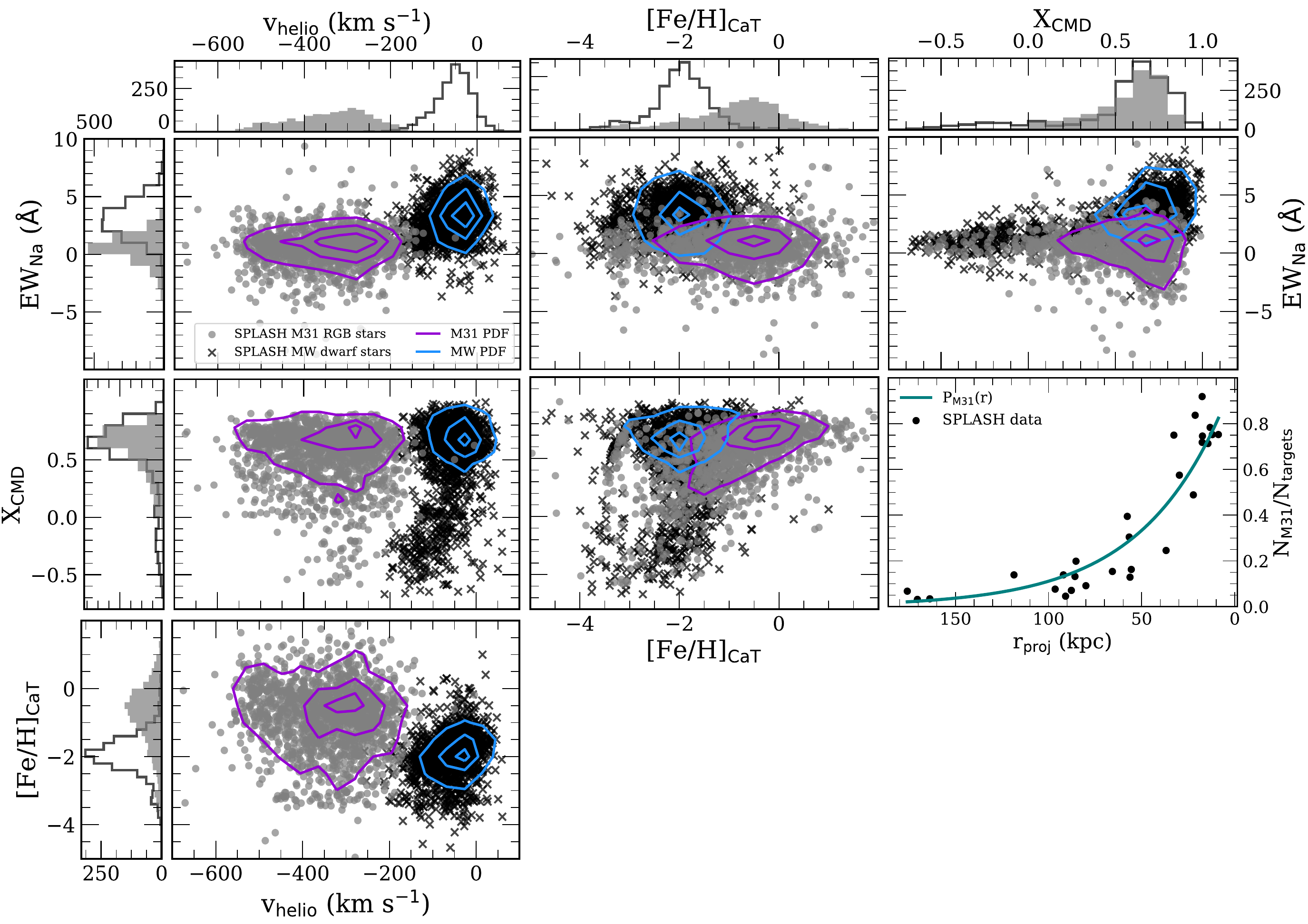}
    \caption{Properties of spectroscopically confirmed M31 ({\it grey circles, purple contours}) and MW ({\it black crosses, blue contours}) member stars from the SPLASH \citep{Guhathakurta2005,Gilbert2006} survey. The contour levels correspond to 16$^{\rm th}$, 50$^{\rm th}$, and 84$^{\rm th}$ percentiles of the distributions.  Histograms of star counts are also shown for 
    both M31 and MW members. For clarity, we omit showing uncertainties in each parameter. The middle right panel illustrates the number of spectroscopically confirmed M31 members \citep{Gilbert2006} relative to the number of targets with successful radial velocity measurements in a given field as a function of projected radius, \added{as observed by the SPLASH survey}, where we have fit the relationship with an exponential distribution, \added{P$_{\rm M31}$($r$) (\S~\ref{sec:prior})}.
    We used heliocentric radial velocity ($v_{\rm helio}$), a parameterization of color ($X_{\rm CMD}$), Na I $\lambda\lambda8190$ equivalent width (EW$_{\rm Na}$), and calcium triplet metallicity estimates (\feh$_{\rm CaT}$) as diagnostics to determine membership (\S~\ref{sec:diagnostics}, \S~\ref{sec:lkhd}) for our sample of stars with successful radial velocity measurements. 
   }
    \label{fig:splash}
\end{figure*}

We described the CMD position of each star by the parameter $X_{\rm CMD}$, which is analogous to photometric metallicity (\fehphot). The advantage of using $X_{\rm CMD}$ rather than color and magntiude is that we can place
all of our stars on the same scale, despite the diversity of the photometric filter sets (\S~\ref{sec:phot}). Assuming 12.6 Gyr isochrones and a distance modulus of 24.47, we defined $X_{\rm CMD}$ = 0 as the color of the most metal-poor PARSEC \citep{Marigo2017} isochrone (\feh\ = $-$2.2) in the relevant photometric filter and $X_{\rm CMD}$ = 1 as the most metal-rich PARSEC isochrone (\feh\ = +0.5).
Then, we used linear interpolation to map the color and magnitude of a star to a value of $X_{\rm CMD}$. The uncertainty on $X_{\rm CMD}$ was derived from the photometric errors by using a Monte Carlo procedure. This normalization provides the advantage of easily identifying stars that are bluer than the most metal-poor isochrone at a fixed stellar age by negative values of $X_{\rm CMD}$, where stars with $X_{\rm CMD} < 0$ are $\gtrsim$10 times more likely to belong to the MW than M31 \citep{Gilbert2006}. In the sample being evaluated for membership, we classified stars with ($X_{\rm CMD} + \delta$X$_{\rm CMD}$) $<$ 0 as MW dwarf stars. 

By including \deleted{both $X_{\rm CMD}$ and} \feh$_{\rm CaT}$ \added{in combination with $X_{\rm CMD}$} as an independent diagnostic in our model, we can additionally distinguish between foreground stars at unknown distances and distant giant stars. Given that our assumed distance modulus is appropriate only for stars at the distance of M31, \fehphot, and analogously, $X_{\rm CMD}$, measurements are fundamentally incorrect for MW stars. Therefore, the two stellar populations will appear distinct in $X_{\rm CMD}$ versus \feh$_{\rm CaT}$\ space. In order to compute \feh$_{\rm CaT}$ \added{in the sample being evaluated for membership}, we \deleted{similarly} fit Gaussian profiles to 15 \AA\ wide windows centered on Ca II absorption lines at 8498, 8542, and 8662 \AA. Then, we calculated a total equivalent width for the calcium triplet from a linear combination of the individual equivalent widths,

\begin{equation}
    \Sigma{\rm Ca} = 0.5 \times {\rm EW}_{\lambda 8498} + 1.0 \times {\rm EW}_{\lambda 8542} + 0.6 \times {\rm EW}_{\lambda 8662},
\end{equation}
following \citet{Rutledge1997a}. In addition to $\Sigma$Ca, the calibration to determine \feh$_{\rm CaT}$\ depends on stellar luminosity,

\begin{equation}
    {\rm [Fe/H]_{CaT} } = -2.66 + 0.42 \times \Sigma{\rm Ca} + 0.27 \times (V - V_{\rm HB}),
\end{equation}
where $V - V_{\rm HB}$ is the Johnson-Cousins $V$-band apparent magnitude above the horizontal branch, assuming that $V_{\rm HB}$ = 25.17 for M31 \citep{Holland1996} \added{and that a star is within the RGB (($V - V_{\rm HB}$) $>$ $-$5).} For fields with CFHT MegaCam $g^\prime$ and $i^\prime$ band photometry (\S~\ref{sec:phot}), we approximated $V - V_{\rm HB}$ by $g^\prime - g^\prime_{\rm HB}$ using an empirical transformation between SDSS and Johnson-Cousins photometry  \citep{Jordi2006},

\begin{equation}
    g-V = 0.56 \times (g - r + 0.23)/1.05 + 0.12.
\end{equation}
Assuming that $g-r$ = 0.6 and $g^\prime_{\rm HB} \approx g_{\rm HB}$, we obtained $g_{\rm HB} = V_{\rm HB} + (g - V) = 25.73$ for stars in M31. \added{For M31's (the MW's) distribution of $\Sigma$Ca in our model (\S~\ref{sec:lkhd}), we found $\langle\Sigma$Ca$\rangle$ = 5.60 (3.00) \AA, with values spanning $-$4.70--21.66 ($-$4.07--10.37) \AA\ and typical errors of $\langle\delta\Sigma$Ca$\rangle$ = 1.71 (0.88) \AA.}

We measured EW$_{\rm Na}$ \added{in the sample being evaluated for membership} by summing the area under the best-fit Gaussian line profiles fit to the observed spectrum between 8178$-$8190, 8189$-$8200 \AA\ with central wavelengths of 8183, 8195 \AA\ using least-squares minimization. This parameter is sensitive to temperature and surface gravity \citep{Schiavon1997}, thus functioning as a discriminant between MW dwarf stars and M31 RGB stars. \added{The distribution of EW$_{\rm Na}$ for M31 (the MW) in our model (\S~\ref{sec:lkhd}) has $\langle$EW$_{\rm Na}\rangle$ = 0.52 (3.07) \AA, with values spanning $-$10.55--9.38 ($-$3.67--8.88) \AA\ and typical errors of $\langle\delta$EW$_{\rm Na}\rangle$ = 1.06 (0.58) \AA.}




\subsubsection{Prior Probability of M31 Membership}
\label{sec:prior}

\begin{figure*}
    \centering
    \includegraphics[width=\textwidth]{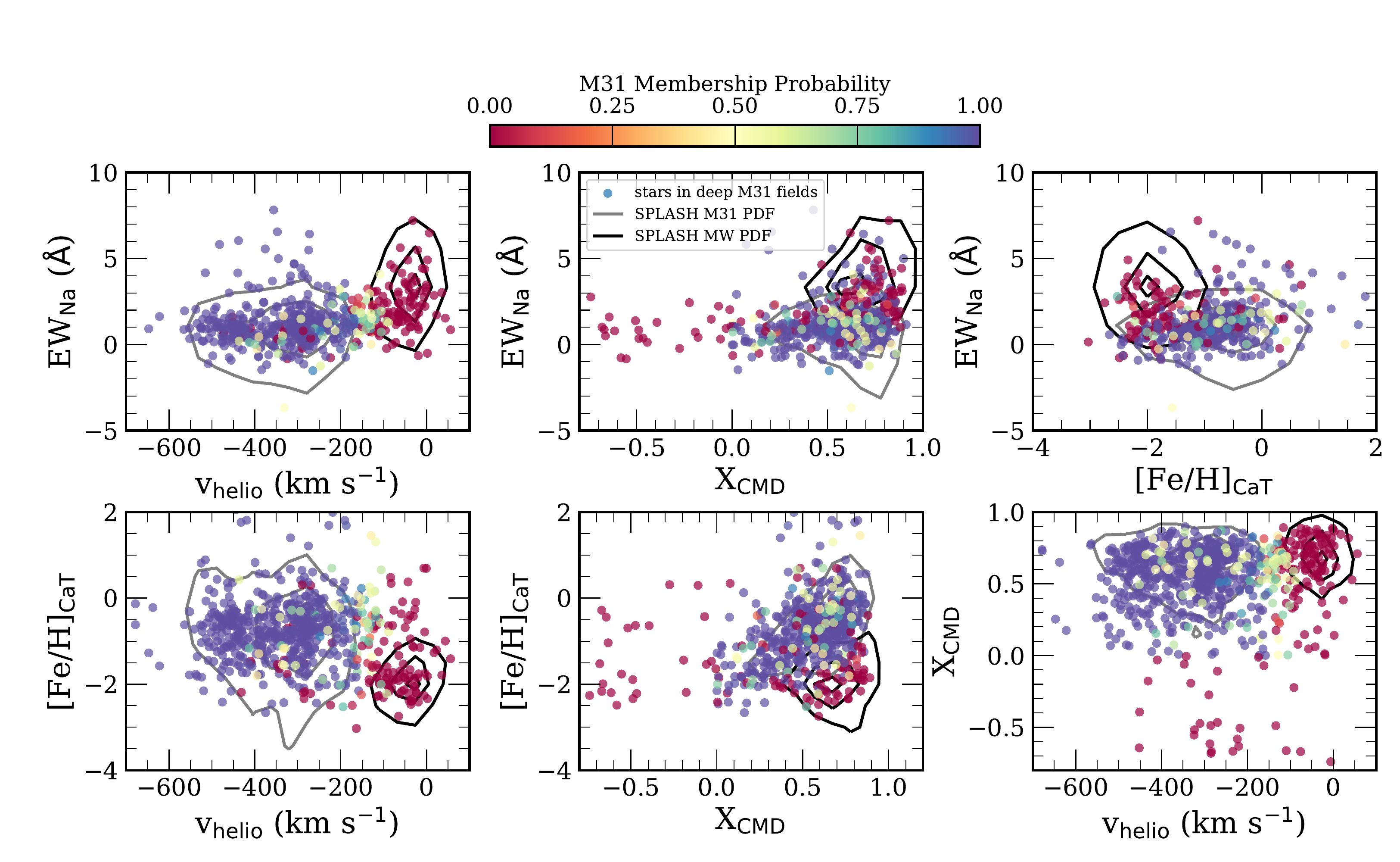}
    \caption{Properties used as membership diagnostics ($v_{\rm helio}$, EW$_{\rm Na}$, $X_{\rm CMD}$, \feh$_{\rm CaT}$) for stars in our spectroscopic fields (Table~\ref{tab:m31_obs}) with successful radial velocity measurements (\S~\ref{sec:spec_synth}). We also include previously observed fields H, S, D, and f130\_2 (Figure~\ref{fig:m31_loc}; \citealt{Escala2020}). For clarity, we omit showing measurement uncertainties.
    Each point is color-coded by its probability of belonging to M31 \added{(Eq.~\ref{eq:pm31}; \S~\ref{sec:lkhd})}.
    The SPLASH distributions utilized in our probability model (\S~\ref{sec:prob_model}) are underlaid as contours. 
    For field D, we used a distinct velocity criterion owing to the presence of M31's disk at $\sim-130$ \kms \citep{Escala2020}. We identified \replaced{329}{328} (\replaced{221}{210}) secure M31 RGB stars and \replaced{91}{92} (75) secure MW dwarf stars across the spectroscopic fields presented in this (prior) work.}
    \label{fig:members}
\end{figure*}

The probability of a given star observed on a DEIMOS slitmask belonging to M31, $P$(M31), increases with decreasing projected radial distance from the center of M31, owing to the increase in M31's stellar surface density. The trend of increasing probability of M31 membership with decreasing projected radius is augmented compared to expectations from M31's surface brightness profile \citep{Courteau2011,Gilbert2012} by our photometric pre-selection of DEIMOS spectroscopic targets. These selection criteria are designed to include M31 members and exclude MW foreground stars. The photometry of the targets spans the magnitude range characteristic of M31 RGB stars (20 $<$ $I_0$ $<$ 22.5). We also considered narrow-band, Washington DDO51 photometry in the fields a0\_1 and a3. The DDO51 filter isolates the Mg b triplet, which acts as discriminant between MW dwarf and M31 giant stars due to its sensitivity to surface gravity. We expect that an enhancement in the probability of observing an M31 RGB star owing to magnitude cuts is fairly uniform at a factor of $\lesssim$ 2 within $r_{\rm proj}$ $\lesssim$ 30 kpc \citep{Gilbert2012}, where the majority of our targets are located. For fields toward the outer halo, such as a0\_1 and a3, DDO51-selection bias increases the likelihood of observing an M31 RGB star by a factor of 3$-$4 \citep{Gilbert2012}.

Thus, we parameterized P(M31) empirically using the ratio of the number of secure M31 RGB stars \citep{Gilbert2006}, $N_{\rm M31}$, to the number of targets with successful radial velocity measurements, $N_{\rm targets}$, from the Spectroscopic and Phomtometric Landscape of Andromeda's Stellar Halo (SPLASH; \citealt{Guhathakurta2005,Gilbert2006}) survey. The SPLASH survey consists of tens of thousands of shallow ($\sim$1 hr exposures) DEIMOS spectra of stars along the line of sight toward M31's halo, disk, and satellite dwarf galaxies (e.g., \citealt{Kalirai2010,Tollerud2012, Dorman2012, Dorman2015, Gilbert2012, Gilbert2014, Gilbert2018}). In addition to $N_{\rm M31}$/$N_{\rm targets}$ as a function of radius, Figure~\ref{fig:splash} presents measurements of EW$_{\rm Na}$, $X_{\rm CMD}$, $v_{\rm helio}$, and \feh$_{\rm CaT}$ for \replaced{1,521}{1,510} secure M31 members and \replaced{1,835}{1,794} secure MW members across 29 spectroscopic fields in M31's stellar halo.  We controlled for differences in methodology between this work and SPLASH by re-determining $X_{\rm CMD}$ homogeneously for the SPLASH data from the original Johnson-Cousins photometry and measuring \feh$_{\rm CaT}$\ for our sample based on the same calibration \citep{Rutledge1997a} used in SPLASH \added{(\S~\ref{sec:diagnostics})}.\footnote{Our equivalent width measurement procedure differs from that utilized in the SPLASH survey. As opposed to summing the flux decrement in a window centered on a given absorption feature, we performed Gaussian fits. However, we do not expect this difference in metholodgy to significantly affect the usability of the SPLASH data to construct the likelihood (\S~\ref{sec:lkhd}), given that our measured EW$_{\rm Na}$ and \feh$_{\rm CaT}$ distributions are consistent with SPLASH (Figure~\ref{fig:members}).}

Based on this data, we approximated P(M31) by an exponential distribution,
\begin{equation}
    P_{\rm M31}(r_j) = \exp(-r_j/r_p),
\end{equation}
where $r_j$ is the projected radius of a star from M31's galactic center and $r_p$ = 45.5 kpc. Figure~\ref{fig:splash} includes $P_{\rm M31}(r)$ overlaid on the SPLASH survey data.

\subsubsection{Likelihood of M31 Membership}
\label{sec:lkhd}

Given the wealth of existing information on the properties of M31 RGB stars (and the MW foreground dwarf stars characteristic of our selection function) from the SPLASH survey, we assigned membership likelihoods to individual stars informed by this extensive data set. We described the likelihood that a star, $j$, with unknown membership belongs to M31 given its diagnostic measurements and uncertainties, ($\vec{x}_j$, $\delta\vec{x}_j$), as,

\begin{equation}
    P(\vec{x_j},\delta\vec{x_j}|\mathrm{M31}) =  \frac{1}{N_i} \sum_{i=1}^{N_i} P(\vec{x_j}|\vec{\theta_i}, \delta\vec{\theta_i}),
\end{equation}
where ($\vec{\theta_i}$, $\delta\vec{\theta_i}$) is a set of four diagnostic measurements and uncertainties for a star, $i$, from the SPLASH survey that is a secure M31 member. The total number of secure SPLASH member stars, $N_i$, equals \replaced{1,521}{1,510} (\replaced{1,835}{1,794}) for M31 (the MW). The likelihood that a star, $j$, with unknown membership belongs to the MW, $P(\vec{x_j},\delta\vec{x_j}|\mathrm{MW})$, is defined analogously. Assuming normally distributed uncertainties, the log likelihood that a star, $j$, is a member of either M31 or the MW given a single set of SPLASH measurements, $i$, is a non-parametric, $N_k$-dimensional Gaussian distribution,

\begin{multline}
    \ln{P(\vec{x_j},\delta\vec{x_j}|\vec{\theta_i},\delta\vec{\theta_i})} = -\frac{N_k}{2} \ln{2\pi} - \frac{1}{2} \sum_{k=1}^{N_k} \ln(\delta x_{j,k}^2 + \delta\theta_{i,k}^2)\\- \frac{1}{2} \sum_{k=1}^{N_k} \frac{(x_{j,k} - \theta_{i,k})^2}{\delta x_{j,k}^2 + \delta\theta_{i,k}^2},
\end{multline}
where $k$ corresponds to a given membership diagnostic (EW$_{\rm Na}$, $X_{\rm CMD}$, $v_{\rm helio}$, or \feh$_{\rm CaT}$) and $N_k$ is the number of diagnostics utilized for a given star. For some stars in our sample, we were unable to measure EW$_{\rm Na}$ and/or \feh$_{\rm CaT}$\ as a consequence of factors such as weak absorption, low S/N, or convergence failure in the Gaussian fit.
For such stars, we excluded EW$_{\rm Na}$ and/or \feh$_{\rm CaT}$\ as a diagnostic, such that $N_k$ = 2$-$3. \added{In four dimensions, the separation between the M31 red giants and MW foreground dwarfs (Figure~\ref{fig:splash}) is analogous to $\sim$2.8$\sigma$ ($\sim$7.4$\sigma$) in units of the covariance matrix of M31's (the MW's) distribution.}


Finally, we computed the probability that a star is a M31 RGB candidate, as opposed to a MW dwarf candidate, from the odds ratio of the posterior probabilities (Eq.~\ref{eq:bayes}),

\begin{equation}
    p_{\rm M31,j} = \frac{ P({\rm M31}|\vec{x_j},\delta\vec{x_j}) /  P({\rm MW}|\vec{x_j},\delta\vec{x_j}) }{1 + P({\rm M31}|\vec{x_j},\delta\vec{x_j}) /  P({\rm MW}|\vec{x_j},\delta\vec{x_j})},
\label{eq:pm31}
\end{equation}
where the proportionality factor in Eq.~\ref{eq:bayes} is equivalent for  $P$(M31$|\vec{x}_j,\delta\vec{x_j}$) and $P$(MW$|\vec{x}_j,\delta\vec{x_j}$).

\subsubsection{Results of Membership Determination}
\label{sec:member_results}

\begin{figure*}
    \centering
    \includegraphics[width=0.8\textwidth]{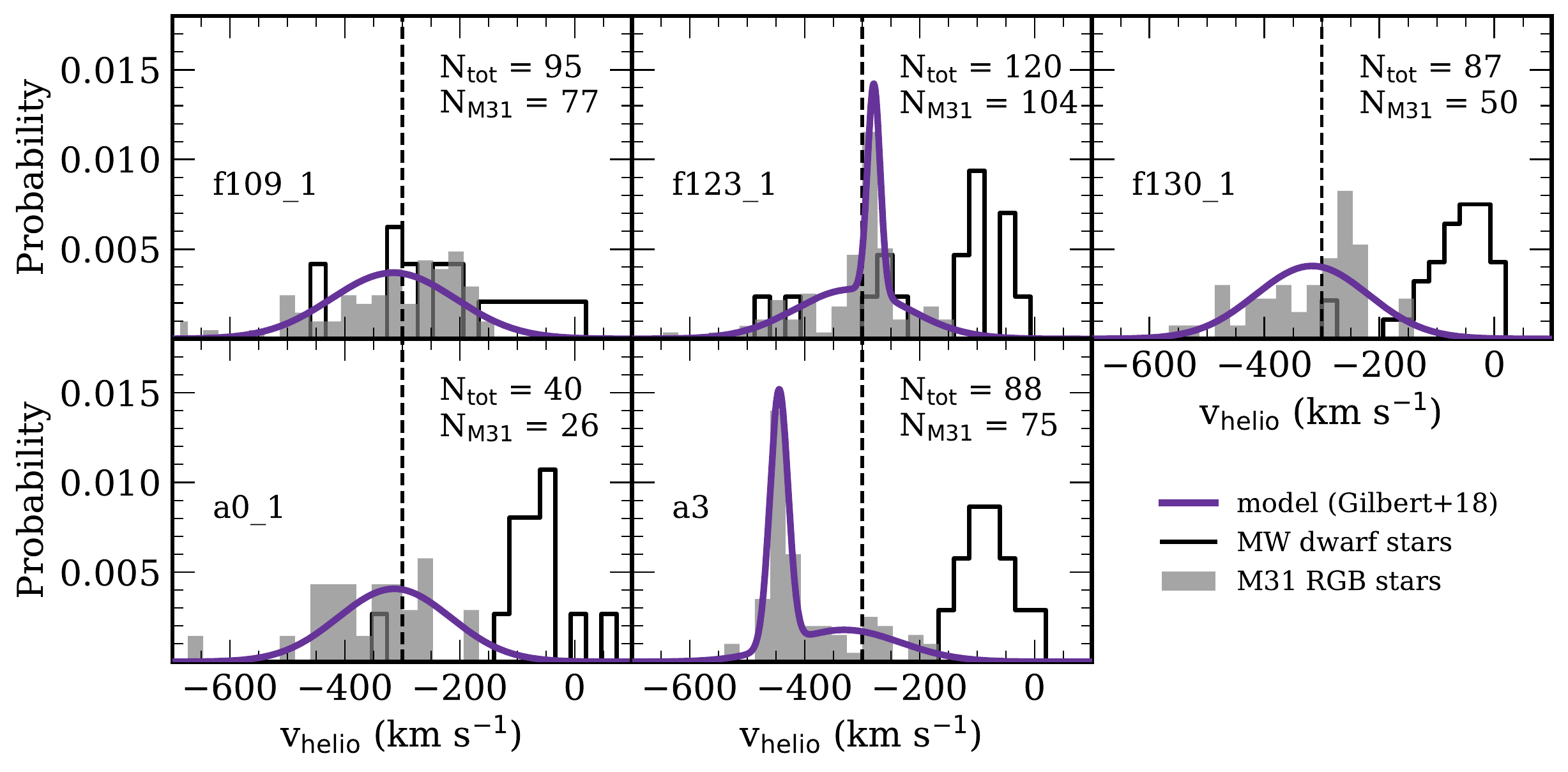}
    \caption{Heliocentric radial velocity distributions for stars with successful velocity measurements (\S~\ref{sec:spec_synth}) in each of the spectroscopic fields (Table~\ref{tab:m31_obs}). Grey histograms represent distributions for likely M31 RGB stars (\S~\ref{sec:prob_model}), whereas black outlined histograms represent likely MW foreground dwarf stars. \added{The total number of stars with velocity measurements ($N_{\rm tot}$) and the subset of M31 RGB stars ($N_{\rm M31}$) are indicated for each field.} The systemic velocity of M31 ($v_{\rm sys} = -300$ km s$^{-1}$) is indicated by a dashed vertical line. The adopted velocity model for M31 RGB stars in each field (\S~\ref{sec:vel}; \citealt{Gilbert2018}) is overplotted as a purple curve. Halo stars are present in each field, where they are distributed in a kinematically hot component centered near the systemic velocity. The velocity distributions of \deleted{the 9, 23, and 31 kpc fields} f109\_1, f130\_1, a0\_1 are fully consistent with a ``smooth'' stellar halo.
    The Southeast shelf, a tidal feature potentially originating from the GSS progenitor \citep{Fardal2007,Gilbert2007}, is evident in \deleted{the 18 kpc halo field} f123\_1, whereas the substructure in \deleted{the 33 kpc} fields a3\_1 \added{and} a3\_2 correspond to the GSS.}
    \label{fig:vel}
\end{figure*}

Figure~\ref{fig:members} summarizes our membership determination for 426 total stars with successful radial velocity measurements across the six spectroscopic fields first presented in this work. The probability distribution is strongly bimodal, where the majority of stars are either secure ($p_{{\rm M31},j}$ $\gtrsim$ 0.75) M31 RGB ($N_{\rm M31} = \replaced{329}{328}$) or MW dwarf ($N_{\rm MW}$ = \replaced{91}{92}) candidates, excepting 6 stars with intermediate properties (0.5 $<$ $p_{{\rm M31},j}$ $\lesssim$ 0.75).\footnote{Our M31 membership yield is high, given that we had prior knowledge of the velocities of individual stars in each field from existing $\sim$1 hr DEIMOS observations (\S~\ref{sec:fields}). When designing our slitmasks for 5+ hr exposures, we prioritized targets known to have a high likelihood M31 membership based on this information.}

Figure~\ref{fig:members} also includes a homogeneously re-evaluated membership determination for fields H, S, D, and f130\_2 (Figure~\ref{fig:m31_loc}; \citealt{Escala2020}), which we further analyze in this work. Across these four fields, 346 stars have successful radial velocity measurements, \replaced{221}{210} (75) of which are classified as secure M31 (MW) stars. In total, 61 stars in these fields have intermediate properties, most of which (60) are located in field D. Owing to the presence of M31's northeastern disk at MW-like line-of-sight velocity ($v_{\rm helio} \sim -130$ \kms; \citealt{Escala2020}), we calculated M31 membership probabilities in D without the use of radial velocity as a diagnostic. Then, we classified all stars with ($v_{\rm helio} - \delta v_{\rm helio}$) $>$ $-$100 \kms\ as MW contaminants, as in \citet{Escala2020}. 

In order to maximize our sample size across all spectroscopic fields, we considered stars that are more likely to belong to M31 than the MW ($p_{{\rm M31},j} > 0.5$) to be M31 members in the following analysis. For stars in common between our dataset and SPLASH, we recovered \replaced{91.7}{91.1}\% of stars classified as M31 RGB stars by \citet{Gilbert2006}, where we used an equivalent definition of membership ($\langle L_{{\rm splash},j}\rangle$ $>$ 0). The excess MW contamination is 0.30\%, in addition to the expected 2-5\% from \citeauthor{Gilbert2006}'s method. The \replaced{8.3}{8.6}\% discrepancy results from stars at MW-like heliocentric velocities that we conservatively classified as MW stars, where these stars are considered M31 RGB stars in SPLASH.

\subsection{Kinematics of M31 RGB Stars}
\label{sec:vel}

Figure~\ref{fig:vel} illustrates the heliocentric radial velocity distributions for all stars with successful measurements (\S~\ref{sec:abund}), including both M31 RGB stars and MW dwarf stars (\S~\ref{sec:prob_model}), across the spectroscopic fields. We also show the adopted Gaussian mixture models \citep{Gilbert2018} describing the velocity distribution for each field, which were computed using over 5,000 spectroscopically confirmed M31 RGB stars across 50 fields in M31's stellar halo. \citeauthor{Gilbert2018} omitted radial velocity as a membership diagnostic (\S~\ref{sec:membership}) in their analysis and simultaneously fit for contributions from M31 and MW components to obtain kinematically unbiased models for each field. Table~\ref{tab:kinematic_decomp} presents the parameters characterizing the velocity model for each field, where we assumed 50$^{\rm th}$ percentile values of \citeauthor{Gilbert2018}'s marginalized posterior probability distribution functions. For the stellar halo components, we transformed the mean velocity from the Galactocentric to heliocentric frame using the median right ascension and declination of all stars in a given field \citep{Gilbert2018}.

We confirmed that the observed velocity distribution for each field is consistent with its velocity model using a two-sided Kolmogorov-Smirnov test. As discussed in \S~\ref{sec:fields}, f109\_1, f130\_1, and a0\_1 probe the ``smooth'' stellar halo of M31 with no detected substructure, whereas fields f123\_1 and a3 show clear evidence of substructure known to be associated with the Southeast shelf \citep{Fardal2007,Gilbert2007,Gilbert2019,Escala2020} and the GSS. Owing to the spatial proximity (Figure~\ref{fig:m31_loc}) and kinematical similarity \citep{Gilbert2007,Gilbert2009a,Gilbert2018} between fields f130\_1 (this work) and f130\_2 \citep{Escala2020,Escala2019} and fields a3 (this work), we consider them together in our subsequent abundance analysis (\S~\ref{sec:abund}).

\subsubsection{Substructure Probability}
\label{sec:psub}

\begin{figure*}
    \centering
    \includegraphics[width=0.8\textwidth]{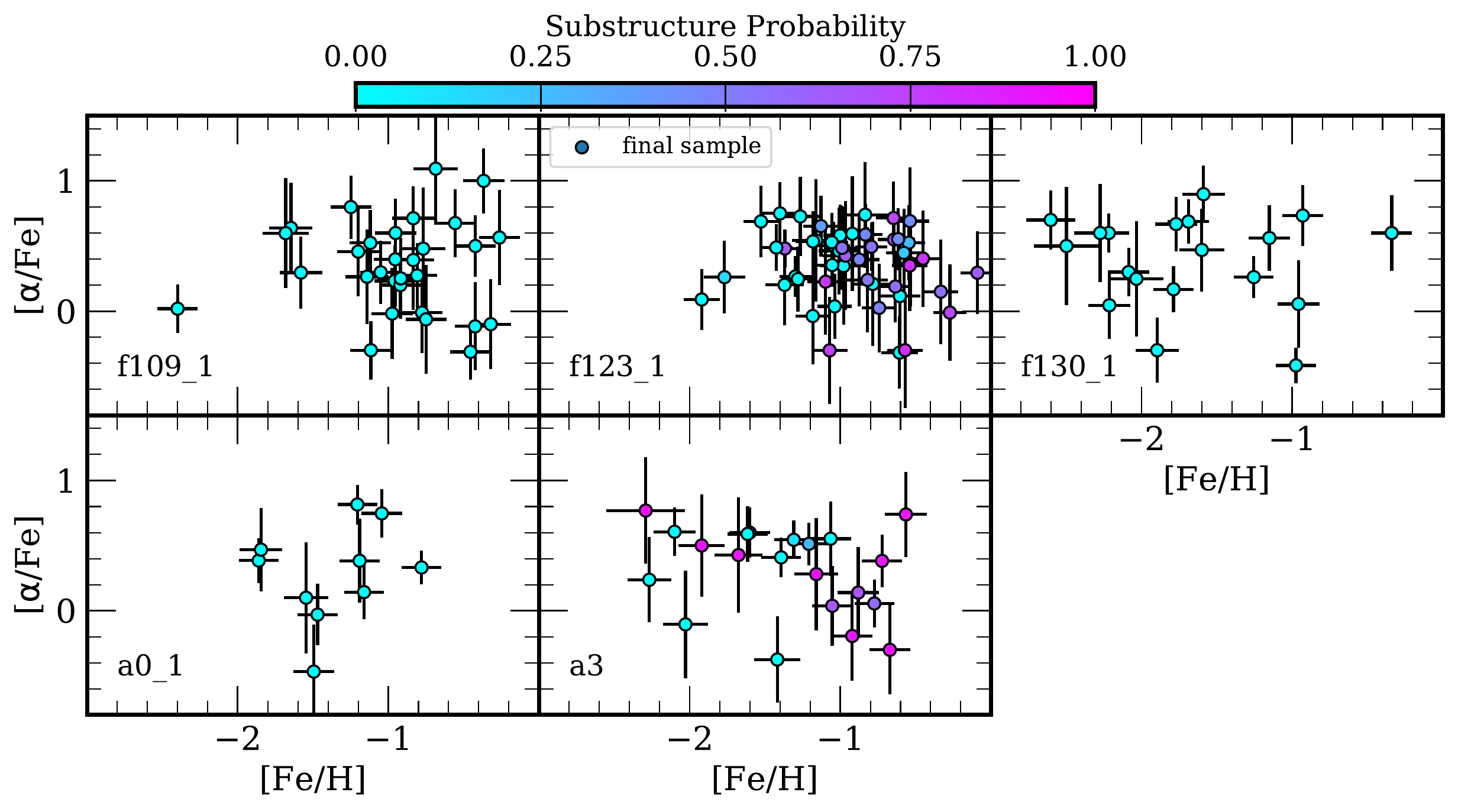}
    \caption{\alphafe\ versus \feh\ for M31 RGB stars in the spectroscopic fields (Table~\ref{tab:m31_obs}), color-coded according to the probability of belonging to substructure (Eq.~\ref{eq:psub}; \S~\ref{sec:psub}).\deleted{as in Figure~\ref{fig:cmd}.} The final sample selection criteria are described in \S~\ref{sec:sample}. 
    Stars belonging to the stellar halo are $\alpha$-enhanced on average, but span a wide range of \alphafe. Most notably, $\alpha$-poor stars (\alphafe\ $\lesssim$ 0) are consistently present in the stellar halo of M31.}
    \label{fig:alphafe_vs_feh}
\end{figure*}

Based on the velocity model for each field, we assigned a probability of belonging to substructure in M31's stellar halo, $p_{\rm sub}$, to every star identified as a likely RGB candidate ($p_{\rm M31}$ $>$ 0.5; Eq.~\ref{eq:pm31}). For smooth stellar halo fields, $p_{\rm sub}$ = 0. For fields with substructure, we computed $p_{\rm sub}$ using an equation analogous to Eq.~\ref{eq:pm31}, where the Bayesian odds ratio is substituted with the relative likelihood that a M31 RGB star belongs to substructure versus the halo,
\begin{equation}
    \mathcal{L}_{{\rm sub},j} = \frac{ f_{\rm sub} \mathcal{N}(v_j|\mu_{\rm sub}, \sigma_{\rm sub}^2)}{f_{\rm halo} \mathcal{N}(v_j|\mu_{\rm halo}, \sigma_{\rm halo}^2)},
    \label{eq:psub}
\end{equation}
where $v_j$ is the heliocentric velocity of an individual star and $\mu$, $\sigma$, and $f$ are the mean, standard deviation, and fractional contribution of the substructure or halo component in the Gaussian mixture describing the velocity distribution for a given field. In contrast, \citet{Escala2020} calculated $p_{\rm sub}$ based on the full posterior distributions, as opposed to 50$^{\rm th}$ percentiles alone, of their velocity models for fields with kinematical substructure. Given that we included abundance measurements of M31 RGB stars from \citet{Escala2020} in this work, we re-calculated $p_{\rm sub}$ for such fields (H, S, and D; Figure~\ref{fig:m31_loc}) based on their 50$^{\rm th}$ percentiles, as above. In the following abundance analysis (\S~\ref{sec:abund}), we incorporated $p_{\rm sub}$ as a weight when determining the chemical properties of the stellar halo.

\section{The Chemical Properties of the Inner Stellar Halo} \label{sec:abund}

We measured \feh\ and \alphafe\ for \replaced{129}{128} M31 RGB stars across six spectroscopic fields spanning $8-34$ kpc in the stellar halo. We measured \feh\ for 80 additional RGB stars for which we could not measure \alphafe. In combination with measurements from previous work by our collaboration \citep{Escala2020,Gilbert2019,Gilbert2020}, we have increased the sample size of individual \alphafe\ and \feh\ measurements in M31 to \replaced{230}{229} RGB stars. 

Figure~\ref{fig:alphafe_vs_feh} shows \feh\ and \alphafe\ measurements for each field, \deleted{including TiO stars that have been discarded from the final sample (\S~\ref{sec:sample}),} where we have color-coded each star according to its probability of belonging to substructure (Eq.~\ref{eq:psub}; \S~\ref{sec:psub}). Table~\ref{tab:kinematic_decomp} summarizes the chemical properties for each kinematical component present in a given field, including previously published inner halo fields with abundance measurements \citep{Escala2019,Escala2020,Gilbert2019}. We calculated each average chemical property from a bootstrap resampling of 10$^{4}$ draws of the final sample for each field, including weighting by the inverse square of the measurement uncertainty and the probability that a star belongs to a given kinematical component (\S~\ref{sec:psub}).

Hereafter, we predominantly restricted our abundance analysis to RGB stars (\S~\ref{sec:prob_model}) from all six fields that are likely to be dynamically associated with the kinematically hot stellar halo of M31. Additionally, we incorporated likely M31 halo stars from inner halo fields with existing abundance measurements \citep{Escala2020,Gilbert2019} into our final sample. We will analyze the chemical composition of RGB stars likely belonging to substructure, such as the Southeast shelf (f123\_1) and outer GSS (a3), in a companion study (I.\@ Escala et al., in preparation). A catalog of stellar parameters and abundance measurements for M31 RGB stars in the six spectroscopic fields is presented in Appendix~\ref{sec:appendix}. 

\added{This section is organized as follows. \S~\ref{sec:sample} defines our final spectroscopic sample and discusses potential observational biases, and \S~\ref{sec:abund_individual} presents a brief overview of our results pertaining to each newly published spectroscopic field. In \S~\ref{sec:abund_gradients}, we measure spectral synthesis based radial abundance gradients for M31's inner stellar halo and compare to metallicity gradient predictions from photometry. Lastly, \S~\ref{sec:low_alpha} compares our enlarged sample of abundance measurements in M31's stellar halo to its dSphs.}

\subsection{Sample Selection and Potential Biases} \label{sec:sample}

\begin{figure*}
    \centering
    \includegraphics[width=0.8\textwidth]{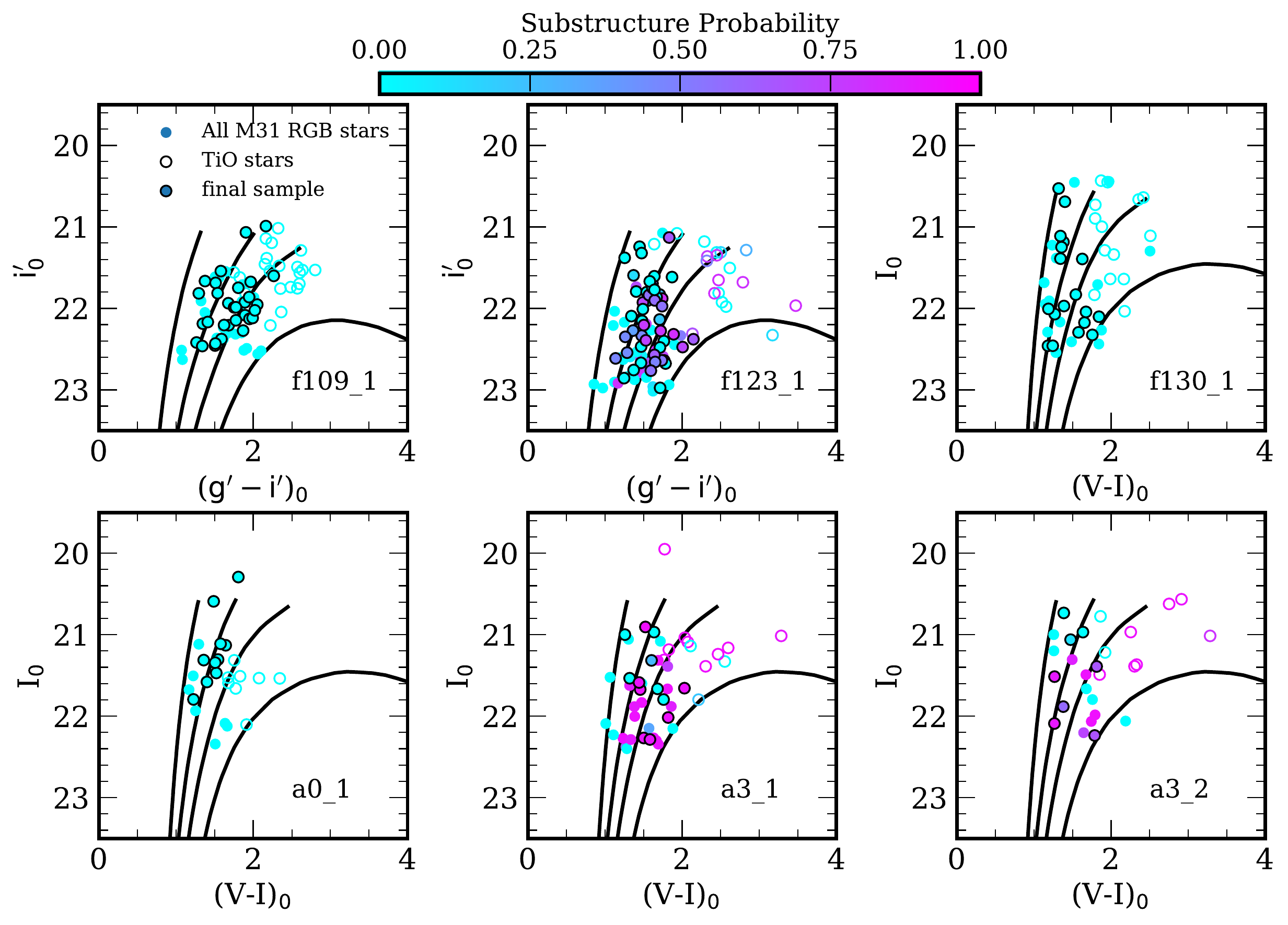}
    \caption{Same as Figure~\ref{fig:cmd}, except showing only M31 RGB stars for each field. Points are color-coded according to their probability of belonging to kinematical substructure in the stellar halo (\S~\ref{sec:psub}, Eq.~\ref{eq:psub}). Open circles indicate M31 RGB stars with strong TiO absorption, which we excluded from our final sample (\S~\ref{sec:sample}).}
    \label{fig:cmd_psub}
\end{figure*}

\begin{table*}
\centering
\begin{threeparttable}
\caption{Kinematical\tnote{a} and Chemical\tnote{b} Properties of M31 Fields}
\label{tab:kinematic_decomp}
\begin{tabular*}{\textwidth}{lcccccccccc}
\hline
\hline
Field & \multicolumn{1}{p{1.0cm}}{\centering $r_{\rm proj}$\\(kpc)} & Comp. &
\multicolumn{1}{p{1.3cm}}{\centering $\mu$\\(\kms)}
& \multicolumn{1}{p{1.3cm}}{\centering $\sigma$\\(\kms)}
& $f$ & $\langle$[Fe/H]$\rangle$ & $\sigma$([Fe/H]) & $\langle$[$\alpha$/Fe]$\rangle$ & $\sigma$([$\alpha$/Fe]) & N$_{\rm [\alpha/Fe] }$ \\[0.4em]
\hline

f109\_1 & 9 & Halo & $-$315.7  & 108.2 & 1.0 &  $-$0.93$^{+0.08}_{-0.09}$ & 0.45 $\pm$ 0.09 & 0.32 $\pm$ 0.08 & 0.36$^{+0.04}_{-0.05}$ & 30 \\[0.3em]

H & 12 & Halo & $-$315.1 & 108.2 & 0.44 &$-$1.30 $\pm$ 0.11 & 0.45$^{+0.07}_{-0.08}$ &0.45$^{+0.12}_{-0.13}$ & 0.42$^{+0.09}_{-0.14}$ & 16\\[0.3em]

& & SE Shelf & $-$295.4 & 65.8 & 0.56 & $-$1.30$^{+0.13}_{-0.12}$ & 0.49$^{+0.08}_{-0.09}$ &0.53$^{+0.08}_{-0.10}$ & 0.36$^{+0.09}_{-0.11}$ & \\[0.3em]

f207\_1 & 17 & Halo & $-$319.6 & 98.1 & 0.35 & $-$1.04$^{+0.09}_{-0.07}$ & 0.26 $\pm$ 0.04 & 0.53 $\pm$ 0.07 & 0.16 $\pm$ 0.04 & 21 \\[0.3em]

& & GSS & $-$529.4 & 24.5 & 0.33 & $-$0.87$^{+0.09}_{-0.10}$ & 0.31 $\pm$ 0.06 & 0.44$^{+0.04}_{-0.05}$ & 0.16 $\pm$ 0.03 & \\[0.3em]

& & KCC & $-$427.3 & 21.0 & 0.32 & $-$0.79 $\pm$ 0.07 & 0.20 $\pm$ 0.04 & 0.54 $\pm$ 0.06 & 0.14 $\pm$ 0.02 & \\[0.3em]

f123\_1 & 18 & Halo & $-$318.2 & 98.1 & 0.68 &  $-$1.00 $\pm$ 0.06 & 0.36 $\pm$ 0.04 & 0.40 $\pm$ 0.04 & 0.23 $\pm$ 0.03 & 49\\[0.3em]

& & SE Shelf & $-$279.9 & 11.0 & 0.32 & $-$0.71 $\pm$ 0.07 & 0.32$^{+0.04}_{-0.05}$ & 0.41$^{+0.04}_{-0.05}$ & 0.23 $\pm$ 0.04 & \\[0.3em]

S & 22 & Halo & $-$318.8 & 98.1 & 0.28 & $-$0.66$^{+0.16}_{-0.18}$ & 0.44$^{+0.07}_{-0.10}$ &0.49$^{+0.05}_{-0.06}$ & 0.21$^{+0.05}_{-0.04}$ & 20\\[0.3em]

& & GSS & $-$489.0 & 26.1 & 0.49 & $-$1.02$^{+0.15}_{-0.14}$ & 0.45$^{+0.10}_{-0.11}$ &0.38$^{+0.17}_{-0.19}$ & 0.45$^{+0.07}_{-0.08}$ & \\[0.3em]

& & KCC & $-$371.6 & 17.6 & 0.22 & $-$0.71 $\pm$ 0.11 & 0.27 $\pm$ 0.09 & 0.35$^{+0.08}_{-0.09}$ & 0.18$^{+0.04}_{-0.05}$ & \\[0.3em]

f130\tnote{c} & 23 & Halo & $-$317.3 & 98.1 & 1.0 & $-$1.64 $\pm$ 0.11 & 0.59$^{+0.08}_{-0.07}$ & 0.39$^{+0.09}_{-0.10}$ & 0.38$^{+0.06}_{-0.09}$ & 30\\[0.3em]

D & 26 & Halo & $-$317.1 & 98.0 & 0.57 & $-$1.00$^{+0.17}_{-0.19}$ & 0.68$^{+0.12}_{-0.14}$ &0.55 $\pm$ 0.13 & 0.40$^{+0.06}_{-0.08}$ & 23\\[0.3em]

& & Disk & $-$128.4 & 16.2 & 0.43 & $-$0.82 $\pm$ 0.09 & 0.28$^{+0.07}_{-0.09}$ & 0.60$^{+0.09}_{-0.10}$ & 0.28$^{+0.05}_{-0.06}$ & \\[0.3em]

a0\_1 & 31 & Halo & $-$314.0 & 98.0 & 1.0 & $-$1.35 $\pm$ 0.10 & 0.33$^{+0.06}_{-0.07}$ & 0.40 $\pm$ 0.10 & 0.30 $\pm$ 0.07 & 10 \\[0.3em]

a3\tnote{c} & 33 & Halo & $-$331.7 & 98.0 & 0.44 & $-$1.48$^{+0.12}_{-0.13}$ & 0.45$^{+0.06}_{-0.07}$ & 0.41$^{+0.05}_{-0.07}$ & 0.24 $\pm$ 0.06 & 21\\[0.3em]

& & GSS & $-$444.6 & 15.7 & 0.56 &  $-$1.11$^{+0.12}_{-0.13}$ & 0.46$^{+0.06}_{-0.07}$ & 0.34$^{+0.08}_{-0.09}$ & 0.30 $\pm$ 0.05 & \\[0.1em]

\hline
\end{tabular*}
\begin{tablenotes}
\item Note. \textemdash\ The columns of the table correspond to field name, projected radial distance from the center of M31, kinematical component, mean heliocentric velocity, velocity dispersion, fractional contribution of the given kinematical component, mean \feh, spread in \feh, mean \alphafe, spread in \alphafe, and total number of RGB stars in a given field with \alphafe\ measurements (regardless of component association).
Chemical properties were calculated from a bootstrap resampling of the final sample, including weighting by the inverse variance of the measurement uncertainty and the probability that a star belongs to a given kinematical component.
\item[a] The parameters of the velocity model are the 50$^{\rm th}$ percentiles of the marginalized posterior probability distribution functions. These were computed by \citet{Gilbert2018} for all fields except H, S, and D. We have transformed the 50$^{\rm th}$ percentile values for the stellar halo components from the Galactocentric to heliocentric frame, based on the median right ascension and declination of all stars in a given field. For fields H, S, and D, \citet{Escala2020} fixed the stellar halo component to the parameters derived by \citet{Gilbert2018} to independently compute the posterior distributions.
\item[b] Chemical abundances for fields f109\_1, f123\_1, f130\_1, a0\_1, and a3 are first presented in this work. We have included chemical properties of previously published M31 fields H, S, D, and f130\_2 \citep{Escala2019,Escala2020} and f207\_1 \citep{Gilbert2019} for reference. We further analyze the halo populations of H, S, f130\_2, and f207\_1 in this work.
\item[c] We combined the chemical abundance samples for fields f130\_1 (this work) and f130\_2 \citep{Escala2019,Escala2020} given their proximity (Figure~\ref{fig:m31_loc}) and the consistency of their velocity distributions \citep{Gilbert2007,Gilbert2018}. The same is true for fields a3\_1 and a3\_2 (this work), where \citet{Gilbert2009a,Gilbert2018} illustrated the similarity in kinematics between these fields.
\end{tablenotes}
\end{threeparttable}
\end{table*}

We vetted our final sample to consist only of {\it reliable} abundance measurements (\S~\ref{sec:spec_synth}) for M31 RGB stars. Similar to \citet{Escala2019,Escala2020}, we restrict our analysis to M31 RGB stars with $\delta$(\teff)\ $<$ 200 K, $\delta$(\feh) \ $<$ 0.5,  $\delta$(\alphafe) \ $<$ 0.5, and well-constrained parameter estimates based on the 5$\sigma$ $\chi^2$ contours for all fitted parameters (\teff, \feh, and \alphafe). We also require that convergence is achieved in each of the measured parameters (\S~\ref{sec:spec_synth}). Unreliable abundance measurements often result from an insufficient signal-to-noise (S/N) ratio, translating to an effective S/N threshold of $\gtrsim$ 8 \AA$^{-1}$ for robust measurements of \feh\ and \alphafe. Such S/N limitations result in a bias in our final sample against metal-poor stars with low S/N spectra, but do not affect the \alphafe\ distributions. This bias is negligible for our innermost halo fields (f109\_1, f123\_1), whereas it is on the order of 0.10$-$0.15 dex for our remaining fields (f130\_1, a0\_1, a3).

We also manually inspected spectra to exclude stars with clear signatures of strong molecular TiO absorption in the wavelength range 7055$-$7245 \AA\ from our final sample. It is unclear whether abundances measured from TiO stars are accurate owing to the lack of (1) the inclusion of TiO in our synthetic spectral modeling \citep{Kirby2008,Escala2019} and (2) an appropriate calibration sample. For stars with successful radial velocity measurements, 31.5\%, 16.7\%, 34.5\%, 32.5\%, 30.4\%, and 31.3\% of stars in fields f109\_1, f123\_1, f130\_1, a0\_1, a3\_1, and a3\_2 have clear evidence of TiO in their spectra. To be conservative, we excluded 54 M31 RGB stars that showed TiO but otherwise would have made the final sample.  In total, \replaced{129}{128} M31 RGB stars across the six spectroscopic fields pass the above selection criteria, thereby constituting our final sample.\footnote{The final samples for fields H, S, and D are identical between \citet{Escala2020} and this work despite differences in the membership determination (\S~\ref{sec:membership}), whereas the final sample for field f130\_2 contains an additional star. No stars re-classified as nonmembers in the formalism presented in this work were included in the final sample of \citet{Escala2020}. \deleted{Three additional stars across H, S, D, and f130\_2 were re-classified as M31 RGB stars with clear evidence of TiO, but otherwise passing our selection criteria (\S~\ref{sec:sample}), and five stars were re-classified M31 RGB stars with robust measurements of \feh, but not \alphafe.}} \added{Figure~\ref{fig:cmd_psub} presents a CMD of M31 RGB stars in each field, color-coded by the probability that a given star belongs to kinematically identified substructure in the stellar halo (\S~\ref{sec:psub}). We have also highlighted our final sample of stars with reliable abundance measurements and indicated which stars were excluded as a consequence of TiO absorption strength.}


As discussed in detail by \citet{Escala2020}, removing stars on the basis of TiO results in a bias against stars with red colors, which translates to a bias in photometric metallicity (\fehphot) against metal-rich stars. Including the sample presented in this work, this photometric bias ([Fe/H]$_{\rm phot,mem}$ $-$ [Fe/H]$_{\rm phot,final}$) ranges from 0.13$-$0.40 dex per field. For fields a3, more stars kinematically associated with the GSS (\S~\ref{sec:vel}) show evident TiO absorption, whereas for field f123\_1, fewer stars in the Southeast shelf substructure are affected compared to those in the halo. Therefore, the exclusion of TiO stars disproportionately impacts the \fehphot\ bias depending on the given field and kinematical component. However, a bias in \fehphot\ cannot be converted into a bias in spectroscopic \feh, considering that \fehphot\ measurements suffer from degeneracy with stellar age and \alphafe\ from which spectroscopic \feh\ measurements are exempt. 

\subsection{Abundance Distributions for Individual Fields}
\label{sec:abund_individual}

In this section, we provide a brief overview of the chemical abundance properties of the six spectroscopic fields first presented in this work (Table~\ref{tab:m31_obs}; Figure~\ref{fig:m31_loc}). We discuss the global properties of the inner stellar halo in \S~\ref{sec:abund_gradients} and \S~\ref{sec:low_alpha}. 

\begin{itemize}
    \item {\it f109\_1 (9 kpc halo field):} \replaced{The 9 kpc halo field}{Field f109\_1} is the innermost region of the stellar halo of M31 yet probed with chemical abundances \citep{Gilbert2007,Gilbert2014}. It does not contain any detected kinematical substructure (\S~\ref{sec:fields},~\ref{sec:vel}). Excepting fields along the GSS, which have an insufficient number of smooth halo stars to constrain the abundances of the stellar halo component, \replaced{the 9 kpc halo field}{f109\_1} is more metal-rich on average ($\langle$\feh$\rangle$ = $-$0.93$^{+0.08}_{-0.09}$) than the majority of halo components in fields at larger projected distance (Table~\ref{tab:kinematic_decomp}; see also \S~\ref{sec:abund_gradients}). Based on the abundances for this field, the inner halo of M31 may be potentially less $\alpha$-enhanced on average ($\langle$\alphafe$\rangle$ = +0.32 $\pm$ 0.08) than the stellar halo at larger projected distances, although we did not measure a statistically significant \alphafe\ gradient in M31's inner stellar halo (\S~\ref{sec:abund_gradients}). Additionally, we did not find evidence for a correlation between \feh\ and \alphafe\ for this field, where we computed a distribution of correlation coefficients 
    from 10$^{5}$ draws of the measured abundances perturbed by their (Gaussian) uncertainties.
    
    \item {\it f123\_1 (18 kpc halo field):} \replaced{The 18 kpc halo field}{Field f123\_1} is dominated by the smooth stellar halo, but it also has a clear detection of substructure (\S~\ref{sec:vel}; Table~\ref{tab:kinematic_decomp}) known as the Southeast shelf (\S~\ref{sec:fields}; \citealt{Fardal2007,Gilbert2007}). We defer further analysis of this component to a companion paper (I.\@ Escala et al., in preparation) owing to its likely connection to the GSS\@. The stellar halo in this field is metal-rich ($\langle$\feh$\rangle$ = $-$0.98 $\pm$ 0.05), $\alpha$-enhanced ($\langle$\alphafe$\rangle$ = $+$0.41$^{+0.03}_{-0.04}$), and exhibits no statistically significant correlation between \alphafe\ and \feh.
    
    \item {\it f130\_1 (23 kpc halo field):} Similar to \replaced{the 9 kpc halo field}{f109\_1}, \replaced{the 23 kpc field}{f130\_1} does not possess detectable substructure. We combined the chemical abundance samples for this field with f130\_2 \citep{Escala2019,Escala2020} due to their proximity (Figure~\ref{fig:m31_loc}) and the consistency of their velocity distributions \citep{Gilbert2007,Gilbert2018}. The average abundances for the combined sample of \replaced{31}{30} M31 RGB stars ($\langle$\feh$\rangle$ = $-$1.6\replaced{2}{4} $\pm$ 0.1\replaced{0}{1}, $\langle$\alphafe$\rangle$ = \replaced{$+$0.38$^{+0.09}_{-0.10}$}{$+$0.39$^{+0.09}_{-0.10}$}; Table~\ref{tab:kinematic_decomp}) agree within the uncertainties with previous determinations from an 11 star sample by \citet{Escala2019,Escala2020}. The lack of a significant trend between \alphafe\ and \feh\ is also maintained by the larger sample. The inner halo \added{as probed by field f130} at 23 kpc appears to be more metal-poor and $\alpha$-enhanced than \replaced{at 9 kpc}{field f109\_1 at 9 kpc}, possibly representing a population that assembled rapidly at early times \citep{Escala2019}. The star formation history inferred for this region of M31's halo indicates that the majority of the stellar population is over 8 Gyr old \citep{Brown2007}, suggesting that an accretion origin would require the progenitor galaxies to have quenched their star formation at least 8 Gyr ago.
    \item {\it a0\_1 (31 kpc halo field):} The stellar population \replaced{in the 31 kpc halo field}{at the location of a0\_1} is solely associated with the kinematically hot component of the stellar halo (\S~\ref{sec:vel}). The 10 RGB star sample in this field suggests a positive trend between \alphafe\ and \feh\ ($r$ = 0.24$^{+0.12}_{-0.24}$) that is likely a consequence of small sample size. Despite its similar projected distance from the center of M31 and kinematical profile, \replaced{the 33 kpc halo field}{a0\_1} appears to be more metal-rich ($\langle$\feh$\rangle$ = $-$1.35 $\pm$ 0.10) than \replaced{the 23 kpc field}{f130}, although it may be comparably $\alpha$-enhanced (Table~\ref{tab:kinematic_decomp}). Nearby {\it HST}/ACS fields at 35 kpc (Figure~\ref{fig:m31_loc}) imply a mean stellar age of 10.5 Gyr \citep{Brown2008} for the vicinity, compared to a mean stellar age of 11.0 Gyr at 21 kpc \added{(near f130)}. The full age distributions suggest that the stellar populations are in fact distinct between the 21 and 35 kpc ACS fields: the star formation history of the 35 kpc field is weighted toward more dominant old stellar populations and is inconsistent with the star formation history at 21 kpc at more than 3$\sigma$ significance \citep{Brown2008}. If applicable to \replaced{the 31 kpc halo field}{a0\_1}, this suggests that its stellar population is both younger and more metal-rich than that at 23 kpc \added{in field f130}.
    \item {\it a3 (33 kpc GSS fields):} \replaced{The 33 kpc field}{Field a3} is dominated by GSS substructure (\S~\ref{sec:fields}, \S~\ref{sec:vel}), such that it may not provide meaningful constraints on the smooth stellar halo in this region. However, the stellar halo at 33 kpc along the GSS \added{in field a3} may be more metal-poor ($\langle$\feh$\rangle$ = $-$1.48$^{+0.12}_{-0.13}$) than at \added{fields f207\_1 and S, which are located at} 17 and 22 kpc along the GSS (Table~\ref{tab:kinematic_decomp}). The abundances in this field clearly show a declining pattern of \alphafe\ with \feh, which is characteristic of dwarf galaxies. A comparison between $\langle$\feh$\rangle$ for the GSS \replaced{at 17 and 22 kpc}{between fields f207\_1 and S} indicates that the GSS likely has a metallicity gradient, as found from photometric based metallicity estimates \citep{Ibata2007,Gilbert2009a,Conn2016,Cohen2018}. We will quantify spectral synthesis based abundance gradients in the GSS and make connections to the properties of the progenitor in future work (I.\@ Escala et al., in preparation).
\end{itemize}

In agreement with previous findings \citep{Escala2019,Escala2020,Gilbert2019}, the M31 fields are $\alpha$-enhanced with a significant spread in metallicity. This implies that stars in the inner stellar halo formed rapidly, such that the timescale for star formation was less than the typical delay time for Type Ia supernovae. For an accreted halo, a spread in metallicity coupled with high $\alpha$-enhancement can indicate contributions from multiple progenitor galaxies or a dominant, massive progenitor galaxy with high star formation efficiency \citep{Robertson2005,Font2006b,Johnston2008,Font2008}. We further discuss formation scenarios for M31's stellar halo in \S~\ref{sec:halo_form}.

\subsection{The Inner vs. Outer Halo}
\label{sec:abund_gradients}

\begin{figure*}
    \centering
    \includegraphics[width=0.9\textwidth]{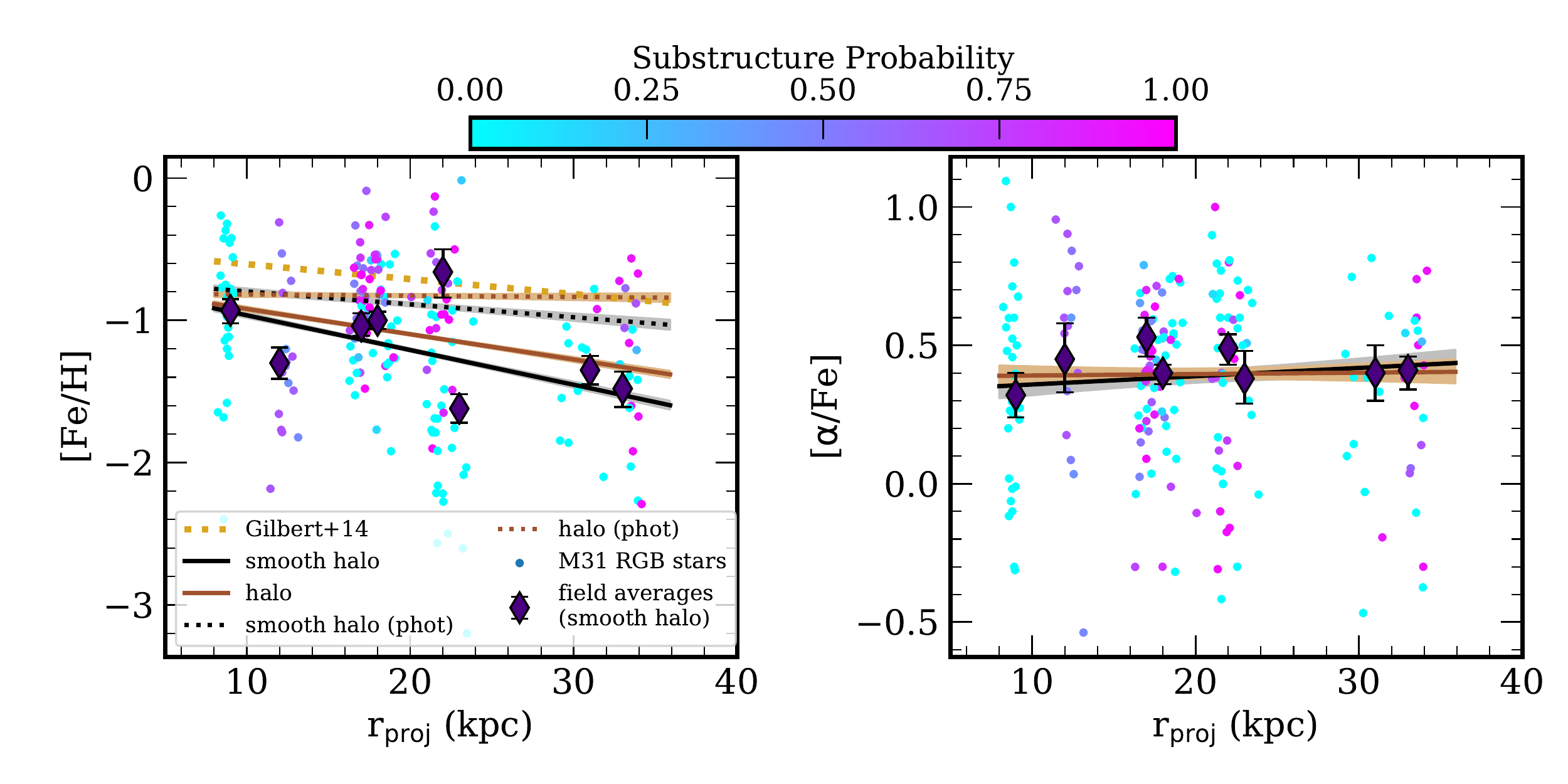}
    \caption{\feh\ (left) and \alphafe\ (right) versus projected distance from the center of M31 for \replaced{198}{197} M31 RGB stars (\citealt{Escala2019,Escala2020,Gilbert2019}; this work) in the inner halo ($r_{\rm proj}$ $\lesssim$ 35 kpc). Stars are color-coded according to their probability of substructure association (\S~\ref{sec:psub}). We omit measurement uncertainties for clarity.
    $\langle$\feh$\rangle$ and $\langle$\alphafe$\rangle$ for the halo component in each spectroscopic field (Table~\ref{tab:kinematic_decomp}) are plotted as indigo diamonds. The photometric metallicity gradient of the stellar halo ($t$ = 10 Gyr, \alphafe\ = 0) measured from spectroscopically confirmed M31 RGB stars 
    \citep{Gilbert2014} is indicated as a dotted gold line, assuming an intercept of \fehphot\ = $-$0.5. This slope is unchanged within the uncertainties with the inclusion/exclusion of substructure. We also display \fehphot\ gradients measured from our \replaced{stellar halo}{final spectroscopic} sample (dotted black and brown lines). The uncertainty envelopes are determined from 16$^{\rm th}$ and 84$^{\rm th}$ percentiles.} 
    \label{fig:abund_gradient}
\end{figure*}

The existence of a steep, global metallicity gradient in M31's stellar halo is well-established from both Ca triplet based \citep{Koch2008} and photometric \citep{Kalirai2006b,Ibata2014,Gilbert2014} metallicity estimates. In addition to radial \feh\ gradients, the possibility of radial \alphafe\ gradients between the inner and outer halo of M31 has recently been explored. From a sample of 70 M31 RGB stars, including an additional 21 RGB stars from \citet{Gilbert2019}, with spectral synthesis based abundance measurements, \citet{Escala2020} found tentative evidence that the inner stellar halo ($r_{\rm proj}$ $\lesssim$ 26 kpc) had higher $\langle$\alphafe$\rangle$ than a sample of four outer halo stars \citep{Vargas2014b} drawn from $\sim$70$-$140 kpc. \citet{Gilbert2020} increased the number of stars in the outer halo (43$-$165 kpc) with abundance measurements from four to nine. In combination with existing literature measurements by \citet{Vargas2014b}, Gilbert et al. found that $\langle$\alphafe$\rangle$ = 0.30 $\pm$ 0.16 for the outer halo. This value is formally consistent with the average $\alpha$-enhancement of M31's inner halo from the 91 M31 RGB star sample ($\langle$\alphafe$\rangle$ = 0.45 $\pm$ 0.06), indicating an absence of a gradient between the inner and outer halo, with the caveat that the sample size in the outer halo is currently limited.

Despite the formal agreement between $\langle$\alphafe$\rangle$ in the inner and outer halo, Gilbert et al. found that \feh\ and \alphafe\ measurements of M31's outer halo are similar to those of M31 satellite dwarf galaxies \citep{Vargas2014a,Kirby2020} and the MW halo \citep[e.g.,][]{Ishigaki2012,Hayes2018}. In comparison, \alphafe\ is higher at fixed \feh\ in M31's inner halo than in its dwarf galaxies \citep{Escala2020}. This suggests that M31's outer halo may have a more dominant population of stars with lower \alphafe\ than the inner halo. This difference implies that the respective stellar halo populations may be, in fact, distinct.\added{\footnote{In order to better constrain $\langle$\alphafe$\rangle$ in the sparsely populated outer halo, abundance measurements from coadded spectra of heterogeneous samples of stars (e.g., obtained using the method of \citealt{Wojno2020arXiv}), will likely be necessary.}}

With the contribution from this work of \replaced{129}{128} M31 individual RGB stars to the existing sample of literature \feh\ and \alphafe\ measurements \citep{Vargas2014b,Escala2019,Escala2020,Gilbert2019,Gilbert2020}, we \added{re-determined the average values for the inner halo,} \replaced{obtained}{obtaining} $\langle$\feh$\rangle$ = $-$1.08 $\pm$ 0.04 ($-$1.17 $\pm$ 0.04) and $\langle$\alphafe$\rangle$ = 0.40 $\pm$ 0.03 (0.39 $\pm$ 0.04) for M31's inner halo when excluding (including) substructure. From \replaced{our}{this} enlarged sample of inner halo abundance measurements, we assessed the presence of \added{spectral synthesis based radial} gradients in \feh\ and \alphafe\ \replaced{in the inner stellar halo of M31}{over the radial range spanned by our spectroscopic sample (8--34 kpc)}. 

Taking stars in the kinematically hot stellar halo to have $p_{\rm sub}$ $<$ 0.5 (Eq.~\ref{eq:psub}), we identified \replaced{123}{122} (75) stars that are likely associated with the ``smooth'' stellar halo (kinematically cold substructure) within a projected distance of $r_{\rm proj}$ $\lesssim$ 35 kpc of M31. We emphasize that these numbers are simply to provide an idea of the relative contribution of each component to the stellar halo--in the subsequent analysis, we {\it do not} employ any cuts on $p_{\rm sub}$, but rather incorporate M31 RGB stars, regardless of halo component association, by using $p_{\rm sub}$ as a weight. We have excluded all M31 RGB stars in field D (Figure~\ref{fig:m31_loc}) from our analysis sample owing to the presence of M31's northeastern disk. To determine the radial abundance gradients of the smooth inner halo, we fit a line using an MCMC ensemble sampler \citep{Foreman-Mackey2013}, weighting each star by $p_{\rm halo} = 1 - p_{\rm sub}$ and the measurement uncertainty. We also determined the radial gradients in the case of including substructure by removing $p_{\rm sub}$ as a weight.

We measured a radial \feh\ gradient of $-$0.025 $\pm$ 0.002 dex kpc$^{-1}$ between 8--34 kpc, with an intercept at $r_{\rm proj} = 0$ of $-$0.72 $\pm$ 0.03, for the smooth halo. The inclusion of substructure results in a shallower \feh\ gradient ($-$0.018 $\pm$ 0.001 dex kpc$^{-1}$) with a similar intercept ($-$0.74 $\pm$ 0.03), reflecting the preferentially metal-rich nature of substructure in M31's halo \citep{Font2008,Gilbert2009b}. We did not find statistically significant radial \alphafe\ gradients between 8--34 kpc in M31's stellar halo, both excluding (0.0029 $\pm$ 0.0027 dex kpc$^{-1}$) and including (0.00048 $\pm$ 0.00261 dex kpc$^{-1}$) substructure. Figure~\ref{fig:abund_gradient} shows \feh\ and \alphafe\ versus projected radial distance \added{within the inner 40 kpc of M31's halo}. \replaced{In Figure~\ref{fig:abund_gradient}}{In this Figure}, we refer to the stellar population including substructure simply by ``halo'', in contrast to ``smooth halo'', which excludes substructure. 

\added{In addition to showing}\deleted{including} our measured radial gradients, \added{we also include}\deleted{and} the photometric metallicity gradient of \citet{Gilbert2014} \added{in Figure~\ref{fig:abund_gradient}}\deleted{measured between 10--90 kpc}. Using a sample of over 1500 spectroscopically confirmed M31 RGB stars, \citet{Gilbert2014} measured a radial \fehphot\ gradient of $-$0.011 $\pm$ 0.001 dex kpc$^{-1}$ between 10$-$90 projected kpc\deleted{in the stellar halo}. The radial \added{photometric metallicity} gradients measured \added{by \citeauthor{Gilbert2014}} with and without kinematical substructure were found to be consistent within the uncertainties. \added{In contrast to \citet{Gilbert2014}}, our results suggest that the radial \feh\ gradient of the smooth halo is inconsistent with that of the halo including substructure over the probed radial range. The substructures in our inner halo fields are likely GSS progenitor debris \citep{Fardal2007,Gilbert2007,Gilbert2009a}, thus the change in slope at its inclusion may reflect a convolution with the distinct metallicity gradient of the GSS progenitor (I. Escala et al., in preparation).

\subsubsection{Spectroscopic vs. Photometric Metallicity Gradients}

In order to control for differences in sample size, target selection, number and locations of spectroscopic fields utilized, and the radial extent of the measured gradient between this work and \citet{Gilbert2014}, we measured a \fehphot\ (\S~\ref{sec:phot}) gradient from our final \added{spectroscopic} sample \added{of 197 halo stars}, assuming $t$ = 10 Gyr and \alphafe\ = 0 (Figure~\ref{fig:abund_gradient}). We obtained a slope of $-$0.00\replaced{89}{91} $\pm$ 0.001\replaced{8}{9} dex kpc$^{-1}$ ($-$0.00070 $\pm$ 0.0016 dex kpc$^{-1}$) with an intercept of $-$0.71 $\pm$ 0.04 ($-$0.81 $\pm$ 0.03) when including (excluding) substructure. These gradients are inconsistent for $r_{\rm proj}$ $\gtrsim$ 20 kpc, where the \fehphot\ gradient including substructure remains flat as the \fehphot\ gradient of the smooth halo declines. Such a difference is not detected from \citeauthor{Gilbert2014}'s sample of over 1500 RGB stars spanning 10$-$90 kpc. However, the \fehphot\ gradients measured in this work provide a more direct comparison to our spectral synthesis based \feh\ gradients.

A possible explanation for the difference in trends with substructure between spectral synthesis and CMD-based gradients is the necessary assumption of uniform stellar age and $\alpha$-enhancement to determine \fehphot. Although we did not measure a significant radial \alphafe\ gradient (Figure~\ref{fig:abund_gradient}), M31's inner stellar halo has a range of stellar ages present at a given location based on {\it HST} CMDs extending down to the main-sequence turn-off \citep{Brown2006,Brown2007,Brown2008}. If the smooth stellar halo is systematically older than the tidal debris toward 30 kpc, this would steepen the relative \fehphot\ gradient between populations with and without substructure. This agrees with our observation that \fehphot$-$\feh\ is increasingly positive on average toward larger projected distances, where the discrepancy is greater for the smooth halo than in the case of including substructure.

Additionally, a comparatively young stellar population at 10 kpc compared to 30 kpc could result in a steeper \fehphot\ gradient in better agreement with our measured \feh\ gradient. The difference between mean stellar age at 35 kpc and 10 kpc is 0.8 Gyr \citep{Brown2008}, though the mean stellar age of M31's halo does not appear to increase monotonically with projected radius. Assuming constant \alphafe, this mean age difference translates to a negligible gradient between 10$-$35 kpc. 
Thus, a more likely explanation for the discrepancy in slope between the \fehphot\ and \feh\ gradients are uncertainties in the stellar isochrone models at the tip of the RGB.

In general, \fehphot\ is offset toward higher metallicity, where the adopted metallicity measurement methodology can result in substantial discrepancies for a given sample (e.g., \citealt{Lianou2011}). For example, the discrepancy between the CMD-based gradient of \citet{Gilbert2014} and literature measurements from spectral synthesis \citep{Vargas2014b,Gilbert2019,Escala2020} decreases when assuming an $\alpha$-enhancement (\alphafe\ = +0.3) in better agreement with the inner and outer halo \citep{Gilbert2020}. Despite differences in the magnitude of the slope, the behavior with substructure, and intercept of the \feh\ gradient from different metallicity measurement techniques, our enlarged sample of spectral synthesis based \feh\ measurements provides further support for the existence of a large-scale metallicity gradient in the inner stellar halo, where this gradient extends out to at least 100 projected kpc in the outer halo \citep{Gilbert2020}. For a thorough consideration of the implications of steep, large-scale negative radial metallicity gradients in M31's stellar halo, we refer the reader to the discussions of \citet{Gilbert2014} and \citet{Escala2020}.

\subsubsection{The Effect of Potential Sources of Bias on Metallicity Gradients}
\label{sec:abund_gradients_bias}

Alternatively, the discrepancy between the \fehphot\ and \feh\ gradients could be partially driven by the bias against red, presumably metal-rich, stars incurred by the omission stars with strong TiO absorption from our final sample (\S~\ref{sec:sample}). We investigated the impact of potential bias from both (1) the exclusion of TiO stars ($b_{\rm TiO}$) and (2) S/N limitations ($b_{\rm S/N}$; \S~\ref{sec:sample}) on our measured radial \feh\ gradients by shifting each \feh\ measurement for an M31 RGB star in a given field by $b_{\rm TiO} + b_{\rm S/N}$. As in \citet{Escala2020}, we estimated $b_{\rm TiO}$ from the discrepancy in $\langle$\feh$\rangle_{\rm phot}$ between all M31 RGB stars in a given field (including TiO stars) and the final sample (excluding TiO stars). We calculated $b_{\rm S/N}$ from the difference between $\langle$\feh$\rangle$ for the sample of M31 RGB stars with \feh\ measurements in each field (regardless of whether an \alphafe\ measurement was obtained for a given star) and the final sample. These sources of bias do not affect the \alphafe\ distributions.  

Incorporating these bias terms yields a radial \feh\ gradient of $-$0.022 $\pm$ 0.002 dex kpc$^{-1}$ ($-$0.018 $\pm$ 0.001 dex kpc$^{-1}$) and an intercept of $-$0.\replaced{59}{60} $\pm$ 0.03 ($-$0.5\replaced{7}{8} $\pm$ 0.03) without (with) substructure. The primary effect of including bias estimates is a shift toward higher metallicity in the overall normalization of the gradient. 
The gradient slopes calculated including bias estimates are consistent with our previous measurements, which did not account for potential sources of bias. We can therefore conclude that the slopes of radial \feh\ gradients between 8-34 kpc in M31's stellar halo are robust against these two possible sources of bias.

\subsection{Comparing the Stellar Halo to M31 dSphs}
\label{sec:low_alpha}

In $\Lambda$CDM cosmology, simulations of stellar halo formation for M31-like galaxies predict that the chemical abundance distributions of an accreted component should be distinct from the present-day satellite population of the host galaxy \citep{Robertson2005,Font2006b,Tissera2012}, as is observed for the MW \citep{Unavane1996,Shetrone2001,Shetrone2003,Tolstoy2003,Venn2004}. This chemical distinction is driven by the early assembly of the stellar halo, where its progenitors were accreted $\sim$8-9 Gyr ago, as opposed to $\sim$4-5 Gyr ago for surviving satellite galaxies \citep{BullockJohnston2005,Font2006b,Fattahi2020arXiv}. Accordingly, \citet{Escala2020} showed that the \alphafe\ distribution for the metal-rich (\feh\ $>$ $-$1.5) component of M31's smooth, inner stellar halo ($r_{\rm proj}$ $\lesssim$ 26 kpc) is inconsistent with having formed from progenitor galaxies similar to present-day M31 \replaced{satellite}{dwarf spheroidal (dSph)} galaxies with measurements of \feh\ and \alphafe\ ($M_* \sim 10^{5-8} M_\odot$; \citealt{Vargas2014a,Kirby2020}). This sample of M31 \replaced{dwarf galaxies}{dSphs} consisted of NGC 185 and And II \citep{Vargas2014a} and And VII, And I, And V, and And III \citep{Kirby2020}.

However, \citeauthor{Escala2020}\ were unable to statistically distinguish between the \alphafe\ distributions for the low-metallicity (\feh\ $<$ $-$1.5) component of the smooth, inner stellar halo. This could be a consequence of an insufficient sample size at low metallicity, where they identified 29 RGB stars likely belonging to the smooth stellar halo ($p_{\rm halo}$ $>$ 0.5; Eq.~\ref{eq:psub}). Another possibility is that such low-metallicity stars, with lower average $\alpha$-enhancement ($\langle$\alphafe$\rangle$ $\sim$ 0.25; \citealt{Escala2020}), may represent a stellar population in the stellar halo more similar to present-day M31 \replaced{satellite dwarf galaxies}{dSphs}.

\begin{figure}
    \centering
    \includegraphics[width=\columnwidth]{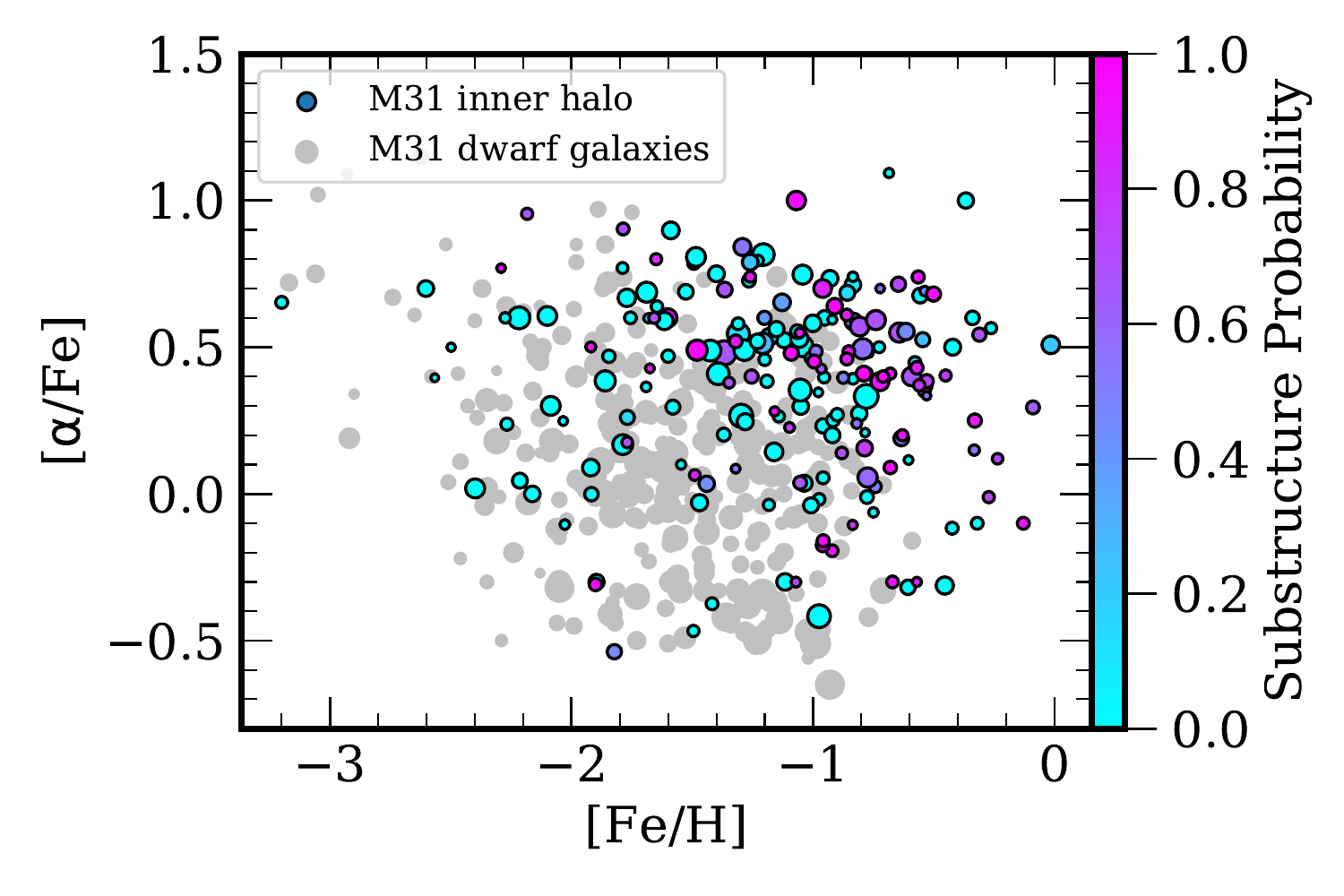}
    \caption{\alphafe\ versus \feh\ for RGB stars in the inner halo of M31 (\citealt{Gilbert2019,Escala2020}; this work) compared to abundance trends for M31 \replaced{dwarf galaxies}{dSphs}. The combined sample of M31 \replaced{dwarf galaxies}{dSphs} consists of NGC 185, NGC 147, And II \citep{Vargas2014a} and And VII, And I, And V, And III, and And X \citep{Kirby2020}. The size of each point is proportional to the inverse variance of the measurement uncertainty, where M31 RGB stars are also color-coded by their probability of belonging to substructure (Eq.~\ref{eq:psub}). The majority of inner halo RGB stars are inconsistent with the stellar populations of present-day M31 \replaced{satellite dwarf galaxies}{dSphs} \citep{Escala2020,Kirby2020}, although some probable halo stars exhibit abundance patterns similar to M31 \replaced{dwarf galaxies}{dSphs} (\S~\ref{sec:low_alpha_cards}).}
    \label{fig:halo_vs_dsphs}
\end{figure}

\begin{figure*}
    \centering
    \includegraphics[width=\textwidth]{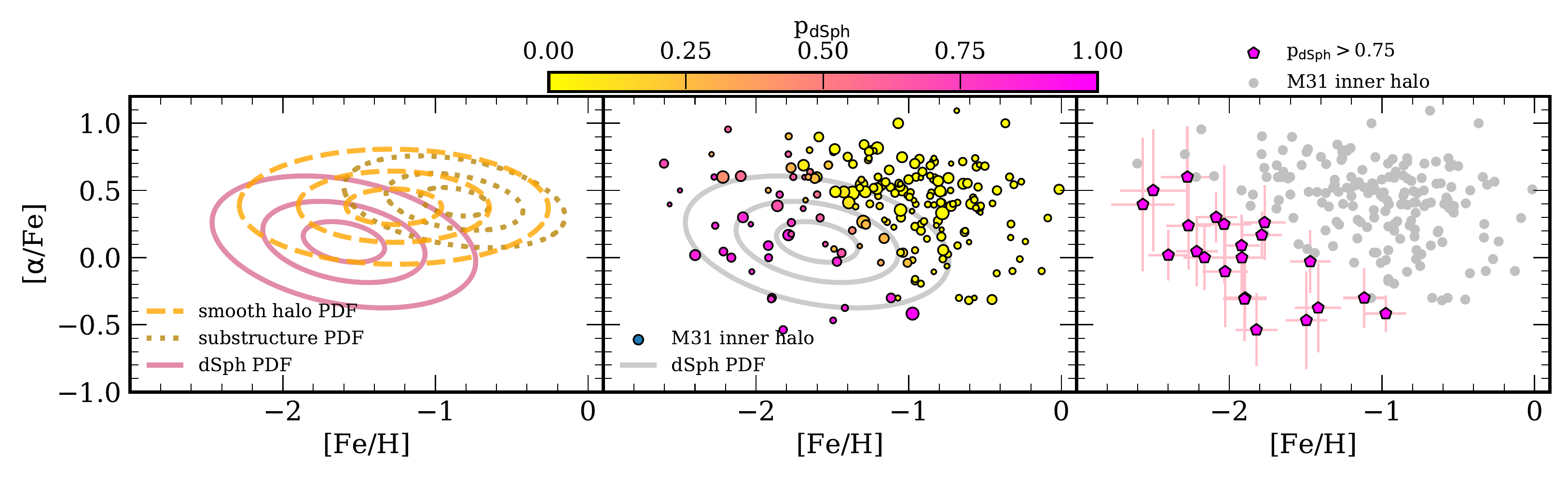}
    \caption{(Left) Probability density functions (PDFs) for the relationship between [$\alpha$/Fe] and [Fe/H] for the smooth component of the inner halo (dashed orange contours; \citealt{Gilbert2019,Escala2020}; this work), substructure in the inner halo (dotted gold contours;  \citealt{Gilbert2019,Escala2020}; this work), and M31 \replaced{satellite dwarf galaxies}{dSphs} (solid pink contours; \citealt{Vargas2014a,Kirby2020}), constructed from bivariate normal distributions (\S~\ref{sec:low_alpha_cards}; Eq.~\ref{eq:bi_normal}; Table~\ref{tab:abund_models}). The contour levels correspond to the 16$^{\rm th}$, 50$^{\rm th}$, and 84$^{\rm th}$ percentiles of the 2D PDFs. 
    (Middle) [$\alpha$/Fe] versus [Fe/H] for individual RGB stars in the inner halo, color-coded by the probability that the star is both kinematically associated with the smooth stellar halo and has abundance patterns similar to M31 \replaced{dwarf galaxies}{dSphs} ($p_{\rm dSph}$; Eq.~\ref{eq:pdwarf}). Point sizes are proportional to the inverse variance of the measurement uncertainties. (Right) [$\alpha$/Fe] versus [Fe/H] for M31 RGB stars identified as secure ($p_{\rm dSph}$ $>$ 0.75; outlined circles) \deleted{and marginal (0.5 $<$ $p_{\rm dSph}$ $<$ 0.75; outlined triangles)} members of the smooth stellar halo that have \replaced{dwarf-galaxy-like}{dSph-like} abundances. \added{The remainder of M31 RGB stars are shown as silver circles.}}
    \label{fig:low_alpha}
\end{figure*}

\subsubsection{1-D Comparisons}
\label{sec:low_alpha_1D}

To investigate whether the \alphafe\ distributions of the stellar halo and \replaced{dwarf galaxies}{dSphs} at low-metallicity are in fact statistically indistinguishable, we repeated the analysis of \citet{Escala2020} with our expanded sample of \replaced{198}{197} inner halo RGB stars (excluding \replaced{the 26 kpc disk field}{field D}; Figure~\ref{fig:m31_loc}) with abundance measurements. 

In contrast to \citet{Escala2020}, we applied corrections to the abundances measured by \citet{Vargas2014a} to place them on the same scale as M31's halo (\citealt{Escala2019,Escala2020,Gilbert2019}; this work) and other \replaced{dwarf galaxy}{M31 dSph} \citep{Kirby2020} measurements. \citet{Kirby2020} found that the \citeauthor{Vargas2014a} measurements were systematically offset toward higher \feh\ by +0.3 dex compared to their measurements, based on an identical sample of spectra of M31 dwarf galaxy stars. \citeauthor{Kirby2020} did not find evidence of a systematic offset between their \alphafe\ measurements and those of \citet{Vargas2014a}. Thus, we adjusted \feh\ values measured by \citet{Vargas2014a} by the mean systematic offset of $-$0.3 dex and did not make any changes to their \alphafe\ values. This correction is reflected for the relevant M31 \replaced{dwarf galaxies}{dSphs} in Figures~\ref{fig:halo_vs_dsphs} and~\ref{fig:low_alpha}.

We constructed \alphafe\ distributions for a mock stellar halo built by destroyed \replaced{dwarf-galaxy-like}{dSph-like} progenitor galaxies, weighting the contribution of each M31 dwarf galaxy \replaced{according to its luminosity function}{by (Eq.~4 of \citealt{Escala2020})},\footnote{We have excluded RGB stars in NGC 185 with \feh\ $>$ $-$0.5 (uncorrected values), owing to the uncertainty in the reliability of \citeauthor{Vargas2014a}'s abundance measurements above this metallicity. \citet{Vargas2014a} calibrated their measurements of bulk \alphafe\ to approximate an $\alpha$-element abundance measured from the arithmetic mean of individual [(Mg,Si,Ca,Ti)/Fe], where their calibration is not valid for \feh\ $>$ $-0.5$. In contrast, we have not applied such a calibration to our \alphafe\ measurements, which are valid for \feh\ $>$ $-0.5$. The exclusion of high-metallicity stars in NGC 185 applies to the analysis in \S~\ref{sec:low_alpha} as well as Figures~\ref{fig:halo_vs_dsphs} and ~\ref{fig:low_alpha}.} 
\begin{equation}
w_{i,j} = \frac{ M_{\star,j} \Phi (L_{V,j}) / N_j }{ \Sigma_{j=1}^{N_{gal}} M_{\star,j} \Phi (L_{V,j})},
\label{eq:p_dsph}
\end{equation}
\added{where $M_{\star, j}$ is the stellar mass and $L_{V,j}$ is the V-band luminosity of the host dSph, $\Phi$ is the V-band luminosity function of present-day M31 dSphs \citep{McConnachie2012}, $N_j$ is the number of RGB stars with abundance measurements in galaxy $j$, and $N_{gal}$ is the total number of M31 dSphs contributing to the abundance distribution of the \replaced{simulated}{mock} stellar halo. We refer the reader to \citet{Escala2020} for further details.}

When re-sampling the observed inner halo abundance distributions, we assigned each star a probability of being drawn according to its $p_{\rm halo}$. Not only did we find that the \alphafe\ distribution of the metal-rich \added{(\feh\ $>$ $-$0.8) and intermediate metallicity ($-$1.5 $<$ \feh\ $<$ $-$0.8) components of} smooth inner halo \replaced{is}{are} inconsistent with that of present-day M31 \replaced{satellite galaxies}{dSphs} at high significance ($p \ll 1$\%), but also that the \alphafe\ distribution of the metal-poor \added{(\feh\ $<$ $-$1.5)} component of the smooth halo disagrees with that of low-metallicity \added{(\feh\ $<$ $-$1.5)} RGB stars in M31 \replaced{satellite galaxies}{dSphs} ($p < 0.3$\%).

Nevertheless, constructing \alphafe\ distributions according to metallicity bins presents a limited, 1-D view of the relationship between \alphafe\ and \feh. Figure~\ref{fig:halo_vs_dsphs} displays \alphafe\ versus \feh\ for the inner halo of M31 compared to M31 \replaced{dwarf galaxies}{dSphs} with abundance measurements, including NGC 147 \citep{Vargas2014a} and And X \citep{Kirby2020}. Evidently, the majority of inner halo RGB stars are inconsistent with the stellar populations of M31 \replaced{dwarf galaxies}{dSphs}, as previously found using the combined 91 RGB star sample of \citet{Escala2020} and \citet{Gilbert2019}, and reinforced by this work. In general, \alphafe\ tends to be higher for M31's inner halo at fixed metallicity than for the \replaced{dwarf galaxies}{dSphs}. \deleted{($\langle$\alphafe$\rangle_{\rm dwarf} = 0.07 \pm 0.02$).} 
Many of the high-metallicity, $\alpha$-enhanced stars in the inner halo have a high probability of being associated with kinematical substructure, such as the GSS. 

In addition to these trends, Figure~\ref{fig:halo_vs_dsphs} suggests the existence of a stellar population preferentially associated with the dynamically hot halo that also has chemical abundance patterns similar to M31 \replaced{dwarf galaxies}{dSphs}. These stars are not contained within a well-defined metallicity bin (such as \feh\ $<$ $-$1.5, the metal-poor bin utilized by \citealt{Escala2020}), but rather span a region of \alphafe\ versus \feh\ space coincident with the mean trend of the \replaced{dwarf galaxies}{dSphs}.

\subsubsection{Modeling 2-D Chemical Abundance Distributions}
\label{sec:low_alpha_cards}

In order to robustly identify M31 halo stars with abundance patterns similar to that of M31 \replaced{satellite dwarf galaxies}{dSphs} (Figure~\ref{fig:halo_vs_dsphs}), we modeled the observed 2-D chemical abundance ratio distributions, as advocated by \citet{Lee2015}. \added{Appendix~\ref{sec:appendix_mw} presents a proof of concept of the statistical method utilized in the following analysis, which is based on the chemical abundance distributions of the MW's stellar halo. Appendix~\ref{sec:appendix_mw} illustrates that differences in populations can be meaningfully distinguished given our typical uncertainties.}

We considered the smooth component of the inner stellar halo, substructure in the inner halo, and M31 \replaced{dwarf galaxies}{dSphs} as distinct \replaced{components}{populations} in our modeling. We expanded upon the M31 \replaced{dwarf galaxy}{dSph} sample analyzed in \S~\ref{sec:low_alpha_1D} by incorporating abundance measurements from galaxies that were previously excluded on the basis of their small sample sizes: And X ($N_{\rm stars}$ = 9 ; \citealt{Kirby2020}) and NGC 147 ($N_{\rm stars}$ = 7 ; \citealt{Vargas2014a}).\footnote{We did not include M32 in our M31 \replaced{dwarf galaxy}{dSph} abundance sample. \citet{Vargas2014a} measured abundances for 3 stars in M32 with \feh\ $<$ $-$0.5 and concluded that they were not representative of the galaxy.} Taking $\delta$([Fe/H]) $<$ 0.5, $\delta$([$\alpha$/Fe]) $<$ 0.5, and \feh\ $<$ $-$0.5, these two additional galaxies result in a sample size of 293 abundance measurements for M31 \replaced{dwarf galaxies}{dSphs}.

Assuming the form of a bivariate normal distribution, the likelihood of observing a given set of abundance measurements ($x_i$, $y_i$) = (\feh$_i$, \alphafe$_i$) and uncertainties ($\delta x_i$, $\delta y_i$) = ($\delta$\feh$_i$, $\delta$\alphafe$_i$)\footnote{We assumed independent errors ($\delta x_i$, $\delta y_i$) in our model. In actuality, we expect that some amount of \deleted{negative} covariance, $\delta_{xy}$, exists between our measurement uncertainties. The net result of such dependent uncertainties \replaced{would}{could} be a perceived \replaced{decrease}{change} in the correlation coefficient, $r$ (Eq.~\ref{eq:bi_normal}). \replaced{Thus, we acknowledge that $r$ may in fact be less negative than suggested by our model (Table~\ref{tab:abund_models})}{However, we found that accounting for $\delta_{xy}$ does not significantly alter our error distribution, so we anticipate that the effect of this covariance on $r$ is minimal.}}
given a model described by means $\vec{\mu} = (\mu_x, \mu_y)$ and standard deviations $\vec{\sigma} = (\sigma_x, \sigma_y)$ is,

\begin{multline}
    \mathcal{L}_i(x_i, y_i, \delta x_i, \delta y_i | \vec{\mu}, \vec{\sigma} ) = \frac{1}{2 \pi \sigma_{x_i} \sigma_{y_i} \sqrt{1 - r^2} } \\\times \exp \left( -\frac{1}{2 (1 - r^2)} \left[ \frac{ (x_i - \mu_x ) }{\sigma_{x_i}^2} + \frac{ (y_i - \mu_y ) }{\sigma_{y_i}^2} \right. \right. \\ \left. \left. - \frac{2 r (x_i - \mu_x) (y_i - \mu_y)}{\sigma_{x_i} \sigma_{y_i}} \right] \right),
    \label{eq:bi_normal}
\end{multline}
where $i$ is an index corresponding to an individual RGB star. We incorporated the measurement uncertainties into Eq.~\ref{eq:bi_normal} via the variable $\sigma_{x_i} = \sqrt{ \sigma_x^2 + \delta_{x_i}^2 }$, which is defined analogously for $\sigma_{y_i}$. The correlation coefficient, $r$, is an additional model parameter that accounts for covariance between $x$ and $y$. 

\begin{table*}
\centering
\begin{threeparttable}
    \caption{2-D Chemical Abundance Ratio Distribution Model Parameters}
    \begin{tabular}{l|ccccc}
    \hline\hline
    
    Model & $\mu_{\rm [Fe/H]}$ & $\sigma_{\rm [Fe/H]}$ & $\mu_{\rm [\alpha/Fe]}$ & $\sigma_{\rm [\alpha/Fe]}$ & $r$ \\\hline
    
    
    Smooth Halo & $-$1.27 $\pm$ 0.05 & 0.54 $\pm$ 0.04 & 0.38 $\pm$ 0.03 & 0.23 $\pm$ 0.03 & $-$0.022 $\pm$ 0.09\\
    
    
    Substructure & $-$0.88 $\pm$ 0.06 & 0.38 $\pm$ 0.05 & 0.42 $\pm$ 0.05 & 0.18 $\pm$ 0.03 & $-$0.320$^{+0.17}_{-0.14}$\\
    
    
   dSphs & $-$1.60 $\pm$ 0.03 & 0.45 $\pm$ 0.02 & 0.12 $\pm$ 0.02 & 0.26 $\pm$ 0.02 & $-$0.297 $\pm$ 0.05\\\hline
    
    \end{tabular}
    \label{tab:abund_models}
\begin{tablenotes}
\item Note. \textemdash\ The columns of the table correspond to the model for the observed 2-D chemical abundance ratio distributions (Figure~\ref{fig:low_alpha}) for a given stellar population, and the parameters describing the bivariate normal probability density functions (Eq.~\ref{eq:bi_normal}).
We fit for the smooth halo and substructure components simultaneously (Eq.~\ref{eq:mixture_model}) using a mixture model. The parameters for the 2-D chemical abundance models are the 50$^{\rm th}$ percentiles of the marginalized posterior probability distribution functions (\S~\ref{sec:low_alpha_cards}), where the uncertainies on each parameter were calculated from the corresponding 16$^{\rm th}$ and 84$^{\rm th}$ percentiles. 
\end{tablenotes}
\end{threeparttable}
\end{table*}

For the full stellar halo sample, we modeled the smooth and substructure components simultaneously by combining the respective likelihood functions (Eq.~\ref{eq:bi_normal}) using a mixture model,

\begin{equation}
    \mathcal{L}_i = (1 - p_{i}^{\rm sub}) \mathcal{L}_{i}^{\rm halo} + p_{i}^{\rm sub} \mathcal{L}_{i}^{\rm sub},
    \label{eq:mixture_model}
\end{equation}
where $p_{i}^{\rm sub}$ is the \added{kinematically-based} probability that a star belongs to substructure in M31's stellar halo (Eq.~\ref{eq:psub}). Thus, the full stellar halo abundance ratio distribution is represented by an eight parameter model ($\vec{\mu}_{\rm halo}$, $\vec{\mu}_{\rm sub}$, $\vec{\sigma}_{\rm halo}$, $\vec{\sigma}_{\rm sub}$), whereas we utilized a four parameter model \replaced{($\vec{\mu}_{\rm dwarf}$, $\vec{\sigma}_{\rm dwarf}$)}{($\vec{\mu}_{\rm dSph}$, $\vec{\sigma}_{\rm dSph}$)} for M31 \replaced{dwarf galaxies}{dSphs}. The likelihood of the entire observed data set for a given stellar population \added{(i.e., M31's stellar halo or the dSphs)} is therefore the product of the individual likelihoods,

\begin{equation}
    \mathcal{L} = \prod_{i=1}^{N} \mathcal{L}_i
    \label{eq:bi_normal_lkhd}
\end{equation}
Using Bayes' theorem (see also Eq.~\ref{eq:bayes}), we evaluated the posterior probability of a particular bivariate model accurately describing a given a set of abundance measurements for a stellar population,

\begin{multline}
    P(\vec{\mu}, \vec{\sigma} | _{i=1}^N x_i, y_i, \delta x_i, \delta y_i) \\\propto P(_{i=1}^N x_i, y_i, \delta x_i, \delta y_i | \vec{\mu}, \vec{\sigma}) P(\vec{\mu}, \vec{\sigma}),
    \label{eq:bayes_abund}
\end{multline}
where $P(_{i=1}^N x_i, y_i, \delta x_i, \delta y_i | \vec{\mu}, \vec{\sigma})$ is the likelihood (Eq.~\ref{eq:bi_normal_lkhd}) and $P(\vec{\mu}, \vec{\sigma})$ represents our prior knowledge regarding constraints on $\vec{\mu}$ and $\vec{\sigma}$. We implemented noninformative priors over the allowed range for each fitted parameter, where we assumed uniform priors and inverse Gamma priors for non-dispersion and dispersion parameters, respectively. 
We allowed the dispersion parameters in our model to vary between
0.0 $<$ $\sigma$ $<$ 1.0, the correlation coefficients between $-$1 $<$ $r$ $<$ 1, and permitted samples of $\mu$ to be drawn from $-3.0 < {\rm [Fe/H]} < 0.0$ and $-0.8 < {\rm [\alpha/Fe]} < 1.2$. 


In the case of the variance, $\vec{\sigma}^2$, the inverse Gamma distribution is described as $\Gamma^{-1}(\alpha,\beta)$, where $\alpha$ is a shape parameter and $\beta$ is a scale parameter. This distribution is defined for $\vec{\sigma}$ $>$ 0 and can account for asymmetry in the posterior distributions for dispersion parameters, which are restricted to be positive. The bootstrap resampled standard deviations \added{(\S~\ref{sec:abund_gradients})}, weighted by the inverse variance of the measurement uncertainty and the probability of belonging to a given kinematical component, are $\sigma$([Fe/H]) = 0.53 $\pm$ 0.03, $\sigma$([$\alpha$/Fe]) = 0.32 $\pm$ 0.02 for the smooth halo, $\sigma$([Fe/H]) = 0.51 $\pm$ 0.03, $\sigma$([$\alpha$/Fe]) = 0.3\replaced{1}{2} $\pm$ 0.02 for halo substructure, and $\sigma$([Fe/H]) = 0.49 $\pm$ 0.02, $\sigma$([$\alpha$/Fe]) = 0.34 $\pm$ 0.02 for M31 \replaced{dwarf galaxies}{dSphs}. Owing to the similarity in the dispersion for each abundance ratio between the various stellar populations, we fixed the priors on $\vec{\sigma}^2$ to $\Gamma^{-1}(3,0.5)$ for all $\sigma^2_{\rm [Fe/H]}$ parameters and $\Gamma^{-1}(3,0.2)$ for all $\sigma^2_{\rm [\alpha/Fe]}$ parameters. These distributions result in priors that peak at $\sigma_{\rm [Fe/H]} \sim 0.50$ and $\sigma_{\rm [\alpha/Fe]} \sim 0.30$ with a standard deviation of 0.3 dex.

With the above formulation, we sampled from the posterior distribution of each model (Eq.~\ref{eq:bayes_abund}) using an affine-invariant Markov Chain Monte Carlo (MCMC) ensemble sampler \citep{Foreman-Mackey2013}. That is, we solved for the parameters describing the two-component stellar halo abundance distribution (Eq.~\ref{eq:mixture_model}), and separately for a model corresponding to M31 \replaced{satellite dwarf galaxies}{dSphs}. We evaluated each model using 100 walkers and ran the sampler for 10$^{4}$ steps, retaining the latter 50\% of the MCMC chains to form the converged posterior distribution. Table~\ref{tab:abund_models} presents the final parameters of each 2-D chemical abundance ratio distribution model. We computed the mean values from the 50$^{\rm th}$ percentiles of the marginalized posterior distributions, whereas the uncertainty on each parameter was calculated relative to the 16$^{\rm th}$ and 84$^{\rm th}$ percentiles.

Figure~\ref{fig:low_alpha} displays the resulting bivariate normal distribution models for the 2-D chemical abundance ratio distributions, reflecting the general trends of the observed abundance distributions (\S~\ref{sec:low_alpha_1D}) for the smooth halo, halo substructure, and \replaced{dwarf galaxy}{dSph} populations. The models predict a larger separation between the mean metallicity of the smooth halo and substructure populations, where $\mu_{\rm [Fe/H]}^{\rm sub} - \mu_{\rm [Fe/H]}^{\rm halo} = 0.39 \pm 0.08$, compared to $\langle$[Fe/H]$\rangle_{\rm sub} - \langle$[Fe/H]$\rangle_{\rm halo} = 0.09 \pm 0.06$ from a weighted boostrap resampling (\S~\ref{sec:abund_gradients}). However, the models predict similar mean $\alpha$-enhancements between the two halo populations. 

Based on these models, we calculated the probability that a star is both kinematically associated with the smooth \added{component of the} stellar halo and has abundance patterns similar to M31 \replaced{dwarf galaxies}{dSphs},

\begin{equation}
    p_{\rm dSph} = \left( \frac{ \mathcal{L}_i^{\rm dwarf} / \mathcal{L}_i^{\rm halo} }{ 1 + \mathcal{L}_i^{\rm dwarf} / \mathcal{L}_i^{\rm halo}} \right) \times (1 - p_{\rm sub,f}),
    \label{eq:pdwarf}
\end{equation}
where $\mathcal{L}_i$ is the likelihood from Eq.~\ref{eq:bi_normal} and $p_{\rm sub,f}$ is the refined probability that a star belongs to kinematical substructure \added{(see Eq.~\ref{eq:psub}).} For every sampling of the posterior distribution for each star, we calculated the odds ratio of the Bayes factor, $K$, for the substructure versus halo models, where $K$ = $p_{\rm sub} \mathcal{L}_i^{\rm sub}$/( ($1 - p_{\rm sub}$) $\mathcal{L}_i^{\rm halo}$). Then, we computed $p_{\rm sub,f}$ for each star from the 50$^{\rm th}$ percentile of these distributions. The outcome of this procedure is an updated determination of the substructure probability for each star that incorporates {\it both} kinematical and chemical information. For stars with nonzero $p_{\rm sub}$, the net effect is $p_{\rm sub,f}$ $>$ $p_{\rm sub}$ for metal-rich stars (\feh\ $\gtrsim$ $-$1.0) and $p_{\rm sub,f}$ $<$ $p_{\rm sub}$ for metal-poor stars. This change occurs because the substructure model (Table~\ref{tab:abund_models}) has a higher mean \feh\ than the smooth halo model, such that metal-rich stars are more likely to belong to substructure.

Figure~\ref{fig:low_alpha} illustrates the \replaced{distribution}{selection} of smooth halo stars with M31 \replaced{dwarf-galaxy-like}{dSph-like} chemical abundances, where we have identified 22 RGB stars that securely fall within this category ($p_{\rm dSph}$ $>$ 0.75).\deleted{in addition to 1\replaced{4}{5} marginally \replaced{dwarf-galaxy-like}{dSph-like} RGB stars (0.5 $<$ $p_{\rm dSph}$ $<$ 0.75).} \added{These stars are at least three times more likely to have abundances consistent with M31 dSphs than the bulk of the smooth halo population.}\deleted{Based on a bootstrap re-sampling weighted by $p_{\rm dwarf}$ and the inverse measurement uncertainty, we computed $\langle$[Fe/H]$\rangle$ = $-$1.73 $\pm$ 0.05 and $\langle$[$\alpha$/Fe]$\rangle$ = 0.16 $\pm$ 0.07 for this population of stars in the smooth stellar halo.} 
Hereafter, we refer to this group of stars as \deleted{the ``low-$\alpha$''}{a dSph-like} population.\deleted{owing to its lower average $\alpha$-enhancement compared to the entire smooth stellar halo ($\langle$\alphafe$\rangle$ = 0.40 $\pm$ 0.03; \S~\ref{sec:abund_gradients})} \added{We refrain from characterizing $\langle$\feh$\rangle$ or $\langle$\alphafe$\rangle$ for these dSph-like stars given that our sample is likely incomplete. On the high metallicity end, we cannot attribute stars to the dSph-like population because the M31 dSph PDF utilized in its detection cuts off at \feh\ $\gtrsim -1.0$ (Figure~\ref{fig:halo_vs_dsphs}) owing to the mass range and metallicity distribution functions of the dSphs.} \added{In part due to this incompleteness, we emphasize that the identification of a subset of M31 RGB stars as dSph-like is \textit{not} akin to a measurement of M31's accreted halo fraction. This is further made the case by (1) our limited sample size, (2) the lack of available higher-dimensional phase space information, and (3) the fact that we have defined the dSph-like sample by de facto excluding stars with a high probability of belonging to kinematically identifiable halo substructure.}

By definition, \replaced{low-$\alpha$}{dSph-like} stars belong to the dynamically hot component of the stellar halo in which we do not detect any kinematical substructure. These stars span the same range of parameter space in line-of-sight velocity ($-$700 km s$^{-1}$ $<$ v$_{\rm helio}$ $\lesssim$ $-$150 km s$^{-1}$) and projected distance (8 kpc $<$ r$_{\rm proj}$ $<$ 34 kpc) as the full sample of RGB stars in M31's inner halo. Unsurprisingly, \replaced{low-$\alpha$ population}{dSph-like} stars are more likely to be found in spectroscopic fields along the minor axis dominated by the smooth halo, such as \replaced{the 23 kpc and 31 kpc halo fields}{f130 and a0\_1}, as opposed to fields dominated by tidal debris along the high surface-brightness core of the GSS. \added{We investigate possible origins for the dSph-like stars in \S~\ref{sec:mcs} and \S~\ref{sec:halo_form}.}

\deleted{Conversely, we can define a ``high-$\alpha$'' population of stars in the smooth halo with chemical abundances distinct from M31 dwarf galaxies by taking the complement of Eq.~\ref{eq:pdwarf}. This population has 66 (16) secure (marginal) smooth halo members, such that $\langle$\feh$\rangle$ = $-$1.05 $\pm$ 0.03 and $\langle$\alphafe$\rangle$ = 0.46 $\pm$ 0.03. Similar to low-$\alpha$ stars, high-$\alpha$ stars span a wide range of heliocentric velocity and projected distance, but are more centrally concentrated along the minor-axis fields, such as the 9 kpc and 18 kpc halo fields, and are more likely to appear in the halo components of GSS fields.}

\subsubsection{The Effect of Potential Sources of Bias on Population Detections}

Potential bias from the omission of red, presumably metal-rich, stars with strong TiO absorption (\S~\ref{sec:sample}) does not affect the robust identification of smooth halo stars with chemical abundance patterns similar to M31 \replaced{dwarf galaxies}{dSphs}. Introducing maximal bias estimates for \feh\ based on this source alone (\S~\ref{sec:abund_gradients_bias}) shifts the mean metallicity of the halo and substructure models (Table~\ref{tab:abund_models}) to $-$1.05 $\pm$ 0.05 and $-$0.64 $\pm$ 0.07, respectively. TiO stars are not a significant source of bias in M31 \replaced{dwarf galaxies}{dSphs} \citep{Kirby2020}, and their abundance measurements are similarly affected by S/N limitations (\S~\ref{sec:sample}) compared to M31 RGB stars. Thus, the net effect of the omission of TiO stars would be an increased separation between the \replaced{dwarf galaxy}{dSph} and halo PDFs. This would result in the classification of \replaced{39}{36} \deleted{(\replaced{4}{6})} secure \deleted{(marginal)} \replaced{low-$\alpha$}{dSph-like} stars, thereby reinforcing the detection of this population.

\deleted{Although possible bias from the exclusion of TiO stars does not impact the detection of a low-$\alpha$ population in M31's stellar halo, the resulting change in $p_{\rm dwarf}$ (Eq.~\ref{eq:pdwarf}) would alter the population means to approximately $\langle$\feh$\rangle$ = $-$1.77 $\pm$ 0.05 and $\langle$\alphafe$\rangle$ = 0.24 $\pm$ 0.07. Analogously, the high-$\alpha$ population would be characterized by $\langle$\feh$\rangle$ = $-$0.98 $\pm$ 0.03 and $\langle$\alphafe$\rangle$ = 0.46 $\pm$ 0.03, with 64 (13) secure (marginal) members. 
These values are formally consistent with the population means for both the low- and high-$\alpha$ populations calculated without bias estimates.
None of the spatial or kinematical properties of either the low- or high-$\alpha$ populations would be altered in this bias scenario.}

\section{Discussion}
\label{sec:discuss}

\subsection{The Stellar Halos of M31 and the MW}
\label{sec:ges}

\begin{figure*}
    \centering
    \includegraphics[width=\textwidth]{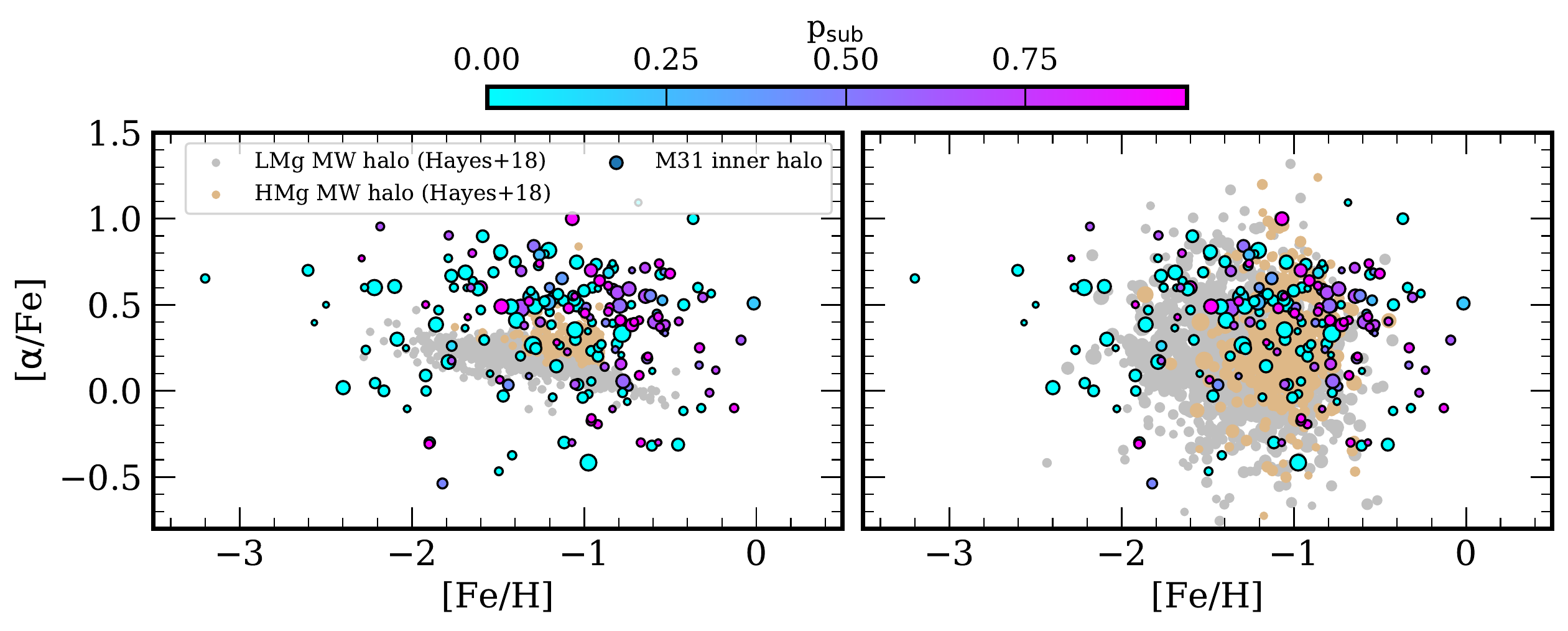}
    \caption{Comparisons between [$\alpha$/Fe] vs. [Fe/H] for the inner halo of M31 (\citealt{Escala2019,Escala2020,Gilbert2019}; this work) and metal-poor (\feh\ $\lesssim$ $-$0.9) stars in the MW halo \citep{Hayes2018}. \added{The size of the data points is inversely proportional to the measurement uncertainty.} The MW stars are separated into low-[Mg/Fe] (LMg; grey points) and high-[Mg/Fe] (HMg; brown points) populations. \added{The M31 RGB stars are color-coded according to their probability of belonging to kinematical substructure in the stellar halo (Eq.~\ref{eq:psub}).}
    \deleted{(Upper left) The color-coding and point scaling for the M31 data is the same as in Figure~\ref{fig:low_alpha}.} \replaced{(Upper right)}{In the right panel, we show} a perturbation of the MW halo abundances by \replaced{our}{M31's} empirical error distribution ($\langle\delta_\textrm{[Fe/H]}\rangle$ $\sim$ 0.12, $\langle\delta_\textrm{[$\alpha$/Fe]}\rangle$ $\sim$ 0.30). \added{This provides a more direct comparison between chemical abundance distributions of M31 and the MW.} 
    \deleted{(Lower left) Empirical PDFs for the LMg (grey dashed lines) and HMg (dotted brown lines) populations in the MW computed from 10$^{6}$ perturbations of the MW abundances by the M31 error distribution. We also show our fitted model for the M31 dwarf galaxy PDF (solid green lines; \S~\ref{sec:low_alpha_cards}, Table~\ref{tab:abund_models}). The contour levels correspond to 16$^{\rm th}$, 50$^{\rm th}$, and 84$^{\rm th}$ percentiles. (Lower right) Low-$\alpha$ population stars ($p_{\rm dSph}$ $>$ 0.5) plotted against the LMg PDF\@. 
    A subset of low-$\alpha$ stars appears similar to the LMg population, but it is unclear if all low-$\alpha$ stars could originate from a LMg-like population.}}
    \label{fig:m31halo_vs_mwhalo}
\end{figure*}

\deleted{In \S~\ref{sec:low_alpha}, we identified populations of low- and high-$\alpha$ stars in the smooth component of M31's stellar halo. In this section, we consider whether similar populations have been identified in the stellar halo of the MW.}

\deleted{Indeed, \citet{NissenSchuster1997} suggested the existence of distinct stellar populations with low- and high-[$\alpha$/Fe] in the MW from a small sample of 29 stars with halo-like kinematics in the solar neighborhood. In a subsequent study, \citet{NissenSchuster2010} used a larger sample of 94 stars to confirm the existence of these low-$\alpha$ and high-$\alpha$ stellar populations, which were further distinguished by their abundances in odd-Z elements \citep{NissenSchuster2010,NissenSchuster2011}. The high-$\alpha$ stars were preferentially bound to the Galaxy on prograde orbits, whereas the low-$\alpha$ stars were less bound with a majority on retrograde orbits. 
Later studies of MW stellar halo populations identified the same chemical abundance trends with respect to \alphafe\ and \feh\ \citep{Navarro2011,Ramirez2012,Sheffield2012,Hawkins2015,Hayes2018,Haywood2018}.
\citet{Schuster2012} found that the low-$\alpha$ stars were clumped at large orbital eccentricities ($>$0.85) and extended out to larger projected distances (30-40 kpc) compared to the high-$\alpha$ stars, which were centrally concentrated ($<$16 kpc) and had more uniformly distributed eccentricities (0.4-1.0). \citet{Schuster2012} also found that the low- and high-$\alpha$ stars had ages $\gtrsim$9 Gyr (see also \citealt{Hawkins2014}, \citealt{Gallart2019} and \citealt{Das2020}). Based on this evidence, \citet{NissenSchuster2010,NissenSchuster2011} and \citet{Schuster2012} associated the low-$\alpha$ population with the preferentially radial accretion of dwarf galaxies(s), and the high-$\alpha$ population with {\it in-situ} star formation, supporting a dual-origin scenario for stellar halo formation in the MW \citep{Zolotov2009,Zolotov2010,Font2011,McCarthy2012,Tissera2012, Tissera2013}.}

\deleted{With the advent of large astrometric and spectroscopic surveys such as {\it Gaia} \citep{GaiaCollaboration2016,GaiaCollaboration2018} and APOGEE \citep{Majewski2017}, the low-$\alpha$ halo population of  \citeauthor{NissenSchuster2010} has come to be associated with the last major accretion event in the MW's history, known as Gaia-Enceladus \citep{Helmi2018}, or Gaia-Sausage \citep{Belokurov2018}. Upon the discovery of a well-separated blue color sequence in the {\it Gaia} DR2 Hertzprung-Russel diagram of stars with high tangential velocities, \citet{GaiaCollaboration2018} suggested that this blue sequence may correspond to \citeauthor{NissenSchuster2010}'s low-$\alpha$ stars. Combining kinematical and chemical information from {\it Gaia} DR2 and APOGEE, \citealt{Helmi2018} showed that the retrograde structure corresponding to \citeauthor{NissenSchuster2010}'s low-$\alpha$ population (e.g., \citealt{Koppelman2018,Belokurov2018}) has chemical abundance patterns characteristic of a massive dwarf galaxy. Using the same data sets, \citet{Haywood2018} independently concluded that the blue {\it Gaia} sequence was indeed accreted from a progenitor with low star formation efficiency during the Galaxy's last significant merger.}

\added{In this section,} \replaced{To compare}{we compare the chemical abundance distributions of} M31's inner halo \replaced{to an analogous population in}{and} the MW. We used the low-metallicity (\feh\ $\lesssim$ $-$0.9) sample from \citet{Hayes2018} based on APOGEE \citep{Majewski2017} data ($R\sim22500$) presented in SDSS-III \citep{Eisenstein2011} Data Release (DR) 13 \citep{Albareti2017}. \citet{Hayes2018} illustrated that their data set is equivalent to the two distinct populations of \citet{NissenSchuster2010,NissenSchuster2011}, and by extension contains Gaia-Enceladus stars \citep{Haywood2018}. Gaia-Enceladus was likely comparable to the Small Magellanic Cloud (SMC; $M_\star \sim 10^{8.5} M_\odot $) in stellar mass at infall, with $M_\star$ $\sim$10$^{8-9}$ $M_\odot$ and was accreted $\sim$10 Gyr ago (e.g., \citealt{Helmi2018,Belokurov2018,Gallart2019,Mackereth2019,Fattahi2019}).\footnote{Via independent estimates, \citet{Helmi2018} and \citet{Gallart2019} infer a 4:1 merger ratio, which \citeauthor{Helmi2018} equates to $M_\star \sim 6 \times 10^8 \ M_\odot$ using a derived star formation rate from \citet{Fernandez-Alvar2018}. \citet{Belokurov2018} only infer a virial mass. \citet{Mackereth2019} and \citet{Fattahi2019} obtain $10^{8.5-9}$ and $10^{9-10} \ M_\odot$, respectively, based on comparisons of Gaia-Enceladus data to hydrodynamical simulations.}

In contrast to earlier work (using DR12) on metal-poor MW field stars \citep{Hawkins2015} that relied on kinematical selection, \citet{Hayes2018} used a larger sample of stars in combination with a data-driven approach to identify stellar populations that were distinct in multi-dimensional chemical abundance space. This resulted in the identification of two kinematically and chemically distinct MW stellar halo populations characterized by low-[Mg/Fe] with negligible Galactic rotation (LMg) and high-[Mg/Fe] with significant Galactic rotation (HMg). Based on these chemodynamical properties, \citeauthor{Hayes2018}\ concluded that the origin of the LMg population is likely massive progenitors similar to the Large Magellanic Cloud (LMC; $M_\star \sim 10^{9.5} M_\odot$), accreted early in the MW's history (i.e., Gaia-Enceladus; \citealt{Haywood2018}). The HMg population was likely formed {\it in-situ}, either via dissipative collapse or disk heating. 
\replaced{Furthermore, the HMg}{In a companion study, \citet{Fernandez-Alvar2018} modeled the chemical evolution of the two populations to find that HMg} stars experienced a \replaced{less}{more} intense and \replaced{short-lived}{long-lived} star formation history than \replaced{HMg}{LMg} stars \added{\citep{Fernandez-Alvar2018}}.
    
\added{The left panel of} Figure~\ref{fig:m31halo_vs_mwhalo} presents a comparison between [$\alpha$/Fe] and [Fe/H] for 1,321 stars from the metal-poor halo (\feh\ $<$ $-$0.9) sample of \citet{Hayes2018} and \replaced{198}{197} RGB stars in the inner stellar halo of M31 (\citealt{Escala2019,Escala2020,Gilbert2019}; this work). The atmospheric [$\alpha$/Fe] from \citet{Hayes2018} is measured from spectral synthesis by the APOGEE Stellar Parameters and Chemical Abundances Pipeline (ASPCAP; \citealt{GarciaPerez2016}) based on a fit to all $\alpha$-elments (O, Mg, Si, Ca, S, and Ti; \citealt{Holtzman2015}) in the infrared H-band ($1.514-1.696$ $\mu$m). We show both the LMg and HMg populations for the MW, whereas M31's inner halo stellar population is color-coded \replaced{to emphasize stars that are likely members of M31's low-$\alpha$ population}{according to kinematically-based substructure probability}.
    
Although the abundance distributions of the MW and M31 halo stars clearly overlap, the spread in \alphafe\ (and perhaps \feh) for the M31 RGB stars is much larger owing to the uncertainties on our measurements. The typical measurement uncertainties for M31 RGB stars are $\delta_\textrm{[Fe/H]}\sim0.12$ and $\delta_\textrm{[$\alpha$/Fe]}\sim0.30$, compared to $\sim$0.04 for MW halo stars in the \citet{Hayes2018} sample. In order to perform a more direct comparison between M31's and the MW's chemical abundance distributions, we perturbed the MW halo abundances by uncertainties from 10$^4$ draws of the empirical error distribution of our measurements.\footnote{
We have assumed that our measured uncertainties in \feh\ and \alphafe\ {\it for an individual star} are independent. \deleted{The net effect of this perturbation by such independent errors is to orthogonally stretch the 2D chemical abundance distributions for the MW halo.} However, as discussed in \S~\ref{sec:low_alpha_cards}, we expect the errors in \alphafe\ and \feh\ on a single measurement to \replaced{be contravariant}{have nonzero covariance}. \replaced{Thus, the MW halo chemical abundances are not necessarily perturbed by the true underlying error distribution for the M31 inner halo stars.}{The effect of incorporating this covariance is negligible for this comparison.}} During each draw, each MW halo star was perturbed by random values sourced from a normal distribution defined by the uncertainties on a single randomly selected M31 RGB star and the MW halo star.\footnote{e.g., $\mathcal{N}$(0, $\sigma_\mathrm{[Fe/H]}$), where $\sigma_\mathrm{[Fe/H]} = \sqrt{(\delta_{\mathrm{[Fe/H]},i}^\mathrm{M31})^2 - (\delta_\mathrm{[Fe/H]}^\mathrm{MW})^2}$} The M31 RGB star is randomly selected from a metallicity bin within 0.2 dex of the MW halo star to preserve the metallicity-dependence of the M31 error distribution: M31 RGB stars with lower \feh\ tend to have larger \feh\ uncertainties. 

We show an example of a single perturbation of the MW halo stars in the upper right panel of Figure~\ref{fig:m31halo_vs_mwhalo}. The primary effect of this perturbation is that the MW halo stars now span a similar range in \alphafe\ as the M31 halo stars. \added{If the MW was observed at the distance of M31 (i.e., if the MW abundances had an M31-like error distribution), we would measure $\langle$\feh$\rangle_{\rm MW} = -1.23 \pm 0.004$, and $\langle$\alphafe$\rangle_{\rm MW} = 0.21 \pm 0.01$ from the full distribution of 10$^{4}$ perturbations ($\langle$\feh$\rangle_{\rm MW} = -1.18 \pm 0.007$, and $\langle$\alphafe$\rangle_{\rm MW} = 0.20 \pm 0.003$ without any perturbations). This suggests that M31's stellar halo ($\langle$\feh$\rangle_{\rm M31} = -1.08 \pm 0.04$, and $\langle$\alphafe$\rangle_{\rm M31} = 0.40 \pm 0.03$; \S~\ref{sec:abund_gradients}) is likely more metal-rich and $\alpha$-rich on average compared to the \citeauthor{Hayes2018}\ sample of the MW's stellar halo (Figure~\ref{fig:m31halo_vs_mwhalo})}.
    
\begin{figure}
    \centering
    \includegraphics[width=\columnwidth]{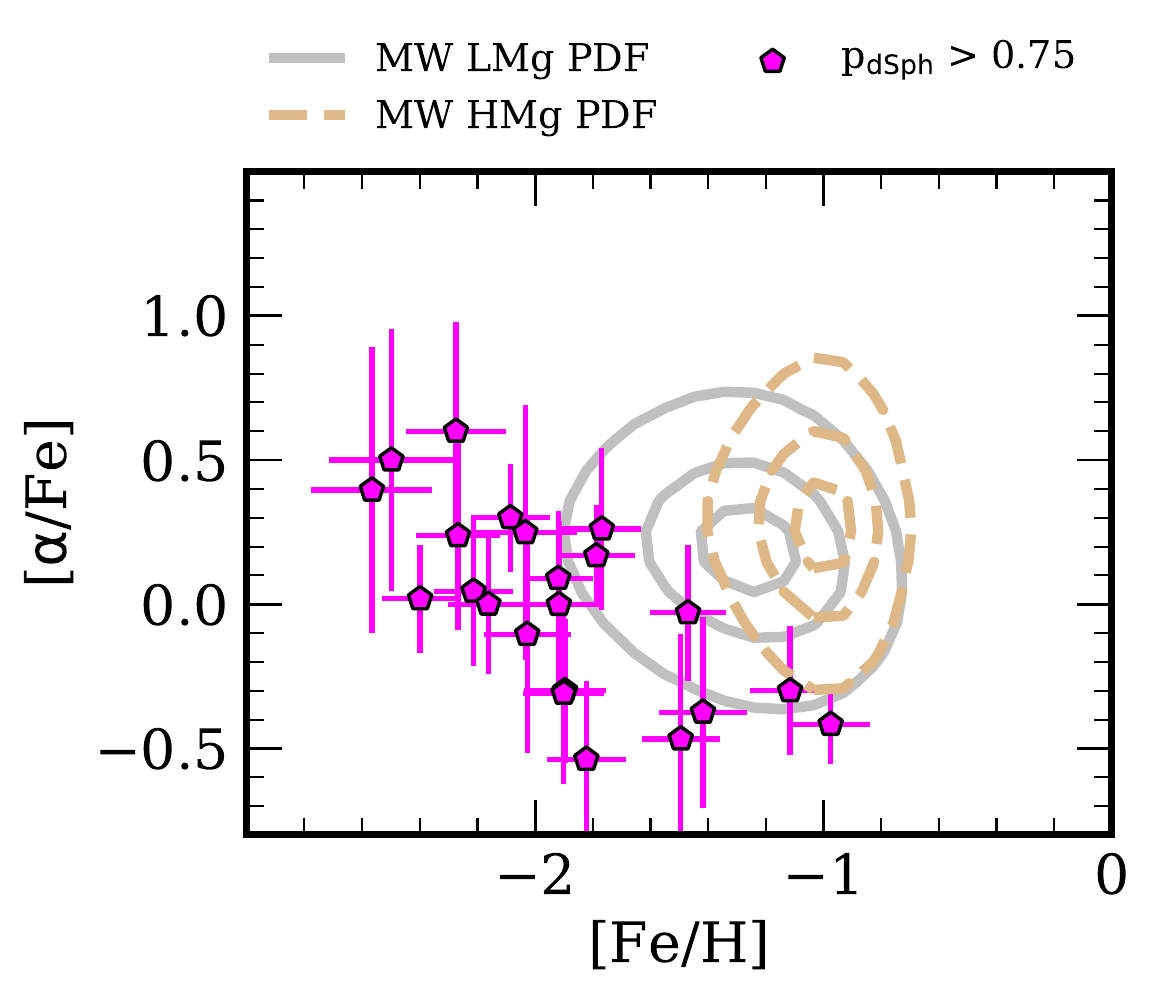}
    \caption{Empirical PDFs for the LMg (grey solid lines) and HMg (dashed brown lines) populations of the MW halo \citep{Hayes2018}, computed from 10$^{4}$ perturbations of the MW abundances by M31's measurement errors (\S~\ref{sec:ges}). The contour levels correspond to 16$^{\rm th}$, 50$^{\rm th}$, and 84$^{\rm th}$ percentiles. We also show RGB stars in the smooth halo of M31 with dSph-like chemical abundances (\S~\ref{sec:low_alpha_cards}) as colored points. A subset of the dSph-like stars appears similar to the LMg population, but it is unclear if all of such stars could originate from a LMg-like population.}
    \label{fig:mwhalo_dsphlike}
\end{figure}

By performing \replaced{this}{the above} exercise, we constructed empirical PDFs in \alphafe\ vs.\ \feh\ space for the LMg and HMg populations \added{of the MW halo} (Figure~\ref{fig:mwhalo_dsphlike}). \deleted{Despite the significantly larger uncertainties, the comparatively narrow spread in \feh\ and higher mean \feh\ and \alphafe\ of the HMg population is preserved relative to the LMg population.} \replaced{In the lower right panel of Figure~\ref{fig:m31halo_vs_mwhalo},}{In this figure,} we \added{also} qualitatively compared M31 RGB stars that likely belong to M31's \replaced{low-$\alpha$}{dSph-like} population ($p_{\rm dSph}$ $>$ \replaced{0.5}{0.75}; Eq.~\ref{eq:pdwarf}) to the \replaced{LMg}{MW} PDFs\@. \added{Given that these M31 stars chemically resemble dSphs, it is interesting to compare to the accreted, LMg population in the MW's stellar halo.} A subset of M31's \replaced{low-$\alpha$}{dSph-like} stars appear similar to the LMg population\deleted{in the MW halo}, but it is unclear if all of M31's \replaced{low-$\alpha$}{dSph-like} stars could originate from a LMg-like population. In particular, M31's halo may have stars that are more metal-poor (\feh\ $\lesssim$ $-$1.8) than those observed in \citeauthor{Hayes2018}'s MW halo sample.\footnote{We do not expect any systematic biases in APOGEE elemental abundances to result in the dearth of data at \feh\ $\lesssim$ $-$1.8 and above the ASPCAP grid edge at \feh\ = $-$2.5. APOGEE is biased against {\it metal-rich} stars in fields that are either distant (i.e., dominated by cool giants) or enriched to supersolar metallicity \citep{Hayden2014,Hayden2015}, but there is no quantified bias for metal-poor stars in APOGEE with \feh\ $\gtrsim$ $-$2.5.} We note that the APOGEE abundances attributed to Gaia-Enceladus from \citet{Helmi2018} extend down to \feh\ $\sim$ $-$2.5 and up to \alphafe\ $\sim$ 0.5, potentially encapsulating the low-metallicity RGB stars in M31's \replaced{low-$\alpha$}{dSph-like} population. \deleted{Regardless, the marginal similarity between M31's \replaced{low-$\alpha$}{dSph-like} and MW LMg populations suggests that they may both originate from the accretion of \replaced{dwarf galaxies}{chemically un-evolved external galaxies} $\gtrsim$8 Gyr ago (e.g., \citealt{BullockJohnston2005,Font2006b,Fattahi2020arXiv}), such that kinematical substructure no longer remains coherent in the stellar halo \added{of M31}. We further investigate this possibility through comparison to the Magellanic Clouds in \S~\ref{sec:mcs} and discuss possible origin scenarios for M31's \replaced{low-$\alpha$}{dSph-like} population in \S~\ref{sec:halo_form}.}\added{We explore origin scenarios for M31's dSph-like stars in \S~\ref{sec:mcs} and \S~\ref{sec:halo_form}.}

\deleted{Figure~\ref{fig:m31halo_vs_mwhalo} also shows the M31 dwarf galaxy PDF (\S~\ref{sec:low_alpha_cards}; Table~\ref{tab:abund_models}), which we used to assign a probability of membership to M31's low-$\alpha$ population to each M31 RGB star in the inner halo. Both the LMg and HMg PDFs overlap the M31 dwarf galaxy PDF for \feh\ $\gtrsim$ $-$1.8, although they seem to extend to higher \alphafe\ at fixed \feh. Based on these PDFs, the M31 dwarf galaxy population appears to be chemically distinct from the  accreted (LMg) and {\it in-situ} (HMg) populations of the MW halo. This difference between M31 dwarf galaxies and the LMg population is in accordance with expectations. \citet{Hayes2018} concluded that the majority of the LMg population cannot be accounted for by the accretion of progenitors similar to present-day MW dwarf galaxies. Because M31 dwarf galaxies have similar abundance patterns as MW dwarf galaxies at fixed stellar mass \citep{Kirby2020}, this conclusion also extends to progenitors similar to present-day M31 dwarf galaxies.}

\deleted{Regarding the HMg population, M31's low-$\alpha$ stars do not appear to be consistent with its empirical PDF\@. However, M31 inner halo stars with $p_{\rm dwarf}$ $<$ $0.5$ (upper right panel of Figure~\ref{fig:m31halo_vs_mwhalo}) do overlap in chemical abundance space with the HMg population. Some of these stars are kinematically associated with substructure, whereas the remainder are the high-$\alpha$, smooth halo population stars defined in \S~\ref{sec:low_alpha_cards}. The high-$\alpha$ population in M31 appears analogous to the {\it in-situ} HMg population based on its \alphafe\ and \feh\ measurements. However, it is unclear if high-metallicity (\feh\ $\gtrsim$ $-$0.5) stars are present in the MW's HMg halo, as is the case for M31's high-$\alpha$ halo, owing to the metallicity cut (\feh\ $\leq$ $-0.9$) used to define the HMg population \citep{Hayes2018}.
We further discuss formation scenarios for M31's high-$\alpha$ population in \S~\ref{sec:halo_form}.}

\begin{figure*}
    \centering
    \includegraphics[width=\textwidth]{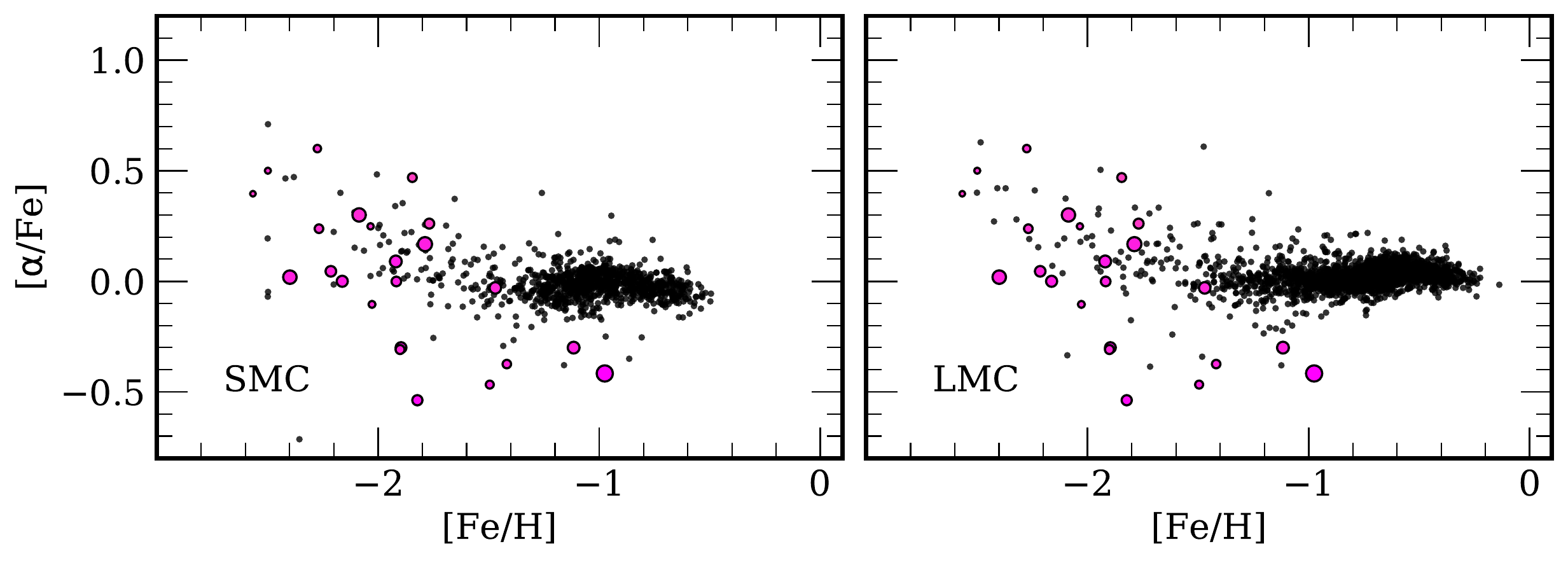}
    \caption{[$\alpha$/Fe] versus [Fe/H] for RGB stars \added{with $p_{\rm dSph} >$ \replaced{0.5}{0.75} (\added{magenta points;} Eq.~\ref{eq:pdwarf})} in the inner halo of M31 compared to the Magellanic Clouds (black points; \citealt{Nidever2019}; \S~\ref{sec:mcs}): the SMC (left) and LMC (right). The \deleted{color-coding and} point scaling for the M31 data is the same as in Figure~\ref{fig:low_alpha}. We do not show the small measurement uncertainties for the MCs ($\delta_{\rm [Fe/H]}\sim0.02$, $\delta_{\rm [\alpha/Fe]}\sim0.03$). \deleted{The abundance patterns of the MCs are largely inconsistent with M31's low-$\alpha$ stars, except for the metal-poor tails that extend to higher \alphafe.} \added{The ancient, metal-poor tails of the relatively isolated MCs follow a similar trajectory in abundance space, which is defined by low star formation efficiency, as the dSph-like stars.}}
    \label{fig:mcs}
\end{figure*}

\subsubsection{Potential Sources of Systematic Offsets}
\label{sec:apogee_systematics}
    
We refrained from \replaced{making quantitative}{drawing detailed} conclusions based on these comparisons owing to potential systematic offsets in the abundance distributions between the MW and M31 data sets. As previously discussed, the APOGEE chemical abundances are based on high-resolution spectroscopy in the near-infrared H-band, in contrast with our measurements, which we derive from low- ($R\sim3000$) and medium- ($R\sim6000$) resolution spectroscopy at optical wavelengths with comparatively low S/N\@. Additionally, the APOGEE abundances are internally calibrated against observations of metal-rich open clusters \citep{Holtzman2015} and externally calibrated to correct for systematic offsets. Although our spectral synthesis method relies on MW globular clusters to determine the systematic uncertainty on our abundance measurements \citep{Kirby2008,Escala2019,Escala2020}, we do not calibrate the stellar parameters and elemental abundances outputted by our abundance pipelines to external observations. The selection functions also differ between \citet{Hayes2018} and this work, where we have avoided selection criteria based on chemical abundances.
Although we acknowledge these caveats, an exploration of the impact of these possible systematic offsets and selection effects between the MW and M31 data sets is beyond the scope of this paper.

\subsection{Comparison of M31's dSph-like Stars to the Magellanic Clouds}
\label{sec:mcs}

\added{In \S~\ref{sec:low_alpha_cards}, we identified a subset of RGB stars in M31's stellar halo with measurements of \feh\ and \alphafe\ that are statistically similar to those of M31's dSphs \citep{Vargas2014a,Kirby2020}. Because the chemical abundance distributions of M31's dSphs are consistent with that of the MW's dSphs at fixed stellar mass \citep{Kirby2020}, these dSph-like stars in M31 are also similar to the MW's dSphs. This implies that these dSph-like stars may have been accreted onto M31's stellar halo from galaxies of similarly low stellar mass and star formation history (LM--SFH) as the dSphs, or of higher stellar mass and similar star formation efficiency (HM--SFE). \added{Although the identified dSph-like stars are statistically likely to have been accreted, they do not represent a complete sample of such stars (\S~\ref{sec:low_alpha_cards}), and are therefore not to be confused with the hypothetical true accreted fraction of M31's stellar halo (\S~\ref{sec:insitu})}

Given that the dSph-like stars could in large part originate from massive dwarf galaxies (\S~\ref{sec:mass_func}),}
\deleted{In \S~\ref{sec:ges}, we compared the abundances of \replaced{M31's inner halo to a low-$\alpha$ population in the MW}{the stellar halos of M31 and the MW} \citep{Hayes2018}. \replaced{which}{where the MW's LMg population} is likely associated with Gaia-Enceladus \citep{Helmi2018,Belokurov2018,Haywood2018}. The \replaced{potential}{possible} similarity \added{(Figure~\ref{fig:mwhalo_dsphlike})} between the \replaced{low-$\alpha$ populations of M31 and the MW}{dSph-like stars in M31 and LMg stars in the MW} \replaced{motivates the hypothesis}{suggests that} that M31's \replaced{low-$\alpha$}{dSph-like} population \replaced{has}{could have} an accretion origin.} \deleted{In this section, we test this possibility by comparing}{we compared} the abundances of M31's \deleted{inner halo}\added{ dSph-like stars} to the Magellanic Clouds (MCs), which have \added{both high} stellar masses ($2.7\times10^9$ $M_\odot$ and $3.1\times10^8$ $M_\odot$ for the LMC and SMC, respectively; \citealt{vanderMarel2002,Stanimirovic2004}) \added{and pronounced metal-poor populations.} \deleted{comparable to the inferred stellar mass of Gaia-Enceladus.} We used the MC abundance sample of \citet{Nidever2019} from APOGEE DR16 \citep{Ahumada2019} obtained as part of SDSS-IV \citep{Blanton2017} through the installation of a second APOGEE spectrograph in the Southern hemisphere \citep{Wilson2019}.\footnote{The discussions on how APOGEE measures abundances, and on the caveats about comparing APOGEE data to our optical spectroscopic survey in M31, in \S~\ref{sec:ges} also applies to the DR16 data of the Magellanic Clouds.} 
Figure~\ref{fig:mcs} shows \alphafe\ versus \feh\ for the MC samples compared to \replaced{the inner halo of M31}{M31's dSph-like stars}. The MCs show the same broad chemical abundance trends defined by low [$\alpha$/Fe], a relatively metal-rich body, and a metal-poor tail extending to higher [$\alpha$/Fe]. \deleted{For both MCs, \citet{Nidever2019} constrained the ``$\alpha$-knee'' in the \alphafe\ versus \feh\ relationship to \feh$\sim-2.2$, where $\langle$\alphafe$\rangle$ is approximately solar. Despite these similarities, there are differences in the abundances between the MCs. The LMC is more metal-rich with $\langle$\feh$\rangle\sim-0.68$, reaching metallicities as high as $-$0.2, compared to the SMC ($\langle$\feh$\rangle\sim-0.97$), which has a maximum metallicity of \feh$\sim-0.5$. 
\citet{Nidever2019} also noted that the SMC abundances may be slightly biased against metal-poor stars owing to the lower S/N of the APOGEE data for the SMC relative to the LMC\@.} \deleted{Compared to M31, the MCs have lower \alphafe\ by $\sim$0.4 dex at fixed \feh\ above ${\rm [Fe/H]} \sim-1$.} \replaced{However, there appears to be considerable overlap between}{Figure~\ref{fig:mcs} further illustrates that} M31 RGB stars with $p_{\rm dSph}>$ \replaced{0.5}{0.75} \added{(Eq.~\ref{eq:pdwarf})} and the metal-poor, high-$\alpha$ tails of the MCs \added{show a similar trajectory in chemical abundance space, defined by a steep decline in [$\alpha$/Fe] vs. [Fe/H].} 

Chemical evolution models for the LMC \added{(and by extension, the SMC, owing to their similar chemical abundance patterns)} predict the dominance of Type Ia supernovae over core-collapse supernovae during this quiescent epoch \citep{BekkiTsujimoto2012,Nidever2019}, resulting in the observed decline in \alphafe\ with \feh\ in the metal-poor tails\deleted{of the LMC's abundance distribution}. \added{These tails are comparably ancient to the dSph-like stars:} the predicted age-metallicity relation from \citet{Nidever2019} has the LMC reaching \feh$\sim-1.5$ at 6 Gyr into its evolution \citep{Nidever2019}, suggesting that the majority of its metal-poor tail was in place by $\sim$7 Gyr ago.\footnote{The models of \citet{BekkiTsujimoto2012} reach ${\rm \feh} \sim-1.5$ by $\sim9-11$ Gyr ago, but were not well-constrained at low metallicity by the available data for the LMC at the time of their study.}

Based on their chemical abundances, \citet{Nidever2019} concluded that the early evolution of the MCs was characterized by low star formation efficiency (gas mass converted into stellar mass). \deleted{Prior studies of the $\alpha$-enhancement of the SMC \citep{Mucciarelli2014} and LMC \citep{Pompeia2008,Lapenna2012,VanderSwaelmen2013} arrived at similar conclusions.} This low efficiency may result from the MCs having evolved in isolation, given that they are likely on first infall into the MW's potential well \citep{Besla2007,Besla2012}. The relatively quiescent star formation history of the MCs until $\sim$4 Gyr ago may \added{also} be a consequence of such isolation (e.g., \citealt{Smecker-Hane2002,HarrisZaritsky2009,Weisz2013}).

\deleted{By inspection, the decline in \alphafe\ with \feh\ for the low-$\alpha$ M31 stars seems to agree with the MCs better than the MW halo LMg population / Gaia-Enceladus (\citealt{Hayes2018}; Figure~\ref{fig:m31halo_vs_mwhalo}). If we perturb the MC abundance distributions by the measurement uncertainties of the M31 data as in \S~\ref{sec:ges}, we find that the MC abundance distributions \textit{taken as a whole} are inconsistent with M31's low-$\alpha$ population. This is because the loci of the MCs' abundance distributions (particularly the LMC) occur at high metallicity, while M31's low-$\alpha$ population is metal-poor. However, the MC abundance distributions include metal-poor tails (\feh\ $\lesssim -1.5$) within the 2$\sigma$ confidence limits of their 2D PDFs. In particular, an abundance distribution similar to the SMCs' metal-poor tail could potentially account for M31's low-$\alpha$ population for \feh\ $\gtrsim -2.0$. This implies that M31's low-$\alpha$ population may have originated from an accreted galaxy (or galaxies) that experienced chemical evolution similar to that of the MCs at low-metallicity (\feh\ $\lesssim -1.5$), or early in their histories.}

Thus, we can infer that the \replaced{dwarf galaxy or dwarf galaxies}{galaxie(s)} that were accreted \added{onto M31's halo} to form \replaced{M31's low-$\alpha$ population}{its dSph-like population}, \replaced{likely had}{which is characterized by} low star formation efficiency\deleted{and slow chemical enrichment}, \added{experienced chemical evolution similar to the early evolution of the isolated MCs}. We evaluate whether \replaced{the multiple or single progenitor scenario}{the scenario of multiple low-mass progenitors (LM-SFH) or a dominant high-mass progenitor (HM-SFE)} is more likely \replaced{for}{to explain} the origin of M31's \replaced{low-$\alpha$ population}{dSph-like stars} in \S~\ref{sec:mass_func}.


\subsection{Formation Scenarios for M31's Inner Stellar Halo}
\label{sec:halo_form}

In this subsection, we explore the implications of our abundance measurements in M31's stellar halo for {\it in-situ} and accreted formation channels (\S~\ref{sec:insitu}). We also investigate possible origins of M31's \replaced{low-$\alpha$}{dSph-like} population in the context of the stellar mass function of accreted galaxies (\S~\ref{sec:mass_func}).

\subsubsection{The In-Situ vs. Accreted Halo}
\label{sec:insitu}

Simulations predict the existence of an {\it in-situ} component of the stellar halo, which formed in the main progenitor of the host galaxy, in addition to an accreted component formed from external galaxies \citep{Zolotov2009,Font2011,Tissera2013,Cooper2015}. In addition to a component formed via dissipative collapse, the {\it in-situ} halo includes contributions from kinematically heated stars originating in the disk \citep{Purcell2010,McCarthy2012,Tissera2013,Cooper2015}. The relative contributions of these two formation channels to stellar halo build-up depend on the stochastic accretion history of the host galaxy, where more active histories correspond to lower {\it in-situ} stellar halo mass fractions \citep{Zolotov2009}. However, the {\it in-situ} fraction also depends on the numerical details of a given simulation \citep{Zolotov2009,Cooper2015}. The more recent simulations of \citet{Cooper2015} have found that the {\it in-situ} fraction typically ranges between $\sim$30-40\%.
Despite variations in the predictions of this fraction, multiple studies have found that the {\it in-situ} component tends to dominate within the inner $\sim$20-30 kpc of the stellar halo \citep{Zolotov2009,Font2011,Tissera2013,Cooper2015}.

Predicted chemical signatures for the {\it in-situ} and accreted stellar halo can also vary depending on the simulation. In some studies \citep{Zolotov2009,Font2011,Tissera2013,Tissera2014}, {\it in-situ} star formation produces more metal-rich stellar halos than accretion alone. In contrast, \citet{Cooper2015} argued that the {\it in-situ} and accreted halo may be indistinguishable based on metallicity alone, assuming that the {\it in-situ} halo forms mostly from gas stripped from accreted galaxies. For galaxies that have not experienced a recent ($z\lesssim1$) major merger, \citet{Zolotov2010} found that the {\it in-situ} halo has higher \alphafe\ at fixed \feh\ than the accreted halo for stars with \feh\ $>-1$ in the solar neighborhood. \citet{Tissera2012,Tissera2013} found the opposite trend, where accreted stars tend to be more $\alpha$-enhanced (but also more metal-poor) than {\it in-situ} stars.

On the observational front, a bimodality in the metallicity of the MW's stellar halo with radius was interpreted as evidence in favor of a two-component halo defined by {\it in-situ} and accreted populations \citep{Carollo2007,Carollo2010}. After {\it Gaia} DR2, it has become apparent that the MW's presumably {\it in-situ} inner stellar halo is in actuality the remnants of Gaia-Enceladus \citep{Helmi2018,Haywood2018,Belokurov2018,Deason2018}. A veritable {\it in-situ} component of the inner stellar halo has recently emerged via its distinct chemical composition and kinematics in relation to Gaia-Enceladus tidal debris (\citealt{Hayes2018,Haywood2018,DiMatteo2019,Gallart2019,Conroy2019}, with earlier suggestions from {\it Gaia} DR1 by \citealt{Bonaca2017}). In particular, \citet{Belokurov2020} identified this component from old, high-eccentricity, $\alpha$-rich stars on retrograde orbits\footnote{While the {\it in-situ} halo component of \citet{Belokurov2020} (dubbed the ``Splash'') most cleanly separates for metal-rich, retrograde stars, it also contains stars on prograde orbits and exhibits low-amplitude net prograde rotation.} where with ${\rm \feh} >-0.7$ in the solar neighborhood, associating it with formation from the MW's proto-disk following its last significant merger.

At this time, the presence of a significant {\it in-situ} component in the inner stellar halo of M31 is less clear than in the case of the MW\@. Along the major axis of M31's northeastern disk, there is compelling kinematical evidence for both a rotating inner spheroid \citep{Dorman2012} and dynamically heated stars originating from the disk \citep{Dorman2013}. An extended disk-like structure possibly formed from M31's last significant merger has also been detected within the inner $\sim$40 kpc of M31's disk plane \citep{Ibata2005}. Although these stellar structures may reasonably be various {\it in-situ} components, they have not been detected within the radial range spanned by our data ($\sim$8-34 kpc in M31's southeastern quadrant, along its minor axis; Figure~\ref{fig:m31_loc}). Even at the distance of our innermost spectroscopic field ($r_{\rm proj}\sim9$ kpc, or $r_{\rm disk}\sim$ 38 kpc assuming $i=77^\circ$), the contribution from the extended disk is expected to be $\lesssim$10\% \citep{Guhathakurta2005,Gilbert2007}.

Interestingly, M31's global stellar halo properties appear to agree with predictions from accretion-only models for stellar halo formation \citep{Harmsen2017}, along with other MW-like, edge-on galaxies in the Local Volume from the GHOSTS survey \citep{Radburn-Smith2011,Monachesi2016}. Indeed, multiple lines of evidence indicate that accretion has played a dominant role in the formation of M31's stellar halo (\S~\ref{sec:mass_func}). 

Based on this alone, both the \replaced{metal-poor, low-$\alpha$ and metal-rich, high-$\alpha$ stars}{dSph-like and non-dSph-like stars} in M31's smooth, inner stellar halo (\S~\ref{sec:low_alpha_cards}) could be accreted populations resulting from chemically distinct groups of progenitors. \added{For example, many of the non-dSph-like stars have high substructure probabilities, and are therefore likely associated with GSS tidal debris. The remainder of the non-dSph-like stars, which are associated with the smooth component of the stellar halo, could plausibly originate from separate massive progentior(s) with high star formation efficiency in a pure accretion scenario.} \added{As for the dSph-like smooth halo stars,}
the similarity \replaced{in the}{of their} chemical abundance patterns \replaced{of the \replaced{low-$\alpha$}{dSph-like smooth halo} population and M31}{to M31's} \replaced{dwarf galaxies}{dSphs} provides strong evidence in favor of an accretion origin \added{in either the LM-SFH or HM-SFE scenarios (\S~\ref{sec:mcs})}.\deleted{in addition to its similarity to metal-poor stars associated with Gaia-Enceladus (\S~\ref{sec:ges}) and the MCs (\S~\ref{sec:mcs}).} \added{However, we acknowledge the possibility that the few low-metallicity (\feh\ $\lesssim$ $-2$) dSph-like stars with high $\alpha$-enhancement (\alphafe\ $\gtrsim$ 0.3) could have formed \textit{in-situ}. In this regime, dSph and stellar halo chemical abundance patterns are generally similar (e.g., \citealt{Venn2004}), such that \alphafe\ is degenerate with star formation history \citep{Lee2015}.}

\replaced{However}{Conversely} \added{to the pure accretion origin scenario}, the overlap between M31's \replaced{high-$\alpha$}{broader stellar halo} population and the MW's {\it in-situ} HMg population in abundance space (\S~\ref{sec:ges}; Figure~\ref{fig:m31halo_vs_mwhalo}) suggests that \replaced{high-$\alpha$ stars could represent}{M31 could reasonably possess} a \added{significant,} ancient {\it in-situ} population, instead of having a \added{predominately} accretion origin in distinct progenitor(s).
In an {\it in-situ} formation scenario,\deleted{M31's 
high-$\alpha$ population}{ the bulk of M31's metal-rich stars in the smooth halo} may have originated from a proto-stellar disk, similar to the MW's {\it in-situ} halo population \citep{Belokurov2020}. Given that we are lacking multi-dimensional kinematical information or higher-dimensional chemical information, it is difficult to distinguish between accreted and {\it in-situ} formation scenarios for the \replaced{high-$\alpha$}{bulk of M31's stellar} population without performing a detailed comparison to simulations, which is beyond the scope of this paper.

\subsubsection{The Mass Function of Destroyed Galaxies}
\label{sec:mass_func}

The general consensus from simulations of MW-like galaxies is that massive dwarf galaxies ($M_\star\sim10^{8-10} M_\odot$) distinct from present-day satellites are the dominant progenitors of the accreted stellar halo \citep{BullockJohnston2005,Robertson2005,Font2006b,Cooper2010,Deason2016,DSouzaBell2018a,Fattahi2020arXiv}. These galaxies have a median accretion time of $\gtrsim$9 Gyr ago \citep{BullockJohnston2005,Font2006b,Fattahi2020arXiv}. Furthermore, a few massive progenitors are predicted to form the bulk of the accreted halo \citep{Cooper2010,Deason2016,DSouzaBell2018a,Fattahi2020arXiv}, where the secondary progenitor is on average half as massive as the primary progenitor \citep{DSouzaBell2018a}. In particular, \citet{Fattahi2020arXiv} found that such massive progenitors dominate the stellar mass budget within the inner 50 kpc of the stellar halo, where contributions from progenitors with $M_\star < 10^8 M_\odot$ become non-negligible around $\sim$100 kpc. \added{The disruption of globular clusters associated with the accretion of low-mass dwarf galaxies and/or ancient systems that formed \textit{in-situ} likely plays a subdominant role in stellar halo formation, with contributions to the stellar mass budget of $\sim$2-5\% based on recent observational \citep{Koch2019,Naidu2020} and theoretical \citep{Reina-Campos2020} studies.}

The details of the stellar mass function of destroyed constituent galaxies are dictated by the accretion history of the host galaxy. \citet{Deason2016} found that MW-like galaxies with more active accretion histories have higher mass accreted progenitors ($M_\star\sim10^{9-10}M_\odot$) and larger accreted halo mass fractions than their quiescent counterparts. The wealth of substructure visible in M31's stellar halo (e.g., \citealt{Ferguson2002,Ibata2007,McConnachie2018}), including the GSS \citep{Ibata2001}, and the lack of an apparent break in its density profile \citep{Guhathakurta2005,Irwin2005,Courteau2011,Gilbert2012} suggests that M31 has likely experienced multiple and continuous contributions to its stellar halo from accreted galaxies \citep{Deason2013,Cooper2013,Font2020arXiv}. Moreover, the total accreted stellar mass of M31 ($1.5\pm0.5\times10^{10}M_\odot$; \citealt{Harmsen2017})
is on the upper end of both observed \citep{Merritt2016,Harmsen2017} and predicted \citep{Cooper2013,Deason2016,Monachesi2019} accreted stellar halo mass fractions for MW-like galaxies. 

The large-scale negative metallicity gradient of M31's stellar halo (\citealt{Gilbert2014,Gilbert2020}; this work) may also lend support to it being dominated by a few massive progenitors, although this trend could also result from a significant {\it in-situ} component in the inner regions of the halo \citep{Font2011,Tissera2014,Monachesi2019}. Thus, M31's accreted mass function is likely weighted toward progenitor galaxies with $M_\star\sim10^{9-10}M_\odot$. 
This agrees with the hypothesis that the main contributor to M31's stellar halo (including substructure) is the GSS progenitor, a massive galaxy with $M_\star\sim10^{9-10}M_\odot$ accreted between $\sim$1-4 Gyr ago, depending on the merger scenario \citep{Fardal2007,Fardal2008,Hammer2018,DSouzaBell2018b}. 

    
The mass spectrum of progenitor galaxies dictates the metallicity of the accreted stellar halo through the stellar mass-metallicity relation for galaxies \citep{Gallazzi2005,Kirby2013} and its redshift evolution \citep{Gallazzi2014,Ma2016,Leethochawalit2018,Leethochawalit2019}. In particular, \citet{Leethochawalit2018,Leethochawalit2019} found that the normalization of the stellar-mass metallicity relation evolves by 0.04 $\pm$ 0.01 dex Gyr$^{-1}$. Assuming that this relation extends to $M_\star < 10^9 \ M_\odot$, this means that an SMC-mass galaxy ($M_\star\sim10^{8.5} \ M_\odot$) accreted 10 Gyr ago would have $\langle$\feh$\rangle = -1.39 \pm 0.04$, compared to $\langle$\feh$\rangle = -0.99 \pm 0.01$ at $z=0$ \citep{Nidever2019}.
Observations of the MW-like GHOSTS galaxies \citep{Harmsen2017} have firmly established such a relationship between stellar halo mass and metallicity, as first suggested by \citet{Mouhcine2005}. The physical driving force for this relation is the fact that the most massive progenitors dominate halo assembly \citep{DSouzaBell2018a,Monachesi2019}. 

In addition to setting the stellar halo metallicity, the most massive progenitors largely determine the full metallicity distribution function of the stellar halo \citep{Deason2016,Fattahi2020arXiv}. For a typical progenitor mass of 10$^{9-10}$ $M_\odot$, these galaxies contribute $\gtrsim$90\% of metal-poor stars (\feh\ $<$ $-$1) and $\sim$20-60\% of stars with \feh\ $<$ $-2$ \citep{Deason2016}. \citeauthor{Deason2016}\ also found that classical dwarf galaxies ($M_\star\sim10^{5-8}M_\odot$) contribute the remainder of  metal-poor stars, with negligible contributions from ultra-faint dwarf galaxies.

Indeed, M31's \replaced{low-$\alpha$}{dSph-like} stars most likely have an accretion origin \added{in either the LM-SFH or HM-SFE scenarios (\S~\ref{sec:mcs})} owing to their defining similarity to the abundance patterns M31 dwarf galaxies (\S~\ref{sec:low_alpha_cards})\deleted{ and their distinction from the predominant abundance patterns of M31's high-$\alpha$ halo (\S~\ref{sec:low_alpha_1D})}. Given that (1) the chemical abundance patterns of M31's \replaced{low-$\alpha$}{dSph-like} population are broadly consistent with the low-metallicity (\feh\ $\lesssim -1.5$), early evolution of the MCs (\S~\ref{sec:mcs}) and (2) we do not detect any kinematical substructure for this population (\S~\ref{sec:low_alpha}), the accreted galaxies(s) that contributed to M31's \replaced{low-$\alpha$}{dSph-like} population \replaced{would need to have low star formation efficiency and occur $\gtrsim$8 Gyr ago}{could have been massive ($M_\star \sim 10^{8-9} M_\odot$) if they evolved in an isolated, low star formation efficiency environment and were accreted $\gtrsim$ 8 Gyr ago.}

Given M31's halo properties and inferred accretion history (based on halo formation models), a \replaced{likely}{possible} scenario for the formation of the \replaced{low-$\alpha$}{dSph-like} population ($-2.5\lesssim$ \feh\ $\lesssim-1.0$) is the accretion of a massive, secondary progenitor \added{with low star formation efficiency (e.g., the HM-SFE scenario)}. Assuming that the GSS progenitor is the dominant progenitor, the stellar mass of this secondary progenitor should be approximately between the mass of the SMC and LMC ($M_\star\sim10^{8.5-9.5}M_\odot$; 
\S~\ref{sec:mcs}) and possibly comparable to Gaia-Enceladus (\S~\ref{sec:ges}). An early, massive accretion event such as this would also deposit its debris closer to the center of the host potential in the inner halo owing to dynamical friction \citep{Cooper2010,Tissera2013,Fattahi2020arXiv}. Alternatively, M31's \replaced{low-$\alpha$}{dSph-like} population could have formed exclusively from the accretion of multiple progenitors similar \added{in stellar mass and star formation history} to classical dwarf galaxies \added{(e.g., the LM-SFH scenario)}. Although this hypothesis cannot be rejected, it is less favored by the predictions of stellar halo formation in a cosmological context \citep{BullockJohnston2005,Font2006b,Deason2016,DSouzaBell2018a}.

To \added{attempt to} discern between \replaced{these two hypotheses}{the LM-SFH and HM-SFE hypotheses}, we consider the accretion of an SMC-mass galaxy onto M31 $\sim$10 Gyr ago. Assuming a significant metallicity dispersion as observed in LG galaxies ($\sim$0.4 dex; e.g., \citealt{Kirby2013,Ho2015,Kirby2020}), such a galaxy would span $-1.9 \ (-2.2) \lesssim \langle{\rm \feh}\rangle \lesssim -0.9 \ (-0.6)$ within 1$\sigma$ (2$\sigma$). It is unclear if such a galaxy could account for the most metal-poor stars observed in M31's \replaced{low-$\alpha$}{dSph-like} population (\feh\ $\lesssim$ $-2.2$). Additional progenitors with $M_\star \sim 10^{7-8} \ M_\odot$ may be necessary to explain these low-metallicity stars \added{in an accretion origin scenario.} \deleted{On the high metallicity end, we cannot attribute stars with \feh\ $\gtrsim -1.0$ to the low-$\alpha$ population because the M31 dwarf galaxy PDF utilized in its detection (\S~\ref{sec:low_alpha_cards}) cuts off at \feh\ $\gtrsim -1.0$ (Figure~\ref{fig:halo_vs_dsphs}). Although low-$\alpha$ stars in M31 appear to exist above this metallicity, and even show overlap with the body of the SMC's abundance distribution (Figure~\ref{fig:mcs}), we cannot meaningfully isolate them from the high-$\alpha$ population with current data.}

\section{Summary}

We have presented measurements of \alphafe\ and \feh\ for \replaced{129}{128} individual M31 RGB stars from low- ($R\sim3000$) and medium- ($R\sim6000$) resolution  Keck/DEIMOS spectroscopy. With a combined sample of \replaced{198}{197} M31 RGB stars with abundance measurements in inner halo fields (\citealt{Escala2019,Escala2020,Gilbert2019}; this work), we have undertaken an analysis of the chemical abundance properties of M31's kinematically smooth stellar halo between 8--34 kpc. Our primary results are the following:
\begin{enumerate}
    \item We measured $\langle$\feh$\rangle$ = $-$1.08 $\pm$ 0.04 ($-$1.17 $\pm$ 0.04) and $\langle$\alphafe$\rangle$ = 0.40 $\pm$ 0.03 (0.39 $\pm$ 0.03) including (excluding) substructure in M31's inner stellar halo (\S~\ref{sec:abund_gradients}).
    \item We measured a radial \feh\ gradient of $-$0.025 $\pm$ 0.002 dex kpc$^{-1}$, with an intercept of \feh\ = $-$0.72 $\pm$ 0.03, for the smooth halo (\S~\ref{sec:abund_gradients}). Including substructure results in a shallower \feh\ gradient ($-$0.018 $\pm$ 0.001). We did not find statistically significant radial gradients in \alphafe, including or excluding substructure.
    \item We reaffirmed previous results based on smaller sample sizes \citep{Escala2020,Kirby2020} that the chemical abundance distribution of M31's inner halo is incompatible with having formed from progenitors similar to present-day M31 \replaced{satellite galaxies}{dSphs} \citep{Vargas2014a,Kirby2020} \textit{when taken as a whole} (\S~\ref{sec:low_alpha_1D}).
    \item \added{Using Bayesian inference techniques,} we robustly identified a subset of stars ($N$ = \replaced{37}{22} with $p_{\rm dSph} >$ \replaced{0.5}{0.75}; Eq.~\ref{eq:pdwarf}) belonging to \added{M31's} smooth halo \added{component} that have \feh\ and \alphafe\ measurements consistent with M31 \replaced{satellite galaxies}{dSphs} (\S~\ref{sec:low_alpha_cards}). \deleted{This ``low-$\alpha$'' population has $\langle$\feh$\rangle$ = $-1.73\pm0.05$ and $\langle$\alphafe$\rangle$ = $0.16\pm0.07$. The complement of the low-$\alpha$ population, M31's high-$\alpha$ population, has $\langle$\feh$\rangle$ = $-1.05\pm0.03$ and $\langle$\alphafe$\rangle$ = $0.46\pm0.03$ with 82 likely members. The remaining 79 M31 RGB stars are kinematically associated with substructure (to be further analyzed in I.\@ Escala et al., in preparation).}
    
    \item We compared \replaced{M31's low-$\alpha$ and high-$\alpha$ populations to analogous low-[Mg/Fe] and high-[Mg/Fe] populations in the stellar halo of the MW (\S~\ref{sec:ges}; \citealt{Hayes2018}), where the low-[Mg/Fe] population is associated with Gaia-Enceladus tidal debris \citep{Helmi2018,Haywood2018,Belokurov2018}.}{the abundance distributions of the stellar halos of M31 and the MW (\citealt{Hayes2018}; \S~\ref{sec:ges}), finding that M31's halo appears to be more metal-rich and $\alpha$-rich on average.} \deleted{\added{Furthermore,} a subset of M31's \replaced{low-$\alpha$}{dSph-like} stars \deleted{with ${\rm \feh} \gtrsim-1.8$} appear \added{potentially} similar to the low-[Mg/Fe] population in the MW's halo \citep{Hayes2018}, which is associated with Gaia-Enceladus tidal debris \citep{Helmi2018,Haywood2018,Belokurov2018}. \deleted{but it is unclear if this applies to the most metal-poor low-$\alpha$ stars in M31.}}
    
    \item We compared M31's \replaced{low-$\alpha$ population}{dSph-like stars} to the Magellanic Clouds \citep{Nidever2019}, finding that \replaced{its}{the population's} chemical evolution is similar to that of the MCs at \added{early times for} \feh\ $\lesssim-1.5$ (\S~\ref{sec:mcs}). This indicates that \replaced{it likely}{the dSph-like stars could have} formed in an \added{isolated} environment, \added{as did the MCs}, \replaced{with}{resulting in} low star formation efficiency.
    
    \item We concluded that M31's \replaced{low-$\alpha$}{dSph-like} population was \added{most} likely accreted (\S~\ref{sec:halo_form}) \added{onto the stellar halo from progenitors either similar in stellar mass ($M_\star\sim10^{6-8}M_\odot$) and star formation history to M31 dSphs (the LM-SFH scenario), or progenitors of higher mass ($M_\star\gtrsim10^{8-9}M_\odot$) and similarly low star formation efficiency (the HM-SFE scenario)}. \deleted{We discussed potential origin scenarios, including massive ($M_\star\gtrsim10^{8-9}M_\odot$) progenitor(s) and progenitor(s) similar to classical dwarf galaxies ($M_\star\sim10^{7-8}M_\odot$).}
    \replaced{The high-$\alpha$ population}{The remaining, dominant population of stars in the smooth component of M31's halo} may result from progenitor(s) distinct from those of the \replaced{low-$\alpha$}{dSph-like} population, \added{and/}or represent an {\it in-situ} stellar halo component.
\end{enumerate}

\acknowledgements

\added{We thank the referee for their comments, which improved this paper.} The authors would additionally like to thank Miles Cranmer and Erik Tollerud for insightful discussions regarding statistics and David Nidever for providing a catalog of APOGEE data for the Magellanic Clouds. We also thank Stephen Gwyn for reducing the photometry for slitmasks f109\_1 and f123\_1 and Jason Kalirai for the reduction of f130\_1. 

I.E. acknowledges support from a National Science Foundation (NSF) Graduate Research Fellowship under Grant No.\ DGE-1745301, \added{in addition to a Carnegie-Princeton Fellowship through the Carnegie Observatories}.  This material is based upon work supported by the NSF under Grants No.\ AST-1614081 (E.N.K., I.E.) AST-1614569 (K.M.G, J.W.), and AST-1412648 (P.G.). E.N.K gratefully acknowledges support from a Cottrell Scholar award administered by the Research Corporation for Science Advancement, as well as funding from generous donors to the California Institute of Technology. E.C.C is supported by a Flatiron Research Fellowship at the Flatiron Institute. The Flatiron Institute is supported by the Simons Foundation. The analysis
pipeline used to reduce the DEIMOS data was developed at UC Berkeley with support from NSF grant AST-
0071048.

We are grateful to the many people who have worked to make the Keck Telescope and its instruments a reality and to operate and maintain the Keck Observatory. The authors wish to recognize and acknowledge the very significant cultural role and reverence that the summit of Maunakea has always had within the indigenous Hawaiian community.  We are most fortunate to have the opportunity to conduct observations from this mountain.

\facilities{Keck II/DEIMOS}
\software{astropy \citep{Astropy2013, Astropy2018},
emcee \citep{Foreman-Mackey2013}}

\newpage
\appendix

\section{Catalog of Stellar Parameters and Abundances}
\label{sec:appendix}

Stellar parameters and elemental abundances of individual M31 RGB stars for fields f109\_1, f123\_1, f130\_1, a0\_1, and a3 are presented in Table~\ref{tab:abund_catalog}. The table includes data for \replaced{129}{128} total M31 RGB stars with reliable \feh \ and \alphafe\ measurements (\S~\ref{sec:sample}), in addition to 80 M31 RGB stars that only have reliable \feh \ measurements. 

\startlongtable
\begin{deluxetable*}{lccccccccccc}
\tablecolumns{12}
\tablewidth{0pc}
\tablecaption{Stellar Parameters and Elemental Abundances of Individual M31 RGB Stars\label{tab:abund_catalog}\tablenotemark{a}}
\tablehead{
        \multicolumn{1}{c}{Object} &
        \multicolumn{2}{c}{Sky Coordinates} & 
        \multicolumn{1}{c}{$v_{\rm{helio}}$} &
        \multicolumn{1}{c}{S/N} &
        \multicolumn{1}{c}{$T_{\rm eff}$} & \multicolumn{1}{c}{$\delta(T_{\rm eff}$)} &\multicolumn{1}{c}{$\log$ $g$} &\multicolumn{1}{c}{\feh} &\multicolumn{1}{c}{$\delta$(\feh)} &\multicolumn{1}{c}{\alphafe} &\multicolumn{1}{c}{$\delta$(\alphafe)}\\
        \multicolumn{1}{c}{ID} &
        \multicolumn{1}{c}{RA} & \multicolumn{1}{c}{Dec} & \multicolumn{1}{c}{(km s$^{-1}$)} & 
        \multicolumn{1}{c}{(\AA\,$^{-1}$)} &
        \multicolumn{1}{c}{(K)} & 
        \multicolumn{1}{c}{(K)} &
        \multicolumn{1}{c}{(dex)} &
        \multicolumn{1}{c}{(dex)} &
        \multicolumn{1}{c}{(dex)} &
        \multicolumn{1}{c}{(dex)} &
        \multicolumn{1}{c}{(dex)}
}
\startdata
\multicolumn{12}{c}{f109\_1 (9 kpc Halo Field)}\\\hline
1092408 & 00$^{\rm{h}}$45$^{\rm{m}}$44.35$^{\rm{s}}$ & +40$^\circ$54$^{'}$15.0$^{''}$ & $-$281.6 & 15 & 3883 & 5 & 0.64 & $-$0.96 & 0.13 & 0.23 & 0.27\\
1095375 & 00$^{\rm{h}}$45$^{\rm{m}}$49.45$^{\rm{s}}$ & +40$^\circ$58$^{'}$53.8$^{''}$ & $-$190.4 & 12 & 4082 & 10 & 1.06 & $-$0.83 & 0.14 & 0.39 & 0.38\\
1092785 & 00$^{\rm{h}}$45$^{\rm{m}}$44.27$^{\rm{s}}$ & +40$^\circ$54$^{'}$52.2$^{''}$ & $-$288.9 & 13 & 3975 & 6 & 0.77 & $-$0.56 & 0.13 & 0.68 & 0.26\\\hline
\multicolumn{12}{c}{f123\_1 (18 kpc Halo Field)}\\\hline
1230048 & 00$^{\rm{h}}$48$^{\rm{m}}$18.89$^{\rm{s}}$ & +40$^\circ$20$^{'}$18.1$^{''}$ & $-$350.2 & 7 & 4825 & 51 & 1.51 & $-$0.81 & 0.17 & \nodata & \nodata\\
1230053 & 00$^{\rm{h}}$48$^{\rm{m}}$05.10$^{\rm{s}}$ & +40$^\circ$20$^{'}$22.8$^{''}$ & $-$158.2 & 22 & 4554 & 41 & 0.73 & $-$1.92 & 0.12 & 0.09 & 0.23\\
1230079 & 00$^{\rm{h}}$48$^{\rm{m}}$18.52$^{\rm{s}}$ & +40$^\circ$20$^{'}$40.1$^{''}$ & $-$324.3 & 13 & 4198 & 39 & 0.91 & $-$1.00 & 0.11 & 0.58 & 0.24\\\hline
\multicolumn{12}{c}{f130\_1 (23 kpc Halo Field)}\\\hline
1300360 & 00$^{\rm{h}}$49$^{\rm{m}}$12.01$^{\rm{s}}$ & +40$^\circ$10$^{'}$00.3$^{''}$ & $-$280.2 & 14 & 3938 & 8 & 1.14 & $-$1.15 & 0.14 & 0.56 & 0.25\\
1300698 & 00$^{\rm{h}}$49$^{\rm{m}}$03.01$^{\rm{s}}$ & +40$^\circ$15$^{'}$05.1$^{''}$ & $-$235.4 & 15 & 3774 & 37 & 0.98 & $-$0.34 & 0.14 & 0.60 & 0.29\\
1300024 & 00$^{\rm{h}}$49$^{\rm{m}}$06.69$^{\rm{s}}$ & +40$^\circ$04$^{'}$54.2$^{''}$ & $-$360.4 & 52 & 4459 & 3 & 0.66 & $-$2.60 & 0.16 & 0.70 & 0.23\\\hline
\multicolumn{12}{c}{a0\_1 (31 kpc Halo Field)}\\\hline
8002454 & 00$^{\rm{h}}$51$^{\rm{m}}$59.62$^{\rm{s}}$ & +39$^\circ$46$^{'}$49.7$^{''}$ & $-$317.4 & 11 & 3991 & 14 & 1.19 & $-$1.18 & 0.14 & \nodata & \nodata\\
7007191 & 00$^{\rm{h}}$51$^{\rm{m}}$30.89$^{\rm{s}}$ & +39$^\circ$55$^{'}$56.4$^{''}$ & $-$278.1 & 19 & 4410 & 14 & 1.13 & $-$1.55 & 0.15 & 0.10 & 0.43\\
7003904 & 00$^{\rm{h}}$51m49.68$^{\rm{s}}$ & +39$^\circ$51$^{'}$43.1$^{''}$ & $-$268.2 & 21 & 3991 & 11 & 0.80 & $-$1.19 & 0.13 & 0.38 & 0.32\\\hline
\multicolumn{12}{c}{a3\_1 (33 kpc Halo Field)}\\\hline
7003496 & 00$^{\rm{h}}$03$^{\rm{m}}$12.68$^{\rm{s}}$ & +39$^\circ$00$^{'}$28.1$^{''}$ & $-$278.5 & 11 & 4261 & 34 & 1.31 & $-$2.90 & 0.26 & \nodata & \nodata\\
7003469 & 00$^{\rm{h}}$03$^{\rm{m}}$12.76$^{\rm{s}}$ & +39$^\circ$03$^{'}$16.7$^{''}$ & $-$462.2 & 9 & 3996 & 14 & 1.14 & $-$0.56 & 0.14 & 0.74 & 0.33\\
7004205 & 00$^{\rm{h}}$03$^{\rm{m}}$11.44$^{\rm{s}}$ & +38$^\circ$59$^{'}$44.0$^{''}$ & $-$355.3 & 11 & 3793 & 9 & 0.83 & $-$1.42 & 0.15 & $-$0.37 & 0.33\\\hline
\multicolumn{12}{c}{a3\_2 (33 kpc Halo Field)}\\\hline
6004645 & 00$^{\rm{h}}$03$^{\rm{m}}$11.76$^{\rm{s}}$ & +39$^\circ$12$^{'}$03.3$^{''}$ & $-$431.1 & 18 & 4372 & 10 & 0.98 & $-$0.92 & 0.13 & $-$0.19 & 0.35\\
6004424 & 00$^{\rm{h}}$03$^{\rm{m}}$12.05$^{\rm{s}}$ & +39$^\circ$10$^{'}$33.0$^{''}$ & $-$343.6 & 48 & 4314 & 5 & 0.64 & $-$2.10 & 0.14 & 0.61 & 0.90\\
7005249 & 00$^{\rm{h}}$03$^{\rm{m}}$11.68$^{\rm{s}}$ & +39$^\circ$03$^{'}$36.6$^{''}$ & $-$412.5 & 19 & 4126 & 9 & 1.03 & $-$0.77 & 0.13 & 0.06 & 0.18\\\hline
\enddata
\tablenotetext{a}{The errors presented for $T_{\rm eff}$ represent only the random component of the total uncertainty. However, the errors for \feh \ and \alphafe\ include systematic components that account for errors propagated by inaccuracies in $T_{\rm eff}$ \citep{Kirby2008,Gilbert2019,Escala2019,Escala2020}}
\tablecomments{(This table is available in its entirety in machine-readable form.).}
\end{deluxetable*}

\section{Identifying Probable Accreted Stars in the Stellar Halo of the MW}
\label{sec:appendix_mw}

\begin{figure*}
    \centering
    \includegraphics[width=\textwidth]{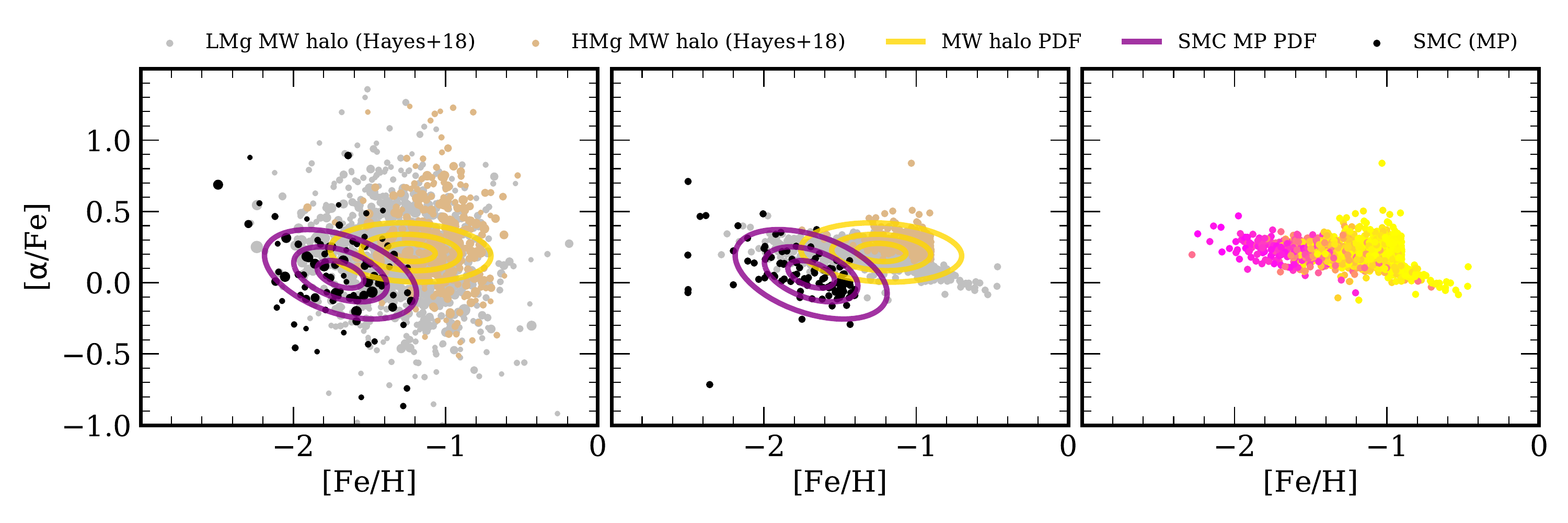}
    \caption{(Left panel) The perturbed abundance distributions of the MW halo \citep{Hayes2018}, separated into LMg (grey points) and HMg (brown points) populations, and the SMC MP tail (black points; \citealt{Nidever2019}). The 2D PDFs fitted to the \textit{entire} MW halo and the SMC MP tail are shown as yellow and magenta contours, respectively. (Middle panel) The 2D PDFs fitted to the perturbed abundance distributions compared against the \textit{unperturbed} abundance distributions. (Right panel) The unperturbed MW halo abundance distribution color-coded according to a star's probability of being SMC-like (defined analogously to Eq.~\ref{eq:pdwarf}), as evaluated from the \textit{perturbed} abundance distributions. Magenta stars are SMC-like, whereas yellow stars are not.}
    \label{fig:mwhalo_smc}
\end{figure*}

In \S~\ref{sec:low_alpha_cards}, we modeled the 2-D chemical abundance distributions of M31's stellar halo and its dSphs in order to statistically identify M31 RGB stars with abundances similar to M31 dSphs. The suggestion that such stars may exist in M31's stellar halo was first made apparent by Figure~\ref{fig:halo_vs_dsphs} through a comparison to abundance measurements in M31 dSphs obtained using the same techniques. In this Appendix, we use APOGEE chemical abundances of the MW's stellar halo \citep{Hayes2018} and the ancient, metal-poor tail of the SMC (\S~\ref{sec:mcs}; \citealt{Nidever2019}), which we perturb by an M31-like error distribution (\S~\ref{sec:ges}),  to illustrate the efficacy of this method for successfully recovering dwarf-galaxy-like stars in a stellar halo population. 

Figure~\ref{fig:mwhalo_smc} outlines the steps in this proof of concept. First, we perturbed the MW's halo abundance distribution by M31-like errors, as described in \S~6.1. Then, we fit a 2D PDF, following \S~5.4.2, to the perturbed MW halo abundances \textit{without separating it into the LMg or HMg populations} of Hayes et al.\@ 2018. The left panel of Figure~\ref{fig:mwhalo_smc} illustrates the resulting PDF (yellow contours). Next, we isolated the metal-poor tail of the SMC abundance distribution ([Fe/H] $\lesssim -1.4$). As discussed in \S~6.2, chemical evolution models for the Large Magellanic Cloud \citep{BekkiTsujimoto2012,Nidever2019} indicate that its metal poor tail was in place by $\sim$7-9 Gyr ago at the latest. Given the similarly low star formation efficiency of the SMC \citep{Nidever2019}, we can assume that its metal-poor tail is similarly ancient. The more ancient, metal-poor portion of the SMC abundance distribution is the relevant portion to compare to the predominately old stellar halo of the MW (e.g., \citealt{Gallart2019,Bonaca2020}). We perturbed the SMC metal-poor tail by M31-like errors and fit a 2D PDF to resulting abundance distribution (left panel of Figure~\ref{fig:mwhalo_smc}, magenta contours). 

The middle panel of Figure~\ref{fig:mwhalo_smc} shows the original unperturbed MW halo and SMC abundance distributions compared to the 2D PDFs calculated \textit{from the perturbed distributions}. We used the MW halo PDF and the SMC metal-poor (MP) PDF to calculate a probability that each MW halo star has abundances similar to the SMC (analogously to Eq.~\ref{eq:pdwarf}). To calculate this probability, we used the \textit{perturbed values} of the abundances for a given star. The right panel of Figure~\ref{fig:mwhalo_smc} shows the \textit{unperturbed} MW halo abundances color-coded according to this probability calculated using the \textit{perturbed} equivalent.

\begin{figure*}
    \centering
    \includegraphics[width=0.49\textwidth]{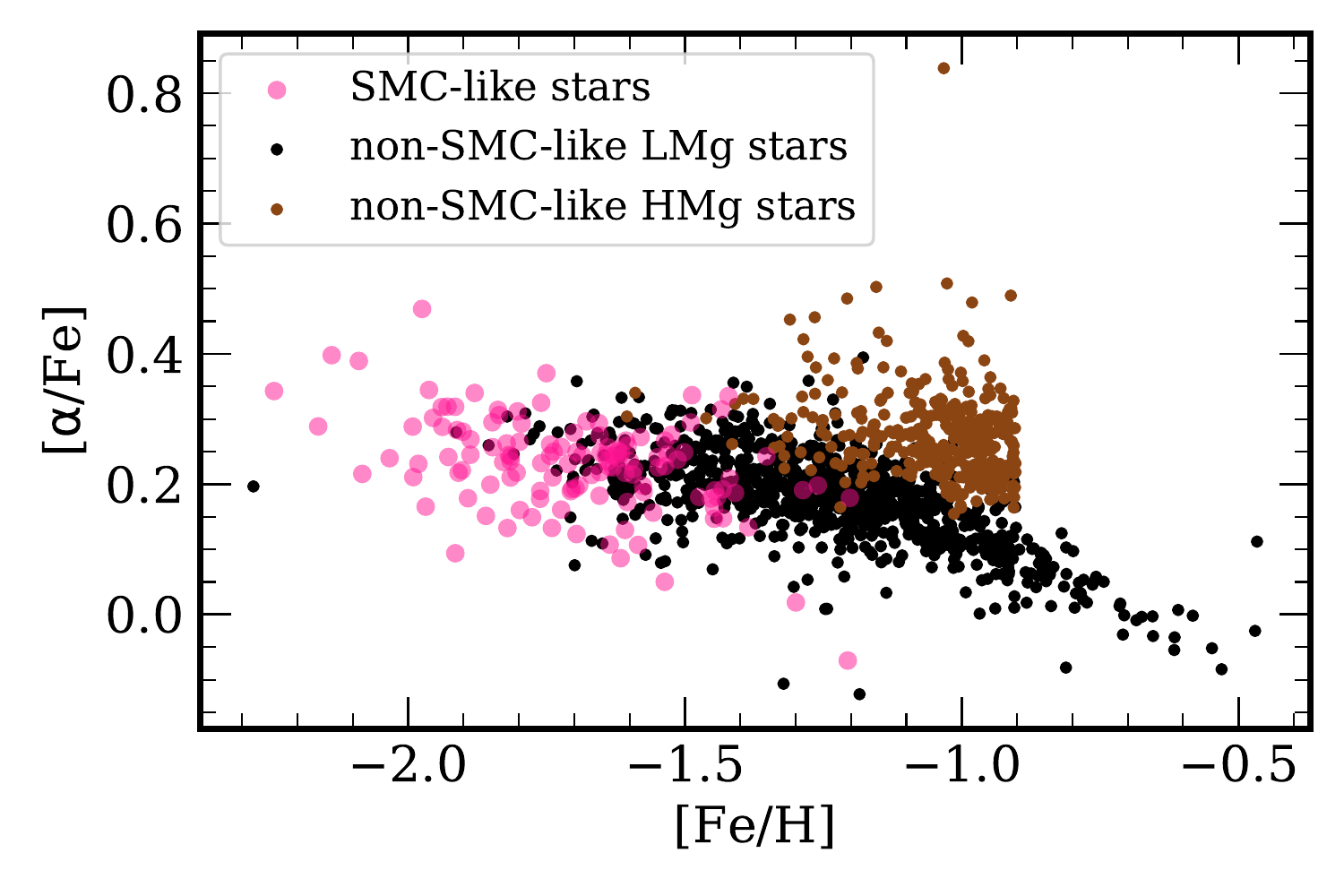}
    \includegraphics[width=0.49\textwidth]{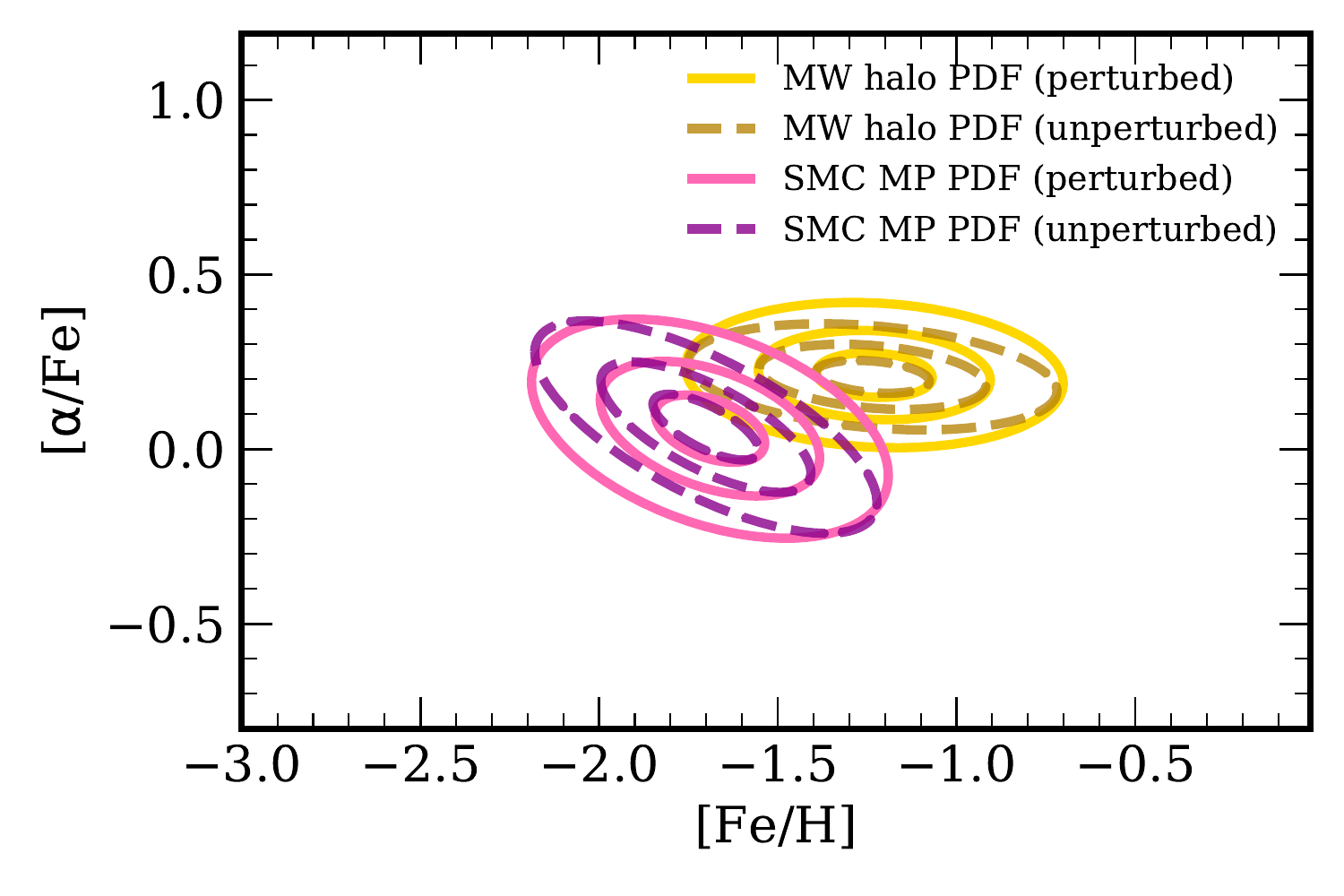}
    \caption{Stars identified as SMC-like ($p_{\rm SMC} >$ \replaced{0.5}{0.75}; pink circles) compared to MW halo stars that were not classified as SMC-like separated according their membership in the LMg or HMg population (black and brown points, respectively; \citealt{Hayes2018}). Some stars classified as SMC-like reside \replaced{in}{close to} the HMg region of the MW halo abundance distribution because they were classified based on their \textit{perturbed} abundances and uncertainties. \replaced{98\%}{99\%} of SMC-like stars are correctly classified as originating from the accreted LMg population. (Right panel) 2D PDFs fit to abundance distributions of the perturbed MW halo (solid yellow contours) and SMC MP tail (solid pink contours), compared to 2D PDFs fit to the \textit{unperturbed} equivalents (dashed gold and purple contours, respectively). The true, underlying abundance distribution is constrained by both the perturbed and unperturbed abundance distributions.}
    \label{fig:mwhalo_smc_part2}
\end{figure*}

We compared the stars classified as ``SMC-like'' ($p_{\rm SMC} >$ \replaced{0.5}{0.75}) by the perturbation analysis to the stars that are truly SMC-like (e.g., the LMg stars). The LMg population corresponds to stars that were accreted during the Gaia-Enceladus merger event (\S~\ref{sec:ges}; \citealt{Helmi2018,Belokurov2018,Haywood2018}) of a galaxy at least as massive as the SMC. Assuming that stars with $p_{\rm SMC} > 0.75$ (0.5) are SMC-like, this results in the correct classification of 99\% (98\%) of SMC-like stars as belonging to the LMg population. If we assume that stars with $p_{\rm SMC} < 0.25$ (0.5) are \textit{not} SMC-like, this results in the correct classification of 38\% (32\%) of not-SMC-like stars as belonging to the HMg population. This is because the classification of SMC-like stars is incomplete at high metallicity beyond the bounds of the SMC MP PDF. However, the relevant point for our analysis is that the classification of stars as SMC-like is highly accurate. Repeating this exercise for different perturbations of the MW halo and SMC abundance distributions does not alter these percentages by more than 0.5\%.

Figure~\ref{fig:mwhalo_smc_part2} further supports this analysis. The left panel shows stars that we have identified as SMC-like overlaid on the full MW halo abundance distribution, where it is evident that the overwhelming majority of SMC-like stars correspond to the LMg population. The right panel shows PDFs fit to the perturbed and unperturbed abundance distributions for the MW halo and SMC MP tail. The primary difference between the perturbed and unperturbed PDFs is that the unperturbed PDFs are narrower in the [$\alpha$/Fe] direction, whereas the shapes, angles, and metallicity spreads between each case are similar. Both the perturbed and unperturbed distributions represent approximations of the true, underlying parent distribution. The fact that they show significant overlap demonstrates that we can in fact reasonably constrain the underlying PDFs describing the abundance distributions in spite of our uncertainties in \alphafe.

In this example, the statistical methodology developed in \S~\ref{sec:low_alpha_cards} is successful at identifying MW stars that are similar in their 2D [$\alpha$/Fe] vs.\ [Fe/H] abundances to stars in the metal-poor tail of the SMC abundance distribution, using abundances perturbed by the typical uncertainties in our M31 sample. This gives us confidence that we can statistically identify M31 halo stars that are similar in their 2D [$\alpha$/Fe] vs. [Fe/H] abundances to stars in M31's dSph galaxies, and thus may have formed in similar environments and been accreted onto M31's halo. This is true despite the fact that we did not assume any knowledge about the true underlying abundance distributions, and despite the size of the uncertainties [$\alpha$/Fe]. Thus, this statistical method can successfully identify dwarf-galaxy-like stars not only in the MW's halo, but also in the smooth component of M31's stellar halo.\\


\begin{thebibliography}

\bibitem[Albareti et al.(2017)]{Albareti2017} Albareti, F.~D., Allende Prieto, C., Almeida, A., et al.\ 2017, \apjs, 233, 25

\bibitem[Ahumada et al.(2019)]{Ahumada2019} Ahumada, R., Allende Prieto, C., Almeida, A., et al.\ 2019, arXiv e-prints, arXiv:1912.02905

\bibitem[Astropy Collaboration et al.(2013)]{Astropy2013} Astropy Collaboration, Robitaille, T.~P., Tollerud, E.~J., et al.\ 2013, \aap, 558, A33

\bibitem[Astropy Collaboration et al.(2018)]{Astropy2018} Astropy Collaboration, Price-Whelan, A.~M., Sip{\H{o}}cz, B.~M., et al.\ 2018, \aj, 156, 123

\bibitem[Bekki \& Tsujimoto(2012)]{BekkiTsujimoto2012} Bekki, K., \& Tsujimoto, T.\ 2012, \apj, 761, 180

\bibitem[Belokurov et al.(2018)]{Belokurov2018} Belokurov, V., Erkal, D., Evans, N.~W., et al.\ 2018, \mnras, 478, 611

\bibitem[Belokurov et al.(2020)]{Belokurov2020} Belokurov, V., Sanders, J.~L., Fattahi, A., et al.\ 2020, \mnras, doi:10.1093/mnras/staa876

\bibitem[Besla et al.(2007)]{Besla2007} Besla, G., Kallivayalil, N., Hernquist, L., et al.\ 2007, \apj, 668, 949

\bibitem[Besla et al.(2012)]{Besla2012} Besla, G., Kallivayalil, N., Hernquist, L., et al.\ 2012, \mnras, 421, 2109

\bibitem[Blanton et al.(2017)]{Blanton2017} Blanton, M.~R., Bershady, M.~A., Abolfathi, B., et al.\ 2017, \aj, 154, 28

\bibitem[Bonaca et al.(2017)]{Bonaca2017} Bonaca, A., Conroy, C., Wetzel, A., et al.\ 2017, \apj, 845, 101

\bibitem[Bonaca et al.(2020)]{Bonaca2020} Bonaca, A., Conroy, C., Cargile, P.~A., et al.\ 2020, arXiv e-prints, arXiv:2004.11384

\bibitem[Brown et al.(2007)]{Brown2007} Brown, T.~M., Smith, E., Ferguson, H.~C., et al.\ 2007, \apj, 658, L95.

\bibitem[Brown et al.(2006)]{Brown2006} Brown, T.~M., Smith, E., Ferguson, H.~C., et al.\ 2006, \apj, 652, 323.

\bibitem[Brown et al.(2008)]{Brown2008} Brown, T.~M., Beaton, R., Chiba, M., et al.\ 2008, \apjl, 685, L121

\bibitem[Brown et al.(2009)]{Brown2009} Brown, T.~M., Smith, E., Ferguson, H.~C., et al.\ 2009, \apjs, 184, 152

\bibitem[Bullock \& Johnston(2005)]{BullockJohnston2005} Bullock, J.~S., \& Johnston, K.~V.\ 2005, \apj, 635, 931

\bibitem[Carollo et al.(2007)]{Carollo2007} Carollo, D., Beers, T.~C., Lee, Y.~S., et al.\ 2007, \nat, 450, 1020

\bibitem[Carollo et al.(2010)]{Carollo2010} Carollo, D., Beers, T.~C., Chiba, M., et al.\ 2010, \apj, 712, 692

\bibitem[Chapman et al.(2006)]{Chapman2006} Chapman, S.~C., Ibata, R., Lewis, G.~F., et al.\ 2006, \apj, 653, 255

\bibitem[Clementini et al.(2011)]{Clementini2011} Clementini, G., Contreras Ramos, R., Federici, L., et al.\ 2011, \apj, 743, 19

\bibitem[Cohen et al.(2018)]{Cohen2018} Cohen, R.~E., Kalirai, J.~S., Gilbert, K.~M., et al.\ 2018, \aj, 156, 230

\bibitem[Collins et al.(2011)]{Collins2011} Collins, M.~L.~M., Chapman, S.~C., Ibata, R.~A., et al.\ 2011, \mnras, 413, 1548

\bibitem[Conn et al.(2016)]{Conn2016} Conn, A.~R., McMonigal, B., Bate, N.~F., et al.\ 2016, \mnras, 458, 3282

\bibitem[Conroy et al.(2019)]{Conroy2019} Conroy, C., Naidu, R.~P., Zaritsky, D., et al.\ 2019, \apj, 887, 237

\bibitem[Cooper et al.(2010)]{Cooper2010} Cooper, A.~P., Cole, S., Frenk, C.~S., et al.\ 2010, \mnras, 406, 744

\bibitem[Cooper et al.(2012)]{Cooper2012} Cooper, M.~C., Griffith, R.~L., Newman, J.~A., et al.\ 2012, \mnras, 419, 3018

\bibitem[Cooper et al.(2013)]{Cooper2013} Cooper, A.~P., D'Souza, R., Kauffmann, G., et al.\ 2013, \mnras, 434, 3348

\bibitem[Cooper et al.(2015)]{Cooper2015} Cooper, A.~P., Parry, O.~H., Lowing, B., et al.\ 2015, \mnras, 454, 3185

\bibitem[Courteau et al.(2011)]{Courteau2011} Courteau, S., Widrow, L.~M., McDonald, M., et al.\ 2011, \apj, 739, 20


\bibitem[D'Souza \& Bell(2018a)]{DSouzaBell2018a} D'Souza, R., \& Bell, E.~F.\ 2018, \mnras, 474, 5300

\bibitem[D'Souza \& Bell(2018b)]{DSouzaBell2018b} D'Souza, R., \& Bell, E.~F.\ 2018, Nature Astronomy, 2, 737

\bibitem[Das et al.(2020)]{Das2020} Das, P., Hawkins, K., \& Jofr{\'e}, P.\ 2020, \mnras, doi:10.1093/mnras/stz3537

\bibitem[Deason et al.(2013)]{Deason2013} Deason, A.~J., Belokurov, V., Evans, N.~W., et al.\ 2013, \apj, 763, 113

\bibitem[Deason et al.(2016)]{Deason2016} Deason, A.~J., Mao, Y.-Y., \& Wechsler, R.~H.\ 2016, \apj, 821, 5

\bibitem[Deason et al.(2018)]{Deason2018} Deason, A.~J., Belokurov, V., Koposov, S.~E., et al.\ 2018, \apjl, 862, L1

\bibitem[Demarque et al.(2004)]{Demarque2004} Demarque, P., Woo, J.-H., Kim, Y.-C., et al.\ 2004, \apjs, 155, 667

\bibitem[de Vaucouleurs(1958)]{deVaucouleurs1958} de Vaucouleurs, G.\ 1958, \apj, 128, 465

\bibitem[Di Matteo et al.(2019)]{DiMatteo2019} Di Matteo, P., Haywood, M., Lehnert, M.~D., et al.\ 2019, \aap, 632, A4

\bibitem[Dorman et al.(2012)]{Dorman2012} Dorman, C.~E., Guhathakurta, P., Fardal, M.~A., et al.\ 2012, \apj, 752, 147

\bibitem[Dorman et al.(2013)]{Dorman2013} Dorman, C.~E., Widrow, L.~M., Guhathakurta, P., et al.\ 2013, \apj, 779, 103

\bibitem[Dorman et al.(2015)]{Dorman2015} Dorman, C.~E., Guhathakurta, P., Seth, A.~C., et al.\ 2015, \apj, 803, 24

\bibitem[Escala et al.(2019)]{Escala2019} Escala, I., Kirby, E.~N., Gilbert, K.~M., et al.\ 2019, \apj, 878, 42

\bibitem[Escala et al.(2020)]{Escala2020} Escala, I., Gilbert, K.~M., Kirby, E.~N., et al.\ 2020, \apj, 889, 177

\bibitem[Eisenstein et al.(2011)]{Eisenstein2011} Eisenstein, D.~J., Weinberg, D.~H., Agol, E., et al.\ 2011, \aj, 142, 72


\bibitem[Faber et al.(2003)]{Faber2003} Faber, S.~M., Phillips, A.~C., Kibrick, R.~I., et al.\ 2003, Instrument Design and Performance for Optical/infrared Ground-based Telescopes, 1657.

\bibitem[Fardal et al.(2007)]{Fardal2007} Fardal, M.~A., Guhathakurta, P., Babul, A., et al.\ 2007, \mnras, 380, 15.

\bibitem[Fardal et al.(2008)]{Fardal2008} Fardal, M.~A., Babul, A., Guhathakurta, P., et al.\ 2008, \apjl, 682, L33

\bibitem[Fattahi et al.(2019)]{Fattahi2019} Fattahi, A., Belokurov, V., Deason, A.~J., et al.\ 2019, \mnras, 484, 4471

\bibitem[Fattahi et al.(2020)]{Fattahi2020arXiv} Fattahi, A., Deason, A.~J., Frenk, C.~S., et al.\ 2020, arXiv e-prints, arXiv:2002.12043

\bibitem[Ferguson et al.(2002)]{Ferguson2002} Ferguson, A.~M.~N., Irwin, M.~J., Ibata, R.~A., et al.\ 2002, \aj, 124, 1452

\bibitem[Fern{\'a}ndez-Alvar et al.(2018)]{Fernandez-Alvar2018} Fern{\'a}ndez-Alvar, E., Carigi, L., Schuster, W.~J., et al.\ 2018, \apj, 852, 50

\bibitem[Font et al.(2006b)]{Font2006b} Font, A.~S., Johnston, K.~V., Bullock, J.~S., et al.\ 2006, \apj, 638, 585

\bibitem[Font et al.(2006c)]{Font2006c} Font, A.~S., Johnston, K.~V., Bullock, J.~S., et al.\ 2006, \apj, 646, 886

\bibitem[Font et al.(2008)]{Font2008} Font, A.~S., Johnston, K.~V., Ferguson, A.~M.~N., et al.\ 2008, \apj, 673, 215

\bibitem[Font et al.(2011)]{Font2011} Font, A.~S., McCarthy, I.~G., Crain, R.~A., et al.\ 2011, \mnras, 416, 2802

\bibitem[Font et al.(2020)]{Font2020arXiv} Font, A.~S., McCarthy, I.~G., Poole-Mckenzie, R., et al.\ 2020, arXiv e-prints, arXiv:2004.01914

\bibitem[Foreman-Mackey et al.(2013)]{Foreman-Mackey2013} Foreman-Mackey, D., Hogg, D.~W., Lang, D., et al.\ 2013, \pasp, 125, 306.

\bibitem[Garc{\'\i}a P{\'e}rez et al.(2016)]{GarciaPerez2016} Garc{\'\i}a P{\'e}rez, A.~E., Allende Prieto, C., Holtzman, J.~A., et al.\ 2016, \aj, 151, 144

\bibitem[Gaia Collaboration et al.(2016)]{GaiaCollaboration2016} Gaia Collaboration, Brown, A.~G.~A., Vallenari, A., et al.\ 2016, \aap, 595, A2

\bibitem[Gaia Collaboration et al.(2018)]{GaiaCollaboration2018} Gaia Collaboration, Brown, A.~G.~A., Vallenari, A., et al.\ 2018, \aap, 616, A1

\bibitem[Gaia Collaboration et al.(2018)]{GaiaCollaboration2018a} Gaia Collaboration, Babusiaux, C., van Leeuwen, F., et al.\ 2018, \aap, 616, A10

\bibitem[Gallart et al.(2019)]{Gallart2019} Gallart, C., Bernard, E.~J., Brook, C.~B., et al.\ 2019, Nature Astronomy, 3, 932

\bibitem[Gallazzi et al.(2005)]{Gallazzi2005} Gallazzi, A., Charlot, S., Brinchmann, J., et al.\ 2005, \mnras, 362, 41

\bibitem[Gallazzi et al.(2014)]{Gallazzi2014} Gallazzi, A., Bell, E.~F., Zibetti, S., et al.\ 2014, \apj, 788, 72

\bibitem[Geha et al.(2017)]{Geha2017} Geha, M., Wechsler, R.~H., Mao, Y.-Y., et al.\ 2017, \apj, 847, 4

\bibitem[Gilbert et al.(2006)]{Gilbert2006} Gilbert, K.~M., Guhathakurta, P., Kalirai, J.~S., et al.\ 2006, \apj, 652, 1188

\bibitem[Gilbert et al.(2007)]{Gilbert2007} Gilbert, K.~M., Fardal, M., Kalirai, J.~S., et al.\ 2007, \apj, 668, 245.

\bibitem[Gilbert et al.(2009a)]{Gilbert2009a} Gilbert, K.~M., Guhathakurta, P., Kollipara, P., et al.\ 2009, \apj, 705, 1275

\bibitem[Gilbert et al.(2009b)]{Gilbert2009b} Gilbert, K.~M., Font, A.~S., Johnston, K.~V., et al.\ 2009, \apj, 701, 776

\bibitem[Gilbert et al.(2012)]{Gilbert2012} Gilbert, K.~M., Guhathakurta, P., Beaton, R.~L., et al.\ 2012, \apj, 760, 76

\bibitem[Gilbert et al.(2014)]{Gilbert2014} Gilbert, K.~M., Kalirai, J.~S., Guhathakurta, P., et al.\ 2014, \apj, 796, 76

\bibitem[Gilbert et al.(2018)]{Gilbert2018} Gilbert, K.~M., Tollerud, E., Beaton, R.~L., et al.\ 2018, \apj, 852, 128

\bibitem[Gilbert et al.(2019)]{Gilbert2019} Gilbert, K.~M., Kirby, E.~N., Escala, I., et al.\ 2019, \apj, 883, 128

\bibitem[Gilbert et al.(2020)]{Gilbert2020} Gilbert, K.~M., Wojno, J., Kirby, E.~N., et al.\ 2020, \aj, 160, 41

\bibitem[Gilmore \& Wyse(1998)]{GilmoreWyse1998} Gilmore, G., \& Wyse, R.~F.~G.\ 1998, \aj, 116, 748

\bibitem[Girardi et al.(2002)]{Girardi2002} Girardi, L., Bertelli, G., Bressan, A., et al.\ 2002, \aap, 391, 195

\bibitem[Guhathakurta et al.(2005)]{Guhathakurta2005} Guhathakurta, P., Ostheimer, J.~C., Gilbert, K.~M., et al.\ 2005, arXiv e-prints, astro-ph/0502366


\bibitem[Gwyn(2008)]{Gwyn2008} Gwyn, S.~D.~J.\ 2008, \pasp, 120, 212

\bibitem[Harmsen et al.(2017)]{Harmsen2017} Harmsen, B., Monachesi, A., Bell, E.~F., et al.\ 2017, \mnras, 466, 1491

\bibitem[Hammer et al.(2018)]{Hammer2018} Hammer, F., Yang, Y.~B., Wang, J.~L., et al.\ 2018, \mnras, 475, 2754

\bibitem[Harris \& Zaritsky(2009)]{HarrisZaritsky2009} Harris, J., \& Zaritsky, D.\ 2009, \aj, 138, 1243

\bibitem[Hawkins et al.(2014)]{Hawkins2014} Hawkins, K., Jofr{\'e}, P., Gilmore, G., et al.\ 2014, \mnras, 445, 2575

\bibitem[Hawkins et al.(2015)]{Hawkins2015} Hawkins, K., Jofr{\'e}, P., Masseron, T., et al.\ 2015, \mnras, 453, 758

\bibitem[Hayden et al.(2014)]{Hayden2014} Hayden, M.~R., Holtzman, J.~A., Bovy, J., et al.\ 2014, \aj, 147, 116

\bibitem[Hayden et al.(2015)]{Hayden2015} Hayden, M.~R., Bovy, J., Holtzman, J.~A., et al.\ 2015, \apj, 808, 132

\bibitem[Hayes et al.(2018)]{Hayes2018} Hayes, C.~R., Majewski, S.~R., Shetrone, M., et al.\ 2018, \apj, 852, 49

\bibitem[Haywood et al.(2018)]{Haywood2018} Haywood, M., Di Matteo, P., Lehnert, M.~D., et al.\ 2018, \apj, 863, 113

\bibitem[Helmi et al.(1999)]{Helmi1999} Helmi, A., White, S.~D.~M., de Zeeuw, P.~T., et al.\ 1999, \nat, 402, 53

\bibitem[Helmi et al.(2018)]{Helmi2018} Helmi, A., Babusiaux, C., Koppelman, H.~H., et al.\ 2018, \nat, 563, 85

\bibitem[Ho et al.(2015)]{Ho2015} Ho, N., Geha, M., Tollerud, E.~J., et al.\ 2015, \apj, 798, 77

\bibitem[Holland et al.(1996)]{Holland1996} Holland, S., Fahlman, G.~G., \& Richer, H.~B.\ 1996, \aj, 112, 1035

\bibitem[Holtzman et al.(2015)]{Holtzman2015} Holtzman, J.~A., Shetrone, M., Johnson, J.~A., et al.\ 2015, \aj, 150, 148

\bibitem[Ibata et al.(2001)]{Ibata2001} Ibata, R., Irwin, M., Lewis, G., et al.\ 2001, \nat, 412, 49

\bibitem[Ibata et al.(2005)]{Ibata2005} Ibata, R., Chapman, S., Ferguson, A.~M.~N., et al.\ 2005, \apj, 634, 287

\bibitem[Ibata et al.(2007)]{Ibata2007} Ibata, R., Martin, N.~F., Irwin, M., et al.\ 2007, \apj, 671, 1591

\bibitem[Ibata et al.(2014)]{Ibata2014} Ibata, R.~A., Lewis, G.~F., McConnachie, A.~W., et al.\ 2014, \apj, 780, 128

\bibitem[Irwin et al.(2005)]{Irwin2005} Irwin, M.~J., Ferguson, A.~M.~N., Ibata, R.~A., et al.\ 2005, \apjl, 628, L105

\bibitem[Ishigaki et al.(2012)]{Ishigaki2012} Ishigaki, M.~N., Chiba, M., \& Aoki, W.\ 2012, \apj, 753, 64

\bibitem[Johnston et al.(2008)]{Johnston2008} Johnston, K.~V., Bullock, J.~S., Sharma, S., et al.\ 2008, \apj, 689, 936

\bibitem[Jordi et al.(2006)]{Jordi2006} Jordi, K., Grebel, E.~K., \& Ammon, K.\ 2006, \aap, 460, 339

\bibitem[Kalirai et al.(2006a)]{Kalirai2006a} Kalirai, J.~S., Guhathakurta, P., Gilbert, K.~M., et al.\ 2006, \apj, 641, 268.

\bibitem[Kalirai et al.(2006b)]{Kalirai2006b} Kalirai, J.~S., Gilbert, K.~M., Guhathakurta, P., et al.\ 2006, \apj, 648, 389

\bibitem[Kalirai et al.(2010)]{Kalirai2010} Kalirai, J.~S., Beaton, R.~L., Geha, M.~C., et al.\ 2010, \apj, 711, 671

\bibitem[Koppelman et al.(2018)]{Koppelman2018} Koppelman, H., Helmi, A., \& Veljanoski, J.\ 2018, \apjl, 860, L11

\bibitem[Kirby et al.(2008)]{Kirby2008} Kirby, E.~N., Guhathakurta, P., \& Sneden, C.\ 2008, \apj, 682, 1217

\bibitem[Kirby et al.(2010)]{Kirby2010} Kirby, E.~N., Guhathakurta, P., Simon, J.~D., et al.\ 2010, \apjs, 191, 352

\bibitem[Kirby et al.(2013)]{Kirby2013} Kirby, E.~N., Cohen, J.~G., Guhathakurta, P., et al.\ 2013, \apj, 779, 102

\bibitem[Kirby et al.(2015)]{Kirby2015} Kirby, E.~N., Simon, J.~D., \& Cohen, J.~G.\ 2015, \apj, 810, 56

\bibitem[Kirby et al.(2020)]{Kirby2020} Kirby, E.~N., Gilbert, K.~M., Escala, I., et al.\ 2020, \aj, 159, 46

\bibitem[Koch et al.(2008)]{Koch2008} Koch, A., Rich, R.~M., Reitzel, D.~B., et al.\ 2008, \apj, 689, 958

\bibitem[Koch et al.(2019)]{Koch2019} Koch, A., Grebel, E.~K., \& Martell, S.~L.\ 2019, \aap, 625, A75

\bibitem[Lapenna et al.(2012)]{Lapenna2012} Lapenna, E., Mucciarelli, A., Origlia, L., et al.\ 2012, \apj, 761, 33

\bibitem[Lee et al.(2015)]{Lee2015} Lee, D.~M., Johnston, K.~V., Sen, B., et al.\ 2015, \apj, 802, 48

\bibitem[Leethochawalit et al.(2018)]{Leethochawalit2018} Leethochawalit, N., Kirby, E.~N., Moran, S.~M., et al.\ 2018, \apj, 856, 15

\bibitem[Leethochawalit et al.(2019)]{Leethochawalit2019} Leethochawalit, N., Kirby, E.~N., Ellis, R.~S., et al.\ 2019, \apj, 885, 100

\bibitem[Lianou et al.(2011)]{Lianou2011} Lianou, S., Grebel, E.~K., \& Koch, A.\ 2011, \aap, 531, A152

\bibitem[Ma et al.(2016)]{Ma2016} Ma, X., Hopkins, P.~F., Faucher-Gigu{\`e}re, C.-A., et al.\ 2016, \mnras, 456, 2140

\bibitem[Mackereth et al.(2019)]{Mackereth2019} Mackereth, J.~T., Schiavon, R.~P., Pfeffer, J., et al.\ 2019, \mnras, 482, 3426

\bibitem[Majewski et al.(2000)]{Majewski2000} Majewski, S.~R., Ostheimer, J.~C., Kunkel, W.~E., et al.\ 2000, \aj, 120, 2550

\bibitem[Majewski et al.(2017)]{Majewski2017} Majewski, S.~R., Schiavon, R.~P., Frinchaboy, P.~M., et al.\ 2017, \aj, 154, 94

\bibitem[Marigo et al.(2017)]{Marigo2017} Marigo, P., Girardi, L., Bressan, A., et al.\ 2017, \apj, 835, 77

\bibitem[McCarthy et al.(2012)]{McCarthy2012} McCarthy, I.~G., Font, A.~S., Crain, R.~A., et al.\ 2012, \mnras, 420, 2245

\bibitem[McConnachie et al.(2005)]{McConnachie2005} McConnachie, A.~W., Irwin, M.~J., Ferguson, A.~M.~N., et al.\ 2005, \mnras, 356, 979

\bibitem[McConnachie et al.(2009)]{McConnachie2009} McConnachie, A.~W., Irwin, M.~J., Ibata, R.~A., et al.\ 2009, \nat, 461, 66

\bibitem[McConnachie(2012)]{McConnachie2012} McConnachie, A.~W.\ 2012, \aj, 144, 4

\bibitem[McConnachie et al.(2018)]{McConnachie2018} McConnachie, A.~W., Ibata, R., Martin, N., et al.\ 2018, \apj, 868, 55

\bibitem[Merritt et al.(2016)]{Merritt2016} Merritt, A., van Dokkum, P., Abraham, R., et al.\ 2016, \apj, 830, 62

\bibitem[Monachesi et al.(2016a)]{Monachesi2016} Monachesi, A., Bell, E.~F., Radburn-Smith, D.~J., et al.\ 2016, \mnras, 457, 1419

\bibitem[Monachesi et al.(2019)]{Monachesi2019} Monachesi, A., G{\'o}mez, F.~A., Grand, R.~J.~J., et al.\ 2019, \mnras, 485, 2589

\bibitem[Mouhcine et al.(2005)]{Mouhcine2005} Mouhcine, M., Ferguson, H.~C., Rich, R.~M., et al.\ 2005, \apj, 633, 821

\bibitem[Mucciarelli(2014)]{Mucciarelli2014} Mucciarelli, A.\ 2014, Astronomische Nachrichten, 335, 79

\bibitem[Naidu et al.(2020)]{Naidu2020} Naidu, R.~P., Conroy, C., Bonaca, A., et al.\ 2020, arXiv e-prints, arXiv:2006.08625

\bibitem[Navarro et al.(2011)]{Navarro2011} Navarro, J.~F., Abadi, M.~G., Venn, K.~A., et al.\ 2011, \mnras, 412, 1203

\bibitem[Nidever et al.(2019)]{Nidever2019} Nidever, D.~L., Hasselquist, S., Hayes, C.~R., et al.\ 2019, arXiv e-prints, arXiv:1901.03448

\bibitem[Nissen \& Schuster(1997)]{NissenSchuster1997} Nissen, P.~E., \& Schuster, W.~J.\ 1997, \aap, 326, 751

\bibitem[Nissen \& Schuster(2010)]{NissenSchuster2010} Nissen, P.~E., \& Schuster, W.~J.\ 2010, \aap, 511, L10

\bibitem[Nissen \& Schuster(2011)]{NissenSchuster2011} Nissen, P.~E., \& Schuster, W.~J.\ 2011, \aap, 530, A15

\bibitem[Newman et al.(2013)]{Newman2013} Newman, J.~A., Cooper, M.~C., Davis, M., et al.\ 2013, \apjs, 208, 5

\bibitem[Ostheimer(2003)]{Ostheimer2003} Ostheimer, J.~C.\ 2003, Ph.D. Thesis

\bibitem[Pomp{\'e}ia et al.(2008)]{Pompeia2008} Pomp{\'e}ia, L., Hill, V., Spite, M., et al.\ 2008, \aap, 480, 379


\bibitem[Purcell et al.(2010)]{Purcell2010} Purcell, C.~W., Bullock, J.~S., \& Kazantzidis, S.\ 2010, \mnras, 404, 1711

\bibitem[Radburn-Smith et al.(2011)]{Radburn-Smith2011} Radburn-Smith, D.~J., de Jong, R.~S., Seth, A.~C., et al.\ 2011, \apjs, 195, 18

\bibitem[Ram{\'i}rez et al.(2012)]{Ramirez2012} Ram{\'i}rez, I., Mel{\'e}ndez, J., \& Chanam{\'e}, J.\ 2012, \apj, 757,
164

\bibitem[Reina-Campos et al.(2020)]{Reina-Campos2020} Reina-Campos, M., Hughes, M.~E., Kruijssen, J.~M.~D., et al.\ 2020, \mnras, 493, 3422

\bibitem[Reitzel, \& Guhathakurta(2002)]{ReitzelGuhathakurta2002} Reitzel, D.~B., \& Guhathakurta, P.\ 2002, \aj, 124, 234

\bibitem[Robertson et al.(2005)]{Robertson2005} Robertson, B., Bullock, J.~S., Font, A.~S., et al.\ 2005, \apj, 632, 872

\bibitem[Rutledge et al.(1997a)]{Rutledge1997a} Rutledge, G.~A., Hesser, J.~E., \& Stetson, P.~B.\ 1997, \pasp, 109, 907

\bibitem[Schiavon et al.(1997)]{Schiavon1997} Schiavon, R.~P., Barbuy, B., Rossi, S.~C.~F., et al.\ 1997, \apj, 479, 902

\bibitem[Schuster et al.(2012)]{Schuster2012} Schuster, W.~J., Moreno, E., Nissen, P.~E., et al.\ 2012, \aap, 538, A21

\bibitem[Searle \& Zinn(1978)]{SearleZinn1978} Searle, L., \& Zinn, R.\ 1978, \apj, 225, 357

\bibitem[Sheffield et al.(2012)]{Sheffield2012} Sheffield, A.~A., Majewski, S.~R., Johnston, K.~V., et al.\ 2012, \apj, 761, 161

\bibitem[Shetrone et al.(2001)]{Shetrone2001} Shetrone, M.~D., C{\^o}t{\'e}, P., \& Sargent, W.~L.~W.\ 2001, \apj, 548, 592

\bibitem[Shetrone et al.(2003)]{Shetrone2003} Shetrone, M., Venn, K.~A., Tolstoy, E., et al.\ 2003, \aj, 125, 684

\bibitem[Sick et al.(2015)]{Sick2015} Sick, J., Courteau, S., Cuillandre, J.-C., et al.\ 2015, Galaxy Masses as Constraints of Formation Models, 82

\bibitem[Simon, \& Geha(2007)]{SimonGeha2007} Simon, J.~D., \& Geha, M.\ 2007, \apj, 670, 313

\bibitem[Smecker-Hane et al.(2002)]{Smecker-Hane2002} Smecker-Hane, T.~A., Cole, A.~A., Gallagher, J.~S., et al.\ 2002, \apj, 566, 239

\bibitem[Smercina et al.(2019)]{Smercina2019arXiv} Smercina, A., Bell, E.~F., Price, P.~A., et al.\ 2019, arXiv e-prints, arXiv:1910.14672

\bibitem[Stanimirovi{\'c} et al.(2004)]{Stanimirovic2004} Stanimirovi{\'c}, S., Staveley-Smith, L., \& Jones, P.~A.\ 2004, \apj, 604, 176

\bibitem[Tissera et al.(2012)]{Tissera2012} Tissera, P.~B., White, S.~D.~M., \& Scannapieco, C.\ 2012, \mnras, 420, 255

\bibitem[Tissera et al.(2013)]{Tissera2013} Tissera, P.~B., Scannapieco, C., Beers, T.~C., et al.\ 2013, \mnras, 432, 3391

\bibitem[Tissera et al.(2014)]{Tissera2014} Tissera, P.~B., Beers, T.~C., Carollo, D., et al.\ 2014, \mnras, 439, 3128

\bibitem[Tollerud et al.(2012)]{Tollerud2012} Tollerud, E.~J., Beaton, R.~L., Geha, M.~C., et al.\ 2012, \apj, 752, 45

\bibitem[Tolstoy et al.(2003)]{Tolstoy2003} Tolstoy, E., Venn, K.~A., Shetrone, M., et al.\ 2003, \aj, 125, 707

\bibitem[Unavane et al.(1996)]{Unavane1996} Unavane, M., Wyse, R.~F.~G., \& Gilmore, G.\ 1996, \mnras, 278, 727

\bibitem[VandenBerg et al.(2006)]{VandenBerg2006} VandenBerg, D.~A., Bergbusch, P.~A., \& Dowler, P.~D.\ 2006, \apjs, 162, 375

\bibitem[van der Marel et al.(2002)]{vanderMarel2002} van der Marel, R.~P., Alves, D.~R., Hardy, E., et al.\ 2002, \aj, 124, 2639

\bibitem[Van der Swaelmen et al.(2013)]{VanderSwaelmen2013} Van der Swaelmen, M., Hill, V., Primas, F., et al.\ 2013, \aap, 560, A44

\bibitem[Vargas et al.(2014a)]{Vargas2014a} Vargas, L.~C., Geha, M.~C., \& Tollerud, E.~J.\ 2014, \apj, 790, 73

\bibitem[Vargas et al.(2014b)]{Vargas2014b} Vargas, L.~C., Gilbert, K.~M., Geha, M., et al.\ 2014, \apjl, 797, L2

\bibitem[Venn et al.(2004)]{Venn2004} Venn, K.~A., Irwin, M., Shetrone, M.~D., et al.\ 2004, \aj, 128, 1177

\bibitem[Weisz et al.(2013)]{Weisz2013} Weisz, D.~R., Dolphin, A.~E., Skillman, E.~D., et al.\ 2013, \mnras, 431, 364

\bibitem[White \& Rees(1978)]{WhiteRees1978} White, S.~D.~M., \& Rees, M.~J.\ 1978, \mnras, 183, 341

\bibitem[Wilson et al.(2019)]{Wilson2019} Wilson, J.~C., Hearty, F.~R., Skrutskie, M.~F., et al.\ 2019, \pasp, 131, 055001

\bibitem[Wojno et al.(2020)]{Wojno2020arXiv} Wojno, J., Gilbert, K.~M., Kirby, E.~N., et al.\ 2020, arXiv e-prints, arXiv:2004.03425

\bibitem[Zolotov et al.(2009)]{Zolotov2009} Zolotov, A., Willman, B., Brooks, A.~M., et al.\ 2009, \apj, 702, 1058

\bibitem[Zolotov et al.(2010)]{Zolotov2010} Zolotov, A., Willman, B., Brooks, A.~M., et al.\ 2010, \apj, 721, 738

\end{thebibliography}
\end{document}